\documentclass[11pt]{article}
\usepackage{epsfig}
\usepackage{color}
\usepackage{html}
\usepackage{rotating}
\usepackage{cite}
\newcommand {\sub}[1]{\ensuremath{_\mathrm{#1}}}

\newcommand {\eps}{\ensuremath{\epsilon}}
\newcommand {\epsplat}{\ensuremath{\eps\sub{0}}}
\newcommand {\epssing}{\ensuremath{\eps\sub{single}}}
\newcommand {\epsevent}{\ensuremath{\eps\sub{event}}}

\newcommand {\lstart}{\ensuremath{l\sub{start}}}
\newcommand {\lstop}{\ensuremath{l\sub{stop}}}

\newcommand {\lmean}{\ensuremath{l\sub{mean}}}
\newcommand {\lmeano}{\ensuremath{l\sub{mean}}}

\newcommand{\ra}{\rightarrow}
\newcommand{\ee}{{\rm e^+e^-}}
\newcommand{\grav}{\tilde {\rm G}}

\newcommand{\nt}{\tilde \chi^0}

\newcommand{\qq}{{\rm q \bar q}}

\newcommand{\dm}{\Delta M}

\newcommand {\slepton}       {\tilde{\ell}}
\newcommand {\slept}         {\tilde{\ell}}
\newcommand {\sle  }         {\tilde{\ell}}

\newcommand {\sel}           {\tilde{\rm e}}

\newcommand {\smu}           {\tilde{\mu}}
\newcommand {\stau}          {\tilde{\tau}}

\newcommand {\chpone}        {\tilde \chi^+_1}
\newcommand {\chnone}        {\tilde \chi^-_1}
\newcommand {\chpnone}       {\tilde \chi^{\pm}_1}
\newcommand {\ntone }        {\tilde \chi^0_1}

\newcommand {\emiss}         {\not \!\! E}
\newcommand {\emist}         {${\not \!\! E}$}

\newcommand {\wpair}         {\mbox{$\mathrm{W}^+\mathrm{W}^-$}}
\newcommand {\W}             {\mbox{$\mathrm{W}$}}
\newcommand {\dedx}          {\mathrm{d}E/\mathrm{d}x}
\newcommand {\pb}            {\ensuremath{\,\mathrm{pb}}}
\newcommand {\keVpcm}        {\,\mathrm{ke}\kern -0.12em\mathrm{V}/\mathrm{cm}}  
\newcommand {\bg}            {\beta\gamma}
\newcommand{\cth}{|\cos{\theta_{\rm miss}}|}

\newcommand{\revis}{E_{\rm vis}/\sqrt{s}}
\newcommand{\acop}{\phi_{\rm acop}}
\newcommand{\degree}{^\circ}

\newcommand{\aro}{\rightarrow}

\newcommand{\ellp}{\ell^+}
\newcommand{\ellm}{\ell^-}

\newcommand{\qqln}{{\mathrm q}\bar{\mathrm q}\ell\bar{\nu}_\ell}

\newcommand{\PZz}{{{\rm Z}^0}}
\newcommand{\PWp}{{{\rm W}^+}}
\newcommand{\PWm}{{{\rm W}^-}}

\definecolor{myyellow}{rgb}{1.000,1.000,0.498}
\definecolor{myorange}{rgb}{1.000,0.647,0.000}
\definecolor{mydarkgreen}{rgb}{0.180,0.545,0.341}
\definecolor{mygreen}{rgb}{0.596,0.980,0.596}
\definecolor{myblue}{rgb}{0.529,0.808,0.980}
\definecolor{myrose}{rgb}{1.000,0.753,0.796}
\definecolor{mypink}{rgb}{0.816,0.125,0.565}
\definecolor{mygrey1}{rgb}{0.3,0.3,0.3}
\definecolor{mygrey2}{rgb}{0.5,0.5,0.5}
\definecolor{mygrey3}{rgb}{0.4,0.4,0.4}
\definecolor{mygrey4}{rgb}{0.6,0.6,0.6}
   \definecolor{MyPink}{rgb}{1.000,0.753,0.796}
   \definecolor{MyDRed}{rgb}{0.800,0.000,0.000}
   \definecolor{MyViolet}{rgb}{0.816,0.125,0.565}
   \definecolor{MyLViolet}{rgb}{0.930,0.510,0.933}
   \definecolor{MyDviolet}{rgb}{0.541,0.168,0.886}
   \definecolor{MyDBlue}{rgb}{0.000,0.000,0.800}
   \definecolor{MyBlue}{rgb}{0.392,0.584,0.929}
   \definecolor{MyLBlue}{rgb}{0.529,0.808,0.980}
   \definecolor{MyLPGreen}{rgb}{0.596,0.980,0.596}
   \definecolor{MyPGreen}{rgb}{0.400,0.804,0.667}
   \definecolor{MyDGreen}{rgb}{0.180,0.545,0.341}
   \definecolor{MyLGreen}{rgb}{0.678,1.000,0.184}
   \definecolor{MyPYellow}{rgb}{1.000,1.000,0.498}
   \definecolor{MyOrange}{rgb}{1.000,0.843,0.000}
   \definecolor{MyBrown}{rgb}{1.000,0.647,0.000}
   \definecolor{MyGold}{rgb}{0.855,0.647,0.125}
   \definecolor{MyLSalmon}{rgb}{1.000,0.627,0.478}
   \definecolor{MySalmon}{rgb}{1.000,0.502,0.447}
   \definecolor{MyBeige}{rgb}{0.961,0.871,0.702}
   \definecolor{MyGreen}{rgb}{0.4,0.72,0.67}
   \definecolor{MyLOrange}{rgb}{1.,0.5,0.0}
   \definecolor{MyPaleOr}{rgb}{1.,0.75,0.3}
   \definecolor{MyLGold}{rgb}{1.,1.,0.6}
   \definecolor{MyDYellow}{rgb}{1.,0.94,0.0}
   \definecolor{MyHSRed}{rgb}{0.8710,0.0,0.0}

\begin{document}

\begin{titlepage}

\begin{center}
  {\large   EUROPEAN ORGANIZATION FOR NUCLEAR RESEARCH}
\end{center}\bigskip

\begin{flushright}
 \large
  CERN-PH-EP/2005-025 \\
  13 June 2005
\end{flushright}
\bigskip
\bigskip
%
%
\begin{center}
    \huge\bf\boldmath
  Searches for Gauge-Mediated Supersymmetry Breaking Topologies
  in e$^+$e$^-$ Collisions at LEP2
\end{center}
\bigskip
\bigskip
\begin{center}
  {\LARGE The OPAL Collaboration}
\end{center}
\bigskip\bigskip
%
%
\begin{center}
\bigskip
\bigskip
\begin{abstract}
Searches were performed for topologies predicted by
gauge-mediated Supersymmetry breaking models (GMSB).
All possible lifetimes of the next-to-lightest SUSY particle (NLSP),
either the lightest neutralino or slepton, 
decaying into the lightest SUSY particle, the gravitino,
were considered.
No evidence for GMSB signatures was found in the OPAL data sample
collected at centre-of-mass energies of $\sqrt{s}=189$--$209\,$GeV at
LEP.
Limits on the product of the production cross-sections and branching
fractions are presented for all search topologies.
To test the impact of the searches,
a complete scan over the parameters of the minimal model of
GMSB was performed.
NLSP masses
below 53.5$\,$GeV/$c^2$ in the neutralino NLSP scenario,
below 87.4$\,$GeV/$c^2$ in the stau NLSP scenario and
below 91.9$\,$GeV/$c^2$ in the slepton co-NLSP scenario
are excluded at 95$\,$\% confidence level
for all NLSP lifetimes.
The scan determines constraints on the universal SUSY mass scale
$\Lambda$ from the direct SUSY particle searches
of $\Lambda>40,\,27,\,21,\,17,\,15$\,TeV/$c^2$ for
messenger indices $N=1,\,2,\,3,\,4,\,5$ for all NLSP lifetimes.
\end{abstract}

\bigskip
\end{center}


\begin{center}{\large 
  (Submitted to Eur. Phys. J C)
}\end{center}

\end{titlepage}

\begin{center}{
G.\thinspace Abbiendi$^{  2}$,
C.\thinspace Ainsley$^{  5}$,
P.F.\thinspace {\AA}kesson$^{  3,  y}$,
G.\thinspace Alexander$^{ 22}$,
G.\thinspace Anagnostou$^{  1}$,
K.J.\thinspace Anderson$^{  9}$,
S.\thinspace Arcelli$^{  2}$,
S.\thinspace Asai$^{ 23}$,
D.\thinspace Axen$^{ 27}$,
I.\thinspace Bailey$^{ 26}$
% ...
}\end{center}\bigskip
\bigskip
$^{  1}$School of Physics and Astronomy, University of Birmingham,
Birmingham B15 2TT, UK
\newline
$^{  2}$Dipartimento di Fisica dell' Universit\`a di Bologna and INFN,
I-40126 Bologna, Italy
\newline
$^{  3}$Physikalisches Institut, Universit\"at Bonn,
D-53115 Bonn, Germany
\newline
% ...
$^{  5}$Cavendish Laboratory, Cambridge CB3 0HE, UK
\newline
% ...
$^{  9}$Enrico Fermi Institute and Department of Physics,
University of Chicago, Chicago IL 60637, USA
\newline
% ...
$^{ 22}$Department of Physics and Astronomy, Tel Aviv University,
Tel Aviv 69978, Israel
\newline
$^{ 23}$International Centre for Elementary Particle Physics and
Department of Physics, University of Tokyo, Tokyo 113-0033, and
Kobe University, Kobe 657-8501, Japan
\newline
% ...
$^{ 26}$University of Victoria, Department of Physics, P O Box 3055,
Victoria BC V8W 3P6, Canada
\newline
$^{ 27}$University of British Columbia, Department of Physics,
Vancouver BC V6T 1Z1, Canada
\newline
% ...
$^{  s}$ now at University of Antwerpen, Physics Department,B-2610 Antwerpen, 
Belgium; supported by Interuniversity Attraction Poles Programme -- Belgian
Science Policy
\newline
% ...
$^{  y}$ now at CERN
\newline
% ...

\tableofcontents
\clearpage

\section{Introduction}
\label{s:intro}
Supersymmetry (SUSY)~\cite{susy} provides a method of solving the 
hierarchy problem by introducing
a set of new particles which cancel the large radiative corrections
to the Higgs mass.
The cancellation is achieved by assuming that, for each
Standard Model (SM)
particle chirality state, there is one additional particle identical
to its SM partner except that its spin differs by 1/2 unit.
If SUSY were an exact symmetry, the new SUSY particles would
have the same masses as their SM partners.  Since this scenario is
experimentally excluded, SUSY must be a broken symmetry.  It is
typically assumed that SUSY is broken in some ``hidden'' sector
of new particles, and is ``communicated'' (or mediated) to the
``visible'' sector of SM and SUSY particles by one of the known
interactions.  The two scenarios for this mediation that have been 
most widely investigated are gravity and gauge mediation.

In gauge-mediated SUSY breaking (GMSB), the
hidden sector can lie at
masses as low as about $10^4$~GeV$/c^2$.
In most current GMSB theoretical work
\cite{theo1,theo2,theo3},
it is assumed that this sector is coupled to a messenger
sector, which in turn
couples to the visible sector through normal SM gauge
interactions. In its minimal version five new parameters and a sign are
introduced in addition to the SM 
parameters, usually chosen to be the SUSY breaking scale $\sqrt{F}$, the 
messenger scale $M$, the messenger index $N$, the 
ratio of the vacuum expectation values of the two 
Higgs doublets $\tan{\beta}$, the sign of the Higgs sector mixing 
parameter sign($\mu$), and the mass scale $\Lambda$, which determines the 
SUSY particle masses at the messenger scale.

A feature which distinguishes gravity-mediated
from gauge-mediated SUSY breaking models is the mass of the gravitino,
$\rm \tilde G$.  In gravity-mediated models, the $\rm \tilde G$ is
usually too heavy to have a significant effect on SUSY phenomenology,
while in GMSB models the $\rm \tilde G$ is typically
light ($<$~1~MeV$/c^2$) and is the lightest SUSY particle, the LSP.
While the $\rm \tilde G$ is a spin 3/2 particle, only its $\pm$ 1/2 spin
projections (which have absorbed the goldstino associated with
spontaneous SUSY breaking via the ``superhiggs'' mechanism~\cite{superh})
interact
with weak, rather than gravitational, strength interactions, and
contribute to phenomenology. 

In GMSB, the next-to-lightest SUSY particle (NLSP) is 
either the lightest neutralino, $\tilde \chi^0_1$ (neutralino NLSP
scenario), or the
lightest scalar lepton\,\footnote
{For the first and 
second generations in the minimal GMSB model, the lighter state is 
predominantly right-handed, since the off-diagonal components 
of the mass matrix are small due to small Yukawa couplings. Therefore in the 
following for the selectron and the smuon the symbol $\tilde \ell_R $ 
instead of $\tilde \ell_1$ is often used.}, $\rm \tilde \ell^\pm_1$,
which is either the mass degenerate $\sel_R$, $\smu_R$ and $\stau _1$
(slepton co-NLSP scenario) or 
in a 
significant fraction of the parameter space the lightest scalar
tau lepton, $\rm \tilde \tau_1$ (stau NLSP scenario).
For SUSY particles heavier than the NLSP,  
the coupling to the $\grav$ is very small, and
typically they will decay to the NLSP, which then
decays to the gravitino via $\nt_1\ra\gamma\grav$ or
$\slepton\ra\ell\grav$.

One feature of GMSB is that the decay length $\beta\gamma c\tau$ of the
NLSP depends
on the intrinsic SUSY breaking scale $\sqrt{F}$:
\begin{equation}
\label{eq:lifetime}
\beta\gamma c\tau \propto \left(\frac{100\, \mathrm{GeV}/c^2}
{m_{\rm NLSP}}\right)^5\left(\frac{\sqrt{F}}{100\, \mathrm{TeV}}\right)^4
\left(\frac{E_{\rm NLSP}^2}{m_{\rm NLSP}^2}-1\right)^{1/2}
\mathrm{cm}\, ,
\end{equation}
where $m_{\rm NLSP}$ and $E_{\rm NLSP}$ are mass and energy of
the NLSP. Due to the range allowed for $\sqrt{F}$, from 2000$\,$TeV$/c^2$
down to 100$\,$TeV$/c^2$ \cite{fconstr}, the decay length is
effectively unconstrained and decays inside and outside the detector volume
have to be considered.
If the NLSP decay to the gravitino occurs with a small lifetime,
event signatures will include
energetic leptons or photons, plus
significant missing energy, \emist , due to the undetected gravitinos.
For intermediate lifetimes, the events may include tracks not pointing to
the primary interaction point (tracks with large impact parameters)
or kinked tracks.
For long lifetimes, the signatures can include tracks from heavy long-lived
charged particles.

\renewcommand{\arraystretch}{1.2}
\begin{table}[htp!]
\begin{center}
\begin{tabular}{lcllc}
\hline
               & Sparticle        &        &             &              \\
Scenario       & production       & Signal & NLSP decay  & Reference    \\
               & (and decay)      &        &             &              \\
\hline
\hline
$\stau_1$ NLSP   & $\stau_1^+\stau_1^-$ & $\tau^+\tau^-\emiss$               &
       prompt         & \cite{Abbiendi:2003ji}\\ 
($\stau_1\ra\tau\grav$) & ($\stau_1\ra\tau\grav$) & $\stau\ra\tau\grav$ large I.P.     &
       within detector  & Sect.\ref{s:largeip1}\\ 
               &                  & $\stau\ra\tau\grav$ kinked tracks  &
       within detector  & Sect.\ref{s:kink2}\\ 
               &                  & $\stau$ tracks, high $\dedx$       &
       outside detector & \cite{Abbiendi:2003yd}\\ \cline{2-5}
               & $\ntone\ntone$   & $\tau^+\tau^-\tau^+\tau^-\emiss$   &
       prompt         & Sect.\ref{s:multil}\\ 
               & ($\ntone\ra\stau_1\tau$) & $\stau\ra\tau\grav$ large I.P.     &
       within detector  & Sect.\ref{s:largeip3}\\ 
               &                  & $\stau\ra\tau\grav$ kinked tracks  &
       within detector  & Sect.\ref{s:kink3}\\ 
               &                  & $\stau$ tracks, high $\dedx$       &
       outside detector & Sect.\ref{s:stable2}\\ \cline{2-5}
               & $\chpone\chnone$ & $\tau^+\tau^-\emiss$               &
       prompt         & \cite{Abbiendi:2003ji}\\ 
               & ($\chpnone\ra\stau_1\nu$) & $\stau\ra\tau\grav$ large I.P.     &
       within detector  & Sect.\ref{s:largeip1}\\ 
               &                  & $\stau\ra\tau\grav$ kinked tracks  &
       within detector  & Sect.\ref{s:kink2}\\ 
               &                  & $\stau$ tracks, high $\dedx$       &
       outside detector & Sect.\ref{s:stable2}\\ \cline{2-5}
               & $\smu_R^+\smu_R^-$,  & $\mu^+\mu^-\tau^+\tau^-\tau^+\tau^-\emiss$ &
       prompt         & Sect.\ref{s:multil}\\ 
               &  $\sel_R^+\sel_R^-$  & ${\rm e}^+{\rm e}^-\tau^+\tau^-\tau^+\tau^-\emiss$ &
       prompt         & Sect.\ref{s:multil}\\ 
               & ($\sle_R\ra\ell\ntone$, & $\stau\ra\tau\grav$ large I.P.     &
       within detector  & Sect.\ref{s:largeip4}\\ 
               & $\ntone\ra\stau_1\tau$) & $\stau\ra\tau\grav$ kinked tracks  &
       within detector  & Sect.\ref{s:kink4}\\ 
               &                  & $\stau$ tracks, high $\dedx$       &
       outside detector & Sect.\ref{s:stable2}\\

\hline
$\sle_R$ co-NLSP & $\sle_R^+\sle_R^-$   & $\ell^+\ell^-\emiss$               &
       prompt         & \cite{Abbiendi:2003ji}\\ 
($\sle_R\ra\ell\grav$) & ($\sle_R\ra\ell\grav$) & $\sle\ra\ell\grav$ large I.P.     &
       within detector  & Sect.\ref{s:largeip1}\\ 
               &                  & $\sle\ra\ell\grav$ kinked tracks  &
       within detector  & Sect.\ref{s:kink2}\\ 
               &                  & $\sle$ tracks, high $\dedx$       &
       outside detector & \cite{Abbiendi:2003yd}\\ \cline{2-5}
               & $\ntone\ntone$   & $\ell^+\ell^-\ell^+\ell^-\emiss$   &
       prompt         & Sect.\ref{s:multil}\\ 
               & ($\ntone\ra\sle_R\ell$) & $\sle\ra\ell\grav$ large I.P.     &
       within detector  & Sect.\ref{s:largeip3}\\ 
               &                  & $\sle\ra\ell\grav$ kinked tracks  &
       within detector  & Sect.\ref{s:kink3}\\ 
               &                  & $\sle$ tracks, high $\dedx$       &
       outside detector & Sect.\ref{s:stable2}\\ \cline{2-5}
               & $\chpone\chnone$ & $\ell^+\ell^-\emiss$               &
       prompt         & \cite{Abbiendi:2003ji}\\ 
               & ($\chpnone\ra\sle_R\nu$) & $\sle\ra\ell\grav$ large I.P.     &
       within detector  & Sect.\ref{s:largeip1}\\ 
               &                  & $\sle\ra\ell\grav$ kinked tracks  &
       within detector  & Sect.\ref{s:kink2}\\ 
               &                  & $\sle$ tracks, high $\dedx$       &
       outside detector & Sect.\ref{s:stable2}\\
\hline
\end{tabular}
\end{center}
\caption{
\label{t:analysen1}
GMSB signatures for the stau NLSP and the slepton co-NLSP scenario.
The notation (1) ``prompt'', (2) ``within detector'' and (3) ``outside
detector'' refers to a NLSP decay such that the decay vertex is
(1) close to the interaction region and not measurably displaced,
(2) resolvable by the large impact parameter and kinked track searches and
(3) well outside the detector.
For each signature a reference to the description of the analysis is given,
either to a section of this or to a separate paper.}
\end{table}

\renewcommand{\arraystretch}{1.2}
\begin{table}[ht!]
\begin{center}
\begin{tabular}{lcllc}
\hline
               & Sparticle        &        &             &              \\
Scenario       & Production       & Signal & NLSP decay  & Reference    \\
               & (and decay)      &        &             &              \\
\hline
\hline
$\ntone$ NLSP  & $\ntone\ntone $   & $\gamma\gamma\emiss $ & within detector & \cite{photons}\\ 
($\ntone\ra\gamma\grav$) & ($\ntone\ra\gamma\grav$) & $\emiss$               & outside detector & ---\\
\cline{2-5}
               & $\sle_R^+\sle_R^-$   & $\ell^+\ell^-\gamma(\gamma)\emiss$ &
                within detector  & Sect.~\ref{s:gamgamx}\\ 
               & ($\sle_R\ra\ntone\ell$) & $\ell^+\ell^-\emiss$               &
                outside detector & \cite{Abbiendi:2003ji}\\ \cline{2-5}
       & $\chpone\chnone$ & ${\rm q_i \bar q_j}{\rm q_k \bar q_l}\gamma(\gamma)\emiss$ &
                within detector  & Sect.~\ref{s:gamgamx}\\
       & ($\chpnone\ra\ntone{\rm W^{\pm}}$)& ${\rm q_i \bar q_j}\ell^{\pm}\gamma(\gamma)\emiss$ &
                within detector  & Sect.~\ref{s:gamgamx}\\
       &                  & $\ell^+\ell^-\gamma(\gamma)\emiss$ &
                within detector  & Sect.~\ref{s:gamgamx}\\
       &                  & ${\rm q_i \bar q_j}{\rm q_k \bar q_l}\emiss$ &
                outside detector  & \cite{Abbiendi:2003sc}\\
       &                  & ${\rm q_i \bar q_j}\ell^{\pm}\emiss$ &
                outside detector  & \cite{Abbiendi:2003sc}\\
       &                  & $\ell^+\ell^-\emiss$ &
                outside detector  & \cite{Abbiendi:2003ji}\\\cline{3-5}
       & $\chpone\chnone$ & $\ell^+\ell^-\gamma(\gamma)\emiss$ &
                within detector  & Sect.~\ref{s:gamgamx}\\
       & ($\chpnone\ra\sle_R\nu$, & $\ell^+\ell^-\emiss$ &
                outside detector  & \cite{Abbiendi:2003ji}\\
       & $\sle_R\ra\ntone\ell$) & & & \\

\hline
\end{tabular}
\end{center}
\caption{
\label{t:analysen2}
GMSB signatures for the neutralino NLSP scenario.
The notation (1) ``within detector'' and (2) ``outside
detector'' refers to a NLSP decay such that the decay vertex is
(1) at the interaction region (prompt decay) up to a few tenths of centimetres
inside the detector volume and
(2) well outside. For each signature a reference to the description
of the analysis is given, either to a section of this
or to a separate paper.}
\end{table}

In this paper, we present results on searches
for GMSB signatures with either a $\nt_1$
or a $\tilde\ell_1$ (or $\stau_1$) NLSP using the
data sample acquired by the OPAL detector
at centre-of-mass energies of $\sqrt{s}=189$--$209\,$GeV.
All relevant signatures -- direct NLSP pair-production and its appearance
in the decay chains of heavier SUSY particles like charginos,
neutralinos or sleptons -- for all NLSP lifetimes are considered in our
searches. Topologies and sparticle decay modes together with a reference
to the description of the analyses
are given in Table~\ref{t:analysen1} for the slepton and stau
NLSP scenarios and in Table~\ref{t:analysen2} for the neutralino NLSP scenario.

The paper is organized as follows. The OPAL detector, data and Monte Carlo
samples are introduced in Section~\ref{s:datamc},
followed by a brief description of the event selections of
different topologies in the slepton and stau NLSP scenarios in
Section~\ref{s:searches_staunlsp} and
for the neutralino NLSP scenario in
Section~\ref{s:searches_neutnlsp}.
The strategy to combine such a large number of analyses is described in
Section~\ref{s:strategy}.
Results are presented in Section~\ref{s:expresults},
and finally in Section~\ref{s:interpretation} constraints
on the parameters of the minimal GMSB model are discussed.

\section{Data and Monte Carlo samples}
\label{s:datamc}
A detailed description of the OPAL detector can be found elsewhere
\cite{detector}.
Tracking of charged particles was performed by a central detector, enclosed
in a solenoid which provided a uniform axial magnetic field of 0.435$\,$T. 
The central detector consisted of a two-layer
silicon microvertex detector~\cite{silicon}, a high precision vertex chamber 
with both axial and stereo wire layers, a large volume jet chamber providing
both tracking and ionization energy loss information, and additional chambers
to measure the $z$ coordinate of tracks as they leave the central
detector\footnote{A right-handed coordinate system 
is adopted, where the $x$-axis points toward the centre of the LEP ring and 
the $z$-axis points in the direction of the electron beam. The polar angle 
$\theta$ and the azimuthal angle $\phi$ are defined w.r.t. $z$ and $x$, 
respectively.}.
These detectors provided tracking coverage for polar angles
$|\cos{\theta}|<0.96$, with a typical transverse momentum ($p_T$) resolution
of $\rm\sigma_{p_T}/p^2_T = 1.25\times 10^{-3}\,(GeV/c)^{-1}$. The solenoid coil was
surrounded by a time-of-flight counter array and a barrel lead-glass
electromagnetic calorimeter with a pre-sampler. Together with the
endcap electromagnetic calorimeters, the lead-glass blocks covered the
range $|\cos{\theta}|<0.98$. Outside the electromagnetic calorimetry,
the magnet return yoke was instrumented with streamer tubes to form a
hadronic calorimeter with angular coverage in the range $|\cos{\theta}|<0.91$.
The region $0.91<|\cos{\theta}|<0.99$ was instrumented with an additional 
pole-tip hadronic calorimeter using multi-wire chambers. 
The detector 
was completed with muon detectors outside the magnet return yoke.
These were composed of drift chambers in the barrel region and limited
streamer tubes in the endcaps, and together covered 93$\,$\% of the full
solid angle.

The data used for the new particle searches described here were taken at
centre-of-mass energies between 189$\,$GeV and 209$\,$GeV during the LEP 
running periods from 1998 to 2000. 
The total integrated luminosity is about 
600$\,$pb$^{-1}$, evaluated
using small angle Bhabha scattering
events observed in the forward calorimeters \cite{Abbiendi:2003dh}
with an error of about 0.3\%.  
The data were analyzed in 8 independent
bins in $\sqrt{s}$, summarized in Table~\ref{t:listresults}.
Large samples of Monte Carlo simulated events were generated 
at each $\sqrt{s}$
to optimize
the search algorithms for new particles and were used 
to evaluate their efficiency and
the number of expected events from Standard Model sources.


Signal events were generated for all search topologies
with 10 different NLSP lifetimes ranging from a very short lifetime,
$\tau = 10^{-12}\,$s, 
corresponding to a prompt NLSP decay with a decay length
of 300$\,\mu$m for a particle with $\beta\gamma=p/m=1$, 
up to very long lifetimes, $\tau=\,10^{-6}\,$s,
corresponding to a stable NLSP with a decay length of 300$\,$m 
which always decays outside the detector. 
Mass grids from 45~GeV$/c^2$ to the kinematic limit
with 10--100 points per channel per $\sqrt{s}$ bin
and 1000~events per grid point
were used for the full simulation of signal Monte Carlo events.
The SUSYGEN \cite{susygen} event generator was used to simulate
the signal events for $\slepton_R$ and $\nt_1$ pair-production.
For chargino pair-production
the W and Z boson widths can play an important
role and are not fully treated in SUSYGEN.
The DFGT generator \cite{DFGT} 
was used to simulate these signal events.
It includes spin correlations and
allows for a proper treatment of the W boson and the Z boson 
width effects in the chargino decay.
Both SUSYGEN and DFGT include initial-state radiation (ISR).
The Lund string fragmentation scheme of
PYTHIA in the JETSET package~\cite{pythia}
was used to describe the hadronization.
The gravitino mass was set to zero in the generation,
since a small mass up to $\mathcal{O}$(MeV) favoured by the models
has a negligible effect on the detection efficiencies.
The GEANT3 \cite{geant} software package, which was used to simulate
the transport of particles through the material of the detector,
only allows isotropic two- or three-body decays of long-lived particles.
Thus an interface between GEANT and the event generators was created 
to obtain correct angular and energy distributions for the decay
products of the NLSP, which decays in flight.

The simulated Standard Model background is composed of two-photon
($\ee \ra \ee \gamma \gamma  \ra \ee  f \bar{f}$),
four-fermion ($\ee \ra \ f \bar{f} f \bar{f}$),
multi-photon ($\ee \ra \gamma \gamma (\gamma)$)
and two-fermion
($\ee \ra (Z/\gamma)^* \ra l \bar{l}$
and $\ee \ra (Z/\gamma)^* \ra q \bar{q}$)
final states.
Leptonic two-photon processes were simulated with the BDK \cite{bdk} and
Vermaseren \cite{vermaseren} programs, and
hadronic two-photon processes with the PHOJET \cite{phojet} and 
HERWIG \cite{herwig} programs. To simulate four-fermion processes, 
the KORALW \cite{koralw} generator (which simulates ISR
using a spectrum with transverse momentum) and,
for final states with electrons, 
the grc4f \cite{grc4f} program (with collinear ISR) were used. 
Multi-photon final states were simulated with the RADCOR \cite{radcor} program.
For Bhabha events, the BHWIDE \cite{bhwide} and TEEGG \cite{teegg} generators
were used. 
Muon- and tau-pairs were simulated using the KK2F~\cite{kk2f} program
and $\nu\nu (\gamma)$ events with the NUNUGPV \cite{nunugpv} generator.
Multi-hadronic events were simulated using PYTHIA and KK2F.
The equivalent luminosities of the background Monte Carlo samples
were in most cases much larger than the luminosity of the data;
however, in a few background samples (especially the high cross-section
two-photon processes), the equivalent luminosity of the background
Monte Carlo was only about $2\times$ that of the data.
Final states with six or more
primary fermions
were not included in the background Monte Carlo
samples but they are expected to make a negligible contribution.

The simulated signal and background events were processed through the full 
simulation of the 
detector, and the same analysis chains were 
applied to Monte Carlo events as to the data. 
The statistical errors on the signal and background Monte Carlo
are included in all results.
All search analyses described
in this paper used the same sets of simulated background
events.

\section{\boldmath Searches in the stau NLSP and slepton co-NLSP scenarios} 
\label{s:searches_staunlsp}
In the scenarios where mass-degenerate charged sleptons or the stau
are the next-to-lightest
SUSY particles, they decay to their charged Standard Model partner and the
electrically neutral, lightest SUSY particle, the gravitino.
Depending on the NLSP lifetime, different analyses were
applied.
In the following section all searches for SUSY particles in the
stau NLSP and the slepton co-NLSP scenario are presented. 
In Table~\ref{t:listresults} all search results not described in
other papers are summarized.
 
\subsection{Searches for promptly decaying charged NLSPs}
\label {s:kap4_prompt}
The decay vertex of an
NLSP with a lifetime less than 10$^{-11}\,$s
is close to the interaction region and 
considered not to be displaced from the primary vertex.
Depending on whether the NLSP is directly pair-produced or appears 
in the decay chain of heavier SUSY particles, two or more charged
leptons are expected in an event and different search algorithms were optimized
to select candidate events with different track multiplicities.

\subsubsection{Search for di-lepton events with missing transverse
               momentum}
\label{s:accop}
Candidate events for pair-produced NLSPs, decaying promptly into a lepton and
a gravitino ($\ee\ra\slept_R^+ \slept_R^-$, $\slept_R\ra\ell\grav$),
should have the signature of two leptons plus
missing transverse momentum.
The search for events of this type
is described in \cite{Abbiendi:2003ji}.
The number of di-lepton events, the dependence on centre-of-mass energy
and the event properties are consistent with expectations from Standard Model
processes, predominantly \wpair\ production with both W bosons 
decaying leptonically. No evidence for new phenomena is apparent.
The topology of two leptons plus
missing transverse momentum can also result from the pair-production
of charginos which decay to an invisible neutrino
and the charged NLSP 
($\ee\ra\chpone\chnone$, $\chpnone\ra\slept_R\nu$, $\slept_R\ra\ell\grav$). 
Thus the same analysis as for pair-produced
NLSPs is applied.

\subsubsection{Selection of events with more than two leptons and missing
               energy}
\label{s:multil}
The NLSP might appear in the decay chain of a heavier pair-produced
SUSY particle, a neutralino or a heavy slepton. In such cases
in addition to the two leptons from the prompt decays of both NLSPs,
two or four more leptons are expected in the detector
($\ee\ra\ntone\ntone$, $\ntone\ra\sle_R\ell$, $\slept_R\ra\ell\grav$\ \ or
\ \ $\ee\ra\sle_R^+\sle_R^-$, $\sle_R\ra\ell\ntone$, $\ntone\ra\tau\stau$,
$\stau\ra\tau\grav$).  

An analysis to search for such topologies with multi-leptons and missing
energy which performs no explicit electron or muon
identification was designed
to maintain simplicity, minimize systematic
errors and maximize efficiency for all
lepton flavours.
After the event reconstruction,
double-counting of energy between tracks and calorimeter
clusters was corrected by reducing the calorimeter cluster energy
by the expected energy deposition from aligned
charged tracks \cite{ref:MT}, including particle identification
information.
Charged tracks and neutral
clusters were defined to be of ``good'' quality using
the requirements of \cite{ref:MT}.

The analysis was split into three steps:
(1) preselection of events with low to intermediate
        multiplicity and missing energy;
(2) additional kinematic requirements to remove unmodeled backgrounds;
(3) requirements that each event contains at least four
        jets, each consistent with being an isolated lepton (including a tau).

\paragraph{(1) Preselection}
  \begin{itemize}
    \item[(P1)] The number of good charged tracks
          $N_{\rm trk}$ in the event satisfied
          $4 \leq N_{\rm trk} \leq 10$.
          This cut removed
          a large fraction of the lepton pair and multihadronic
          backgrounds.
    \item[(P2)] Events that may include charged tracks originating from 
          noise in the
          central jet chamber were removed by vetoing events
          containing any tracks with ${\rm d}E/{\rm d}x<5$ keV/cm.
    \item[(P3)] $N_{\rm trk}/N_{\rm trk}^{\rm tot}>0.20$, where
          $N_{\rm trk}^{\rm tot}$ is the
          total number of charged tracks reconstructed in the event.
          This reduced the number of 
          events due to machine backgrounds ({\it eg.}
          beam-gas and beam-wall interactions).
    \item[(P4)] Cosmic rays were vetoed using information from the 
          tracking and time-of-flight systems.
    \item[(P5)] $M_{\rm vis}>5.0$ GeV$/c^2$, where $M_{\rm vis}$ is the visible
          mass of the event.  This cut was required to remove events
          which were not simulated in the untagged two-photon Monte
          Carlo generation.
    \item[(P6)] $P_{\rm T,miss}>0.05\times\sqrt{s}$, where $P_{\rm T,miss}$
          is the magnitude of the component of event momentum
          transverse to the beam axis.  
          This
          greatly reduced the number of two-photon and two-fermion
          events.
    \item[(P7)] Cuts were made on the maximum energy deposited in
          either side of the forward calorimeters
          in order to reduce two-photon and some machine backgrounds.
          These cuts have an associated efficiency loss due to
          activity which may be present in these subdetectors for
          signal (or background) events, which was evaluated with
          random beam-crossing triggers.  The typical efficiency loss
          was about 3\%. 
  \end{itemize}

\begin{figure}[H]
     \begin{minipage}[t]{.1cm}
       \epsfxsize=.1cm
     \end{minipage}
   \hfill
     \begin{minipage}[t]{15.5cm}
       \epsfxsize=15.5cm
       \epsfbox{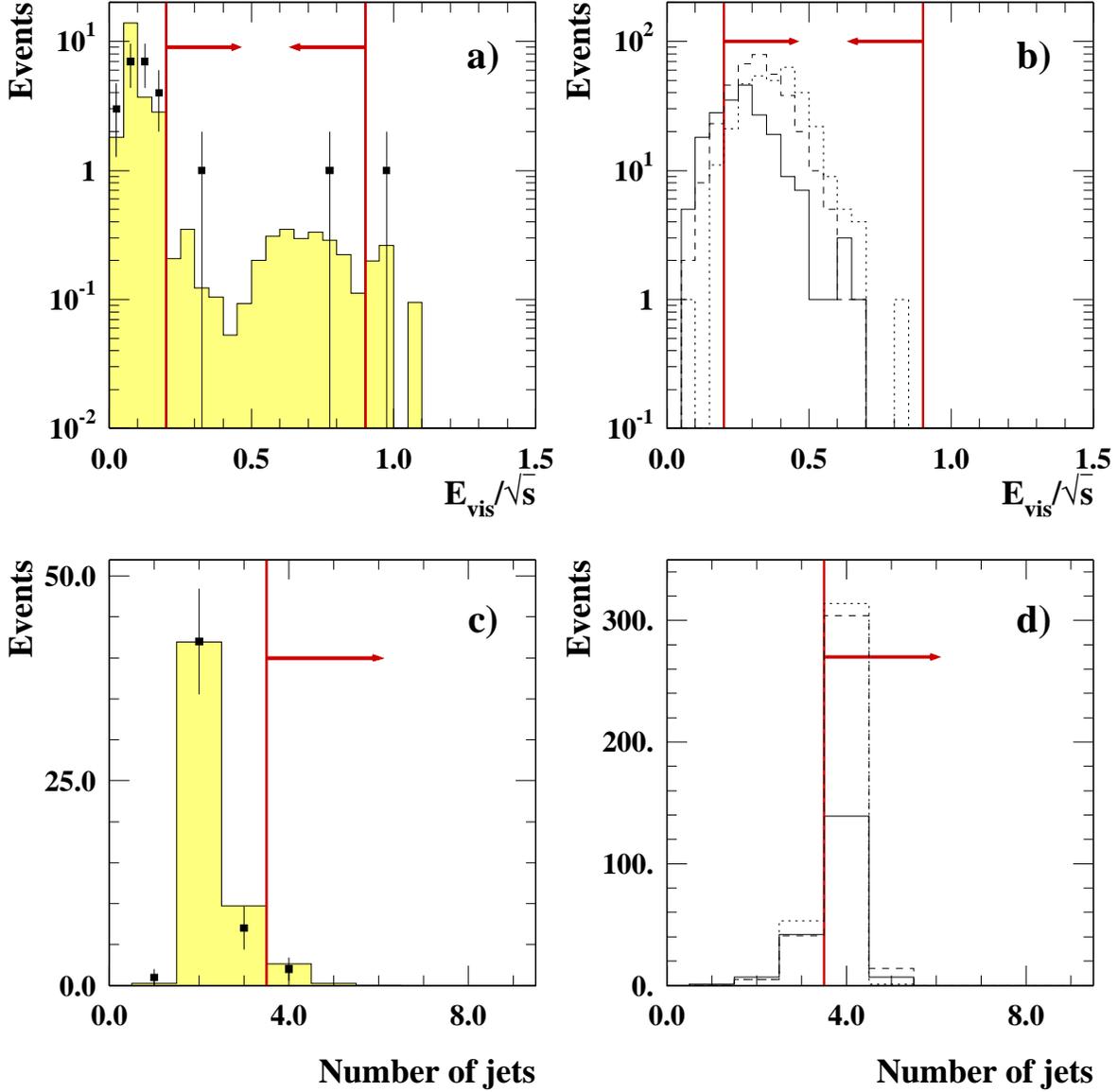}
     \end{minipage}
   \hfill
     \begin{minipage}[t]{.1cm}
       \epsfxsize=.1cm
     \end{minipage}
\caption{
\label{f:mutigcuts}
  \protect{
    Distributions of the
    event visible energy (a, b) and
    the number of low multiplicity jets consistent
    with having originated from leptons (c, d)
    in the search 
    for multi-leptons with missing energy 
    described in Section~\ref{s:multil}.
    The distributions are plotted with all other
    cuts applied.  In (a) and (c) the points represent the selected data
    events from the entire 189--209~GeV data sets,
    while the shaded regions show the expectation
    from Standard Model backgrounds.  
    Plots (b) and (d) show examples of 
    distributions with arbitrary scale for 
    signal events for
    $\ee\ra\nt_1\nt_1\ra\stau\tau\stau\tau$ with the prompt decay
    $\stau\ra\tau\grav$ at $\sqrt{s}=206.0$~GeV .
    The solid, dashed and dotted lines correspond to
    $(\nt_1,\stau)$ mass combinations of
    (50,45), (85,65) and (102,45) GeV$/c^2$, respectively.
    The arrows indicate the cut positions.
  }
}
\end{figure}

\paragraph{(2) Kinematic requirements}
\ \\
    There is an excess of events surviving the preselection.
    This excess does not have
    the characteristics of any of the signal hypotheses considered
    in this paper, and
    it was removed with
    the following requirements.
    \begin{itemize}
      \item[(K1)] $\phi_{\rm acop}>10^\circ$.\  The acoplanarity
            angle $\phi_{\rm acop}$ was calculated by forcing the event into
            two jets using the Durham jet finder \cite{durham}, and
            subtracting the opening angle between the two jets in the
            plane transverse to the beam axis
            from $180^\circ$.\  This cut removed some surviving
            background from lepton pair production and two photon events.
      \item[(K2)] $|\cos{\theta_{\rm miss}}|<0.90$, where
            $\theta_{\rm miss}$ is the polar angle of the event
            missing momentum vector.  This cut removed some residual
            surviving Bhabha scattering and two photon background.
      \item[(K3)] $0.20 < E_{\rm vis}/\sqrt{s} < 0.90$
        where $E_{\rm vis}$ is the event visible energy.
        This cut primarily removed two-photon background.
        The distributions of the event visible energy in data and
        simulated background are illustrated
        in Figure~\ref{f:mutigcuts}\,(a) and (b).

    \end{itemize}

\paragraph{(3) Lepton-jet identification}
\ \\
    The events surviving the preselection and kinematic
    requirements were split up into jets. The jets
    were then required to be consistent with having originated from electrons,
    muons, or tau-leptons.  The signal hypotheses considered
    here are most often dominated by tau-leptons, and no
    explicit electron or muon identification tools are used.
    The event was split into jets using the Durham jet
    finder \cite{durham} with $y_{\rm cut}=0.0004$.
    This $y_{\rm cut}$ was chosen as the smallest value
    (to maximize efficiency for finding
    leptons which are close to each other) which would
    not split single 3-prong tau-lepton jets
    into multiple jets.  The following cuts were then applied.
  \begin{itemize}
    \item[(1)] $N_{\rm\ell-jet}\geq4$, where
          $N_{\ell-jet}$ is the number of jets in the event
          compatible with being from a parton-level lepton:\\
            -- Lepton-jet momentum satisfies $0.01 \leq P_{\rm\ell-jet}/E_{\rm beam} \leq 0.80$;\\
            -- Lepton-jet mass satisfies $M_{\rm\ell-jet} \leq 5$~GeV$/c^2$;\\
            -- Number of charged tracks satisfies $1\leq N_{\rm trk} \leq 3$.\\
          The distributions of the number of lepton-jets in data and
          simulated background 
          are illustrated
          in Figure~\ref{f:mutigcuts}\,(c) and (d).
    \item[(2)] $\theta_{\rm isol}>10^\circ$\ 
          where $\theta_{\rm isol}$ is the smallest opening angle
          between any two lepton-jets.
    \item[(3)] $\Sigma \rm |\cos\theta_{\rm\ell-jet}|\leq 3$, where
          $\Sigma\rm|\cos\theta_{\rm\ell-jet}|$ is the sum of the magnitudes of
          the cosines of the polar angles of the lepton-jet candidates.  
          This cut
          removed a small residual amount of $\rm \ee\ra q\bar q$
          background.
    \item[(4)] $E_{\rm sum}(\rm non~\ell-jet)/E_{\rm vis}<0.20$,
          where $E_{\rm sum}(\rm non~\ell-jet)/E_{\rm vis}$ is the fraction of
          the event visible energy which is not contained in a lepton-jet
          candidate.
  \end{itemize}

The full analysis retains efficiency for the GMSB processes
$\nt_1\nt_1\ra\slepton\ell\slepton^\prime\ell^\prime$, both for
the case where the neutralino decays with equal branching ratios
into all three slepton generations (equal BR),
and also when it decays
to $\stau\tau$ with a 100$\,$\% branching fraction.
Typically when the mass
difference $M(\nt_1)-M(\slepton)$ is above 5$\,$GeV$/c^2$, the
efficiency for the equal BR scenario is 40--50$\,$\%, while
for the 100$\,$\% $\stau\tau$ decay it is 30--40$\,$\%.
For $M(\nt_1)-M(\slepton)=5$~GeV$/c^2$, the corresponding efficiencies
are typically 30$\,$\% and 15$\,$\%.  For $M(\nt_1)-M(\slepton)$ as
low as 2~GeV$/c^2$, the equal BR scenario retains about 10$\,$\% efficiency,
while the very soft recoil tau leptons make the 100$\,$\% $\stau\tau$ BR
have about 1\% efficiency.
The full selection yields two events in the data with
$2.8\pm0.3$(stat)$^{+0.9}_{-0.3}$(syst) expected from the
background.

To search for selectron and smuon pair-production, 
the final analysis cut (on the fraction of energy in the event
which is not included in a lepton candidate jet) has relatively
poor efficiency.  This is due to the 6 leptons in the event,
any of which could fail the lepton-jet classification.
The optimal search for that process therefore excludes
the final cut, and selects 5 events in the data with
$4.2\pm0.3$(stat)$^{+1.4}_{-0.4}$(syst) expected in the
background.
The mass grid in this mode is three dimensional, since
the decay proceeds via $\slepton\ra\ell\nt\ra\ell\stau\tau$,
and therefore the neutralino mass is also important with
the efficiency depending strongly on both $M(\slepton)-M(\nt_1)$
and $M(\nt_1)-M(\stau)$.
The detection efficiency
for $\sel^+\sel^-$ and $\smu^+\smu^-$ are similar, with the
$\smu^+\smu^-$ process always slightly higher.  If all mass
differences in the process are $>$5~GeV$/c^2$, the detection efficiencies
are about 50$\,$\%.  For $M(\slepton)-M(\nt_1)$ as low as 2~GeV$/c^2$
(soft electrons or muons) the efficiency is about 35--45$\,$\%, while
for $M(\nt_1)-M(\stau)$ as low as 2~GeV$/c^2$ (soft taus) the efficiency
is typically reduced to about 15$\,$\%.

\paragraph{3.1.2.1 Systematic uncertainties}
\ \\
\ \\
The dominant background in the selection is from
non-multiperipheral 4-fermion production (which
in our simulations is almost entirely due to the production
of four charged leptons), constituting 80$\,$\% of the total background
expectation.
A smaller contribution of about 15$\,$\% is expected from
the multiperipheral diagrams $\ee\ra\ee\ell^+\ell^-$
(about 95$\,$\% from $\ee\ra\ee\tau^+\tau^-$).
The remaining 5$\,$\% is from $\ee\ra\tau^+\tau^-$.
All other background processes are negligible.
The Monte Carlo modeling of the background was tested
using different event generators.
Of particular
concern is the fact that, in our principal background
samples available at all centre-of-mass energies, the
multiperipheral diagrams are treated with specialized
``two-photon'' event generators, which neglect interference
with non-multiperipheral diagrams.  Special samples of
the full set of $\ee\ra\ee\ell^+\ell^-$ diagrams,
including interference,
were prepared using grc4f2.2~\cite{grc4f}
at $\sqrt{s}=206$~GeV to study this effect.
The background
using the full set of $\ee\ell^+\ell^-$ diagrams including
interference changes by $(-41\pm5)$$\,$\% after the preselection
compared to using our standard set of Monte Carlo generators.
After all cuts, the result is $(+20\pm20)$$\,$\%.
The
systematic error for this effect is taken to be 40$\,$\% on
the $\ee\ell^+\ell^-$ component of the background.
The modeling of the 4-fermion background, excluding
$\ee\ell^+\ell^-$, was tested by comparing grc4f
to our primary event generator KORALW4f.
After the preselection, the background from grc4f
changes by
$(+11\pm4)$$\,$\%, while after all cuts it 
changes by $(-10\pm30)$$\,$\%.
A 10$\,$\% systematic error is assigned to the non $\ee\ell^+\ell^-$
4-fermion background from this check.
The modeling of $\ee\ra\tau^+\tau^-$ was checked comparing
the predictions of KK2f and KoralZ.  The difference
was negligible ($<1$$\,$\%) and is neglected.

The modeling of the cut variables by the detector
simulation can introduce additional systematic errors.
This modeling was checked by comparing data samples with ``relaxed cuts'',
in which any signal contribution is expected to be negligible,
with Monte Carlo simulation to determine the uncertainty on each
cut variable.
Each cut position was varied in each direction by this uncertainty,
and the background level and signal efficiency re-evaluated.  
The lepton-jet identification was checked using different
jet finders and jet resolution parameters, but there
was no indication for any systematics from that procedure
and no additional errors are assigned.
From these studies, the systematic error on the background
due to the Monte Carlo modeling of the cut variables
is taken as +31.5/--5.6$\,$\%, and on the relative signal efficiency
as +4.2/--4.0$\,$\%, which is symmetrised to $\pm$4$\,$\%.


\subsection{Searches for charged NLSPs with short lifetime
            (tracks with large impact parameters)}
\label{s:largeip}

If the NLSP is a slepton and its lifetime is such that the decay happens 
within a few centimetres of the primary interaction point, no NLSP track
can be reconstructed.
Only secondary tracks of its decay products are visible and they have large
impact parameters with respect to the primary interaction point.
In the following section the analyses searching for large impact parameter
tracks consistent with short-lived NLSPs with lifetimes between roughly 
10$^{-11}\,$s and 10$^{-9}\,$s are discussed.

All search topologies expected from SUSY particles in the stau or
slepton co-NLSP scenarios have in common that, for short NLSP lifetimes,
at least two tracks with large impact parameters would be present. The
number of additional tracks depends on the pair-produced SUSY particle.
The number of good tracks \cite{ref:MT}
in the event was determined, and
then all tracks in the event were classified into one of three
categories.

\begin{itemize}
  \item[] {\bf Primary tracks:} Good tracks with a distance from 
          the point of closest approach to the event vertex in the 
          $x$-$y$ plane, the impact parameter $|d_0|$, 
          smaller than 0.05$\,$cm.
  \item[] {\bf Secondary tracks:} Good tracks with an impact parameter 
         $|d_0|$ greater than 0.05$\,$cm, a high impact parameter 
         precision of $\delta d_0/|d_0|<0.1$ and a significant amount 
         of transverse momentum ($p_{\rm T}>1.5\,$GeV$/c$).
  \item[] {\bf Additional tracks:} All remaining tracks.
\end{itemize}

Candidate events must contain at most ten good tracks and 
at least two secondary tracks.

\subsubsection{Event selection for direct NLSP and chargino pair-production}
\label{s:largeip1}

For a smuon or selectron NLSP,
the processes $\ee\rightarrow\slept_{\rm R}^+ \slept_{\rm R}^-$ and 
$\ee\rightarrow\chpone\chnone$ ($\tilde \chi_1^{\pm}\ra\slept_{\rm R}\nu $)
followed by the delayed decay of the NLSP ($\slept_{\rm R}\ra\ell\grav$)
would lead to a signature of two tracks in the
event, both of which can have large impact parameters.
\begin{itemize}
\item[(1)] It was required that exactly
           two secondary tracks, no primary tracks and no more than
           three additional tracks were found.
\end{itemize}
After this topological cut, a selection was applied to separate the signal 
from the background, mainly two--photon or (radiative) Bhabha events,
cosmic particles traversing the detector,
hadronic interactions with the detector material or beam wall and beam gas
events.
\begin{itemize}
\item[(2)] To reduce combinatorial background, the two secondary tracks were
           required to
           have different charges to satisfy charge conservation.
\item[(3)] The barrel time-of-flight detector and the central jet chamber
           were used to reject cosmic ray particles.
\item[(4)] To reduce the background from beam-gas events, which often
           contain tracks coming from an interaction point that is shifted
           by a large amount along the $z$-axis with respect to the main
           event vertex, the longitudinal impact parameter $z_0$ of each
           secondary track was required to satisfy $|z_0|<40\,$cm.
           If both tracks had a $|z_0|$ greater than $6\,$cm,
           the longitudinal impact parameters $z_0$ of the
           two tracks were required to have opposite signs.
\item[(5)] To reject background containing back--to--back tracks,
           especially Bhabha events and muon pairs, the
           angle $\Delta\phi$ in the $x$-$y$ plane between 
           the two secondary tracks
           was required to satisfy $5^\circ<\Delta\phi<175^\circ$.\ 
\item[(6)] To reduce the remaining background from two--photon events,
           each secondary track must have had a total momentum $p>10\,$GeV$/c$.
\item[(7)] 
           To reduce the two-photon background,
           the invariant mass $W$ of
           the two selected secondary tracks, which were assumed to be pions,
           was required to satisfy $W>5\,$GeV$/c^2$.
\end{itemize}
For pair-produced smuons, selection efficiencies in the range of $60-70\,$\%
were achieved. As selectrons would have been
produced in the $t$-channel, leading to an
enhanced production in the direction of the initial particle beams,
selection efficiencies for selectrons range between $30-65\,$\%.
Efficiencies for
chargino pair-production and decays to a smuon or selectron
are in the range of 35--43\,\%.
This efficiency is lower than for directly produced smuons or selectrons
as additional visible energy is taken away by the neutrino from the chargino
decay.
In the whole data set no event was selected by the analysis, in agreement 
with the $0.13\pm 0.03_{\rm stat.}$$\pm$0.73$_{\rm syst.}$
events expected from Standard Model sources. 

If the pair-produced NLSP is a stau or if the pair-produced chargino decays
into a neutrino and the stau, the lepton in the event is a tau.
As the tau decays to one or more charged particles plus a neutrino,
additional energy is missing from the event.
To take this into account, relaxed cuts were used for searches in the 
stau NLSP scenario.
\begin{itemize}
\item[(6$^{*}$)] At least one of the secondary tracks must
have had a momentum $p\ge 5\,$GeV$/c$.
\item[(7$^{*}$)] The invariant mass of the two secondary tracks was
 required to be at
 least $3\,$GeV$/c^2$.
\end{itemize}


The efficiency for pair-produced staus
ranges between 45$\,$\% and 50$\,$\%.
For charginos in the stau NLSP scenario the efficiency ranges
from 38\,\% up to 48\,\% and is a bit lower than for direct stau production
because of additional missing energy from the neutrino from the chargino decay.
After all cuts there were three events left in the data, while
$3.79\pm 0.57_{\rm stat.}$$\pm$1.14$_{\rm syst.}$
events are expected from simulated Standard Model
sources.

\subsubsection{Event selection for neutralino pair-production} 
\label{s:largeip3}

Pair-produced neutralinos are produced and decay via the processes
$\ee\ra\ntone\ntone$, $\ntone\ra\sle_R\ell$ and $\sle_R\ra\ell\grav$
with the lepton a tau, muon or electron.
Thus in this channel there are four leptons in the final state, two of which
might have large impact parameters depending on the slepton NLSP lifetime.
The cuts to search for such events were
\begin{itemize}
\item[(1)] events with at least two secondary tracks
           and at least two
           primary tracks were selected.
           No cut on the number of additional tracks was imposed.
\item[(2)] For all secondary tracks the longitudinal impact parameter $|z_0|$
           was required to satisfy $|z_0|\le 40\,$cm to reduce 
           cosmic particles, beam wall and beam gas events.
\item[(3)] At least two of the primary tracks must have had 
           a transverse momentum
           larger than $1\,$GeV$/c$.
           The transverse momenta of the primary tracks for the
           data and the simulated background before this cut are shown
           in Figure~\ref{f:licuts}. Apart from the poorly
           modeled very low--momentum region, mainly consisting of
           two-photon events, good agreement is found.
           After the cut, 158 events remained in the data samples
           recorded at centre-of-mass energies between
           189$\,$GeV and 209$\,$GeV, while
           186.4 events are expected from Standard Model sources.
           Additionally at least two primary tracks were required to have had
           hits in the silicon microvertex detector and the vertex chamber.
\item[(4)] For Bhabha events the two primary tracks are usually
           back-to-back. To reject this background,
           the opening angle between all pairs of primary tracks $\varphi$
           in the $x$-$y$ plane was required to satisfy
           $\varphi\le 176^{\circ}$.\ 
\item[(5)] To separate the signal from the remaining two-photon background, 
           the invariant mass $W$ of all secondary tracks
           was required to fulfill $W>5\,$GeV$/c^2$.
           In Figure~\ref{f:licuts} the distributions of the
           invariant mass before the cut are shown for data and
           simulated background. After the cut, 12 events remained in the
           entire data sample, with 9.0 events expected from the
           background sources.
\item[(6)] Secondary particles can also be produced in photon conversions.
           In this case their tracks often start in the middle of the
           jet chamber and
           have, due to the boost in the forward direction,
           relatively small impact parameters. For the signal, secondary
           particles arise from the decay of heavy SUSY particles,
           resulting in a much smaller boost. Thus to produce equally 
           small impact parameters the flight lengths of the primary 
           particles must be much shorter than for photons, and the 
           secondary tracks start, in general, at radii smaller than 
           those of the jet chamber. To veto such tracks from photon 
           conversions the following cut was applied: for tracks with an 
           impact parameter $|d_0|<2\,$cm, the first hit wire
           in the jet chamber 
           must have been measured at a radial distance to the primary interaction 
           point of less than $40\,$cm.
\item[(7)] Finally, to reduce remaining background from two-fermion
           and multi-hadronic events, the vector sum of
           the transverse momenta of all secondary tracks
           was required to exceed $3\,$GeV$/c$.
\end{itemize}
Efficiencies of the selection are around $50\,$\% in the slepton co-NLSP
scenario and 35$\,$\% in the stau NLSP scenario. 
In total four events were left in the data in good agreement with
$5.71\pm 1.15_{\rm stat.}$$\pm$0.29$_{\rm syst.}$
events expected from simulated Standard Model background
sources. 

\subsubsection{Event selection for selectron or smuon pair-production (stau NLSP)}
\label{s:largeip4}

The production and decay of selectrons or smuons following
$\ee\ra\sle_R^+\sle_R^-$, $\sle_R\ra\ell\ntone$, $\ntone\ra\tau\stau$
and $\tilde{\tau}\ra\tau\grav$, leads to a final state with 
six leptons.
In the event, two primary electrons or muons, two primary taus and finally
two secondary taus and their 
decay products, which -- depending on the NLSP lifetime -- can 
have large impact parameters, are expected and 
the following cuts are used.
\begin{itemize}
\item[(1)] For a candidate event, at least two secondary and at least 
           three primary
           tracks were required. No cut on the number of additional
           tracks was applied.
\item[(2)] For all secondary tracks the longitudinal impact parameter $|z_0|$
           was required to satisfy $|z_0|\le 40\,$cm to reduce 
           cosmic particles, beam wall and beam gas events.
\item[(3)] At least two of the primary tracks must have had a transverse momentum
           larger than $1\,$GeV$/c$ to reduce the background, mainly from
           two-photon events.
           Additionally at least two primary tracks must have had
           hits in the silicon microvertex detector and the vertex chamber.
\item[(4)] The invariant mass $W$ calculated from all secondary tracks 
           was required to 
           be larger than $5\,$GeV$/c^2$ to remove remaining background
           from two-photon events. 
\item[(5)] To veto photon conversions the radius of the innermost hit wire
           in the central jet chamber associated to tracks with $|d_0|<2\,$cm 
           was required 
           to
           be smaller than $40\,$cm, as described for cut$\,$(6) in
           Section~\ref{s:largeip3}.
\end{itemize}
Selection efficiencies of 45\% for both selectron and smuon 
pair-production are reached.
In the entire data set, 13 events survived the
selection, which is in good agreement with 
$13.72\pm 1.74_{\rm stat.}$$\pm$0.77$_{\rm syst.}$
events that are
expected from Standard Model background sources. 

\subsubsection{Statistical and systematic uncertainties}
\label{s:largeip6}
The statistical uncertainty on the number of expected background
events due to the limited Monte Carlo statistics 
reach 50\,\% for single centre-of-mass energies mainly because
of the limited statistics of the simulated two-photon events,
which are expected to have high production cross-sections
of the order of 10$\,$nb.

Most cuts applied in the search for tracks with large impact parameters
rely on the tracking performance of the detector.
If the resolution of the track parameters in the detector
simulation is different from the one in the recorded data, this can lead to
systematic effects on the number of expected background events.
Therefore, to obtain a conservative
estimation of the size of this effect, the resolution of the track parameters
was degraded (smeared) by 10$\,$\% in the $x$-$y$ plane 
and by 40$\,$\% in the $z$ direction.
With the smeared track parameters all analyses were repeated on the 
simulated background samples. The absolute differences to the
background numbers without smearing were taken as the systematic uncertainties.
For analyses with a very small number of expected SM events,
the uncertainties can reach 100\,\% for single centre-of-mass energies.
But in general systematic effects were found to be smaller than statistical
uncertainties.


\subsection{Searches for charged NLSPs with medium lifetime (kinked tracks)}
\label{s:kink}
The discovery signals for charged NLSPs with a lifetime in the range of
about $10^{-9}\,$s$\, <\tau\, <10^{-7}\,$s 
are spectacular, 
with a kinked track in the tracking chambers. 
A kinked track is defined by the
presence of a primary and at least one secondary track segment which intersect
inside the sensitive volume of the central tracking chamber. 
In this analysis, all tracks reconstructed
in the central tracking system are considered.

\begin{itemize}
  \item[] {\bf Primary tracks: } tracks originating
          at the interaction point of the event, traverse through the 
          tracking detector volume with a minimum
          transverse momentum of $110\,$MeV$/c$ with no signal
          in the outer detectors.
  \item[] {\bf Secondary tracks: } A minimum transverse momentum of 
          $110\,$MeV$/c$ with $|d_0|$ greater than 2.5$\,$cm and 
          no associated hit in the silicon detector.
\end{itemize}

Primary and secondary tracks can be combined into a kink candidate
if there is an intersection point, the kink vertex, between the helices
which describe the curvature of the primary and secondary tracks.
No cut on the number of secondary tracks connected to a
single primary track was made to keep sensitivity for heavy charged particles
which possibly decay into a tau lepton. As a consequence the search algorithms
do not depend on the flavour of the slepton.
Also, no selection cut was placed on the specific ionization energy loss $\dedx$
of the primary heavy charged particle, in order to keep the 
kinked track search independent from the searches for
heavy stable charged particles. For the latter, the measurement of the
$\dedx$ was the main tool.

Depending on whether the NLSP was directly pair-produced or appeared 
in the decay chain of other SUSY particles, the kinked 
tracks would be accompanied by up to four additional charged leptons; thus,
different search algorithms were designed and optimized for the different 
track multiplicities in the event. 

\subsubsection{Event selection for direct NLSP and chargino pair-production}
\label{s:kink2}

The processes $\ee\rightarrow\slept_R^+ \slept_R^-$ and 
$\ee\rightarrow\chpone\chnone$ ($\tilde \chi_1^{\pm}\ra\slept_R\nu $)
followed by the delayed decay of the NLSP
lead to a signature of two tracks in the
event of which one or both can have a kink.
To be selected as an event consistent
with this topology the following criteria have to be satisfied:
\begin{itemize}
  \item[(1)] At least one kink vertex in the event.
  \item[(2)] The sum of the numbers of primary plus secondary
             tracks must be less than or equal to 20 and the number of
             good tracks
             should not exceed five.
  \item[(3)] There must not be more than two primary track candidates
             and the scalar sum of
             their transverse momenta $p_T$ had to be greater than
             $4\,$GeV$/c$ to reject events from two-photon processes.
  \item[(4)] The kink vertex had to be within the geometrical acceptance
             of the central jet chamber: $r\le 181.5\,$cm and
             $|z| \le (0.5\cdot[153.288\,$cm $+(r - 24.5\,$cm$\,)\cdot
             \tan (15^\circ)\cdot\cos (7.5^\circ)] -1.0)\,$.
  \item[(5)] The momentum of the secondary track had to satisfy
             $p\ge 1\,$GeV$/c$ and
             must not be parallel to a wire in the jet
             chamber to within $1^\circ$\ to remove combinatoric background
             and poorly measured tracks.
  \item[(6)] The transverse momentum $p_{\rm T}$ of the primary track had
             to satisfy $p_{\rm T}\ge 1\,$GeV$/c$. 
             Figure~\ref{f:kinkcuts} (a) shows the distribution
             of the transverse momentum of the primary track.
             Unmodeled background events at momenta less than 1$\,$GeV$/c$
             which are expected from two-photon events are removed.
             After the cut, 89 data events were left,  consistent 
             with 88.3 events expected from Standard Model sources.
  \item[(7)] The invariant mass $W_{00}$ of the primary particle was calculated
             using the primary and secondary track momenta and
             assuming the hypothesis of a massive particle decaying into
             two massless particles (one visible, and one invisible).
             Low--mass resonances were rejected by requiring this mass
             to be greater than $4\,$GeV$/c^2$.
             In Figure~\ref{f:kinkcuts}~(b) the distributions of the
             invariant mass for data and simulated background are shown.
             After the cut, 7 data events remained while 5.1 events are expected
             from Standard Model sources.
  \item[(8)] Remaining background caused by shower electrons from
             Bhabha scattering which traverse the detector under small angles
             with respect to the beam axis were reduced by requiring that
             the angle between primary and secondary
             tracks be smaller than $172^\circ$.\ 
             Also, a primary track having an
             angle smaller than $25^\circ$\ 
             with
             respect to
             the beam axis must not point to a cluster in the
             electromagnetic calorimeter with an energy greater than $70\,$\%
             of the beam energy.
  \item[(9)] Hadronic interactions were suppressed by rejecting
             a kinked track with more than two secondaries associated
             to its primary track. If there were two
             secondaries,  the angle $\zeta$ between
             them was required to be smaller than
             $20^\circ$\ ($\cos\zeta < 0.93$)
             to keep sensitivity for decaying tau leptons.
\end{itemize}
After all cuts 
1.83$\pm$0.42$_{\rm stat.}$$\pm$0.47$_{\rm syst.}$
events were expected background for the entire
data set. The background consists of events with hadronic interactions
of particles with the detector material, mainly from two-photon processes.
No data event passed all the cuts, which is consistent with the
expectation from Standard Model sources.
Typical efficiencies of the selection for pair-produced smuons and staus
are around 45$\,$\%, independent of the sparticle mass. For selectrons
the efficiency decreases to 25$\,$\% at masses of 45$\,$GeV/$c^2$
as  selectrons can be produced in the $t$-channel, which leads to a more
forward production. 

\subsubsection{Event selection for neutralino pair-production}
\label{s:kink3}

Searching for pair--produced neutralinos
($\ee\ra\ntone\ntone$, $\ntone\ra\stau_1\tau$ or $\ntone\ra\sle_R\ell$)
which decay into a lepton and
the NLSP,
the data
were searched
for topologies with a few tracks originating from the primary
interaction point plus at least one kinked track.
In the following list all cuts which are changed with respect to the search for
pair--produced NLSPs (Section~\ref{s:kink2}) are given.
\begin{itemize}
  \item[(2$^{*}$)] The sum of the numbers of primary plus secondary
             tracks must be less than or equal to 25 and the number of good tracks
             should not exceed six.
  \item[(3$^{*}$)] There must not be more than four primary track candidates
             and to reject events from two-photon  processes
             the scalar sum of
             their $p_{\rm T}$ has to be greater than $4\,$GeV$/c$.
  \item[(7$^{*}$)] The invariant mass $W_{00}$ of the primary particle has to
             be greater than $10\,$GeV$/c^2$ to compensate for the 
             relaxed cut on the number of tracks in the event.
\end{itemize}
As for all other selections of events with kinked tracks the remaining
expected background of 1.05$\pm$0.35$_{\rm stat.}$$\pm$0.50$_{\rm syst.}$
events in this analysis consists of hadronic interactions
from particles in two-photon events.
One data event survived the complete selection which is consistent with the
expectation.
The efficiency for detecting neutralino pair-production
followed by a delayed decay of the NLSP is typically around
$40\,$\%, independent of the neutralino and NLSP mass.

\subsubsection{Event selection for selectron and smuon pair-production (stau NLSP)}
\label{s:kink4}
In the case of a cascade of pair-produced selectrons or smuons 
($\ee\ra\sle_R^+\sle_R^-$, $\sle_R\ra\ell\ntone$, $\ntone\ra\tau\stau$)
to a stau with its decay inside the tracking chamber,
the events
contain a few tracks originating from the primary vertex plus at
least one kink vertex.
The cut on the number of tracks in the event has to be relaxed
compared to the search for directly produced NLSPs.
This is compensated by a tighter requirement on the invariant mass $W_{00}$
of the primary track. In the following list all cuts which are
changed with respect to the
search for direct NLSP production (Section~\ref{s:kink2}) are given.
\begin{itemize}
  \item[(2$^{**}$)] The sum of the numbers of primary plus secondary
             tracks must be less than or equal to 25 and the number of good tracks
             should not exceed ten.
  \item[(3$^{**}$)] There must not be more than eight primary track candidates,
             and to reject two-photon processes, the
             scalar sum of their transverse momenta $p_T$ should be
             greater than $4\,$GeV$/c$.
  \item[(7$^{* }$)] The invariant mass $W_{00}$ of the primary particle
                    has to be greater than 10$\,$GeV$/c^2$.
\end{itemize}
After all cuts 
2.76$\pm$0.69$_{\rm stat.}$$\pm$0.51$_{\rm syst.}$
events are expected from Standard Model sources,
mainly hadronic interactions with the detector material. 
Three events survived the selection in the entire data set,
consistent with the expectation.
One surviving candidate is the same as the one selected
by the search for pair-produced neutralinos (Section~\ref{s:kink3}).
The efficiency to detect the production and decay of
pair-produced selectrons and smuons in the stau NLSP scenario is
around $30\,$\% for all slepton, neutralino and NLSP masses.

\subsubsection{Statistical and systematic uncertainties}
\label{s:kink5}

The expected Standard Model background for the kinked track search
comes from
hadronic interactions of particles with the detector material.
All simulated processes with charged particles in the final
state can contribute to it. 
Because of a high production cross-section, of the order
of 10$\,$nb, for two-photon
events the number of simulated events of this type is not much higher
than the number of data events expected. Thus selecting such an individual
event contributes significantly to the number of expected background
events and results in a significant statistical error,
reaching up to 100$\,$\% for a single
centre-of-mass energy.

The selections of events with kinked tracks are mainly based on
tracking information. Thus
systematic errors were estimated using the same method
as the
search for tracks with large impact parameters described in
Section~\ref{s:largeip6}. The resolution of the parameters describing
the tracks in the $r$-$\phi$ plane in the simulated background and signal
samples was degraded by 10$\,$\% and the resolution of the $z$ parameters 
by 40$\,$\%. The analyses were repeated on the smeared sample and
differences with respect to the original results were counted
as systematic errors. For a single centre-of-mass energy, differences
up to 100$\,$\% were found.


\subsection{Searches for long-lived charged NLSPs}
\label{s:stable}
If the slepton is the NLSP and the lifetime is such that the decay
happens outside the detector volume the heavy charged particle has to traverse
the central tracking detector. In this gaseous detector the measurement
of the specific ionization energy loss $\dedx$ along the particle's track
allows a powerful identification of signal events. 

\subsubsection{Search for pair-produced stable charged NLSPs}
\label{s:stable1}
The search for pair-produced long-lived massive
particles with a lifetime longer than 10$^{-7}\,$s is reported in
\cite{Abbiendi:2003yd}.
The search is used here to select candidate events for pair-produced charged
NLSPs, $\ee\ra\slept_R^+ \slept_R^-$.
The selection is primarily based on the precise
measurement of the ionization energy loss (d$E$/d$x$) of charged particles
in the jet chamber, using a data set corresponding to a total
integrated luminosity of 693.1~pb$^{-1}$ between centre-of-mass energies
of 130 and 209$\,$GeV.
No candidate event was reported in the entire data set which
is consistent with the expectation from Standard Model background
of 1.1 events.

The typical efficiency of the search is about 95$\,$\%;
however, for particles with a $\beta\gamma = p/m = \sqrt{{s\over{4m^2}}-1}$
between 0.75 and 1.70, the efficiency drops below 5$\,$\%.
Figure~\ref{f:dedxcuts} shows that this effect is due to the fact that in this
particular $\beta\gamma$ region,
the d$E$/d$x$ of tracks of heavy (SUSY) particles is very similar
to that of light (Standard Model) particles.
But as the LEP accelerator operated
at different centre-of-mass energies, particles with a certain mass 
were in different efficiency regions for different centre-of-mass energies.

\subsubsection{Event selection for heavy stable charged particles in multi-track events}
\label{s:stable2}

If long-lived NLSPs are produced as secondary or tertiary 
particles from the decays of neutralinos, charginos or heavier
sleptons, their tracks are no longer back-to-back and
additional tracks or additional missing energy are expected.
The analysis searching for directly pair-produced NLSPs
(Section~\ref{s:stable1}) cannot be applied and
a different analysis
was developed to search for tracks with anomalous
ionization energy loss in events with more than two tracks or
additional missing energy.
For all signal topologies one rather general data selection is chosen 
where at least one track per event with an anomalous specific energy loss is
required. 

The analysis was split into three steps:
(1) preselection;
(2) cuts on full event properties;
(3) selection of high quality tracks with anomalous ionization energy loss.

\paragraph{(1) Preselection}

\begin{itemize}
  \item[(P1)]Events with either more than 20 tracks or with a low total
        visible energy less than  $0.10\,\sqrt{s}$ were rejected
      to reduce the number of two-photon events.
  \item[(P2)] There must be at least one track in the event which
              has at least
              one hit in the precision vertex chamber, 20 hits used for
              $\dedx$ measurements and the specific 
              energy loss should be higher or lower than  that expected 
              from SM particles
             ($\dedx>11\,$keV/cm or $\dedx<9\,$keV/cm for a track 
             momentum $p>10\,$GeV/c).
  \item[(P3)]  Non-simulated backgrounds like cosmic muons and beam gas 
             interactions are removed
             by 
             requiring that the 
             distance between the beam axis and the track at the point of closest approach
             had to be smaller than $1.5\,\mathrm{cm}$ in the $x$-$y$ plane
             and smaller than $10 \,\mathrm{cm}$ along the $z$ coordinate.
             If there were hits in the time-of-flight barrel scintillators
             the measured time of the
             closest barrel hit was allowed to differ by at most $10\,\mathrm{ns}$ 
             from the time expected from a physics event. 
\end{itemize}

\begin{figure}[H]
     \begin{minipage}[t]{.1cm}
       \epsfxsize=.1cm
     \end{minipage}
   \hfill
     \begin{minipage}[t]{15.5cm}
       \epsfxsize=15.5cm
       \epsfbox{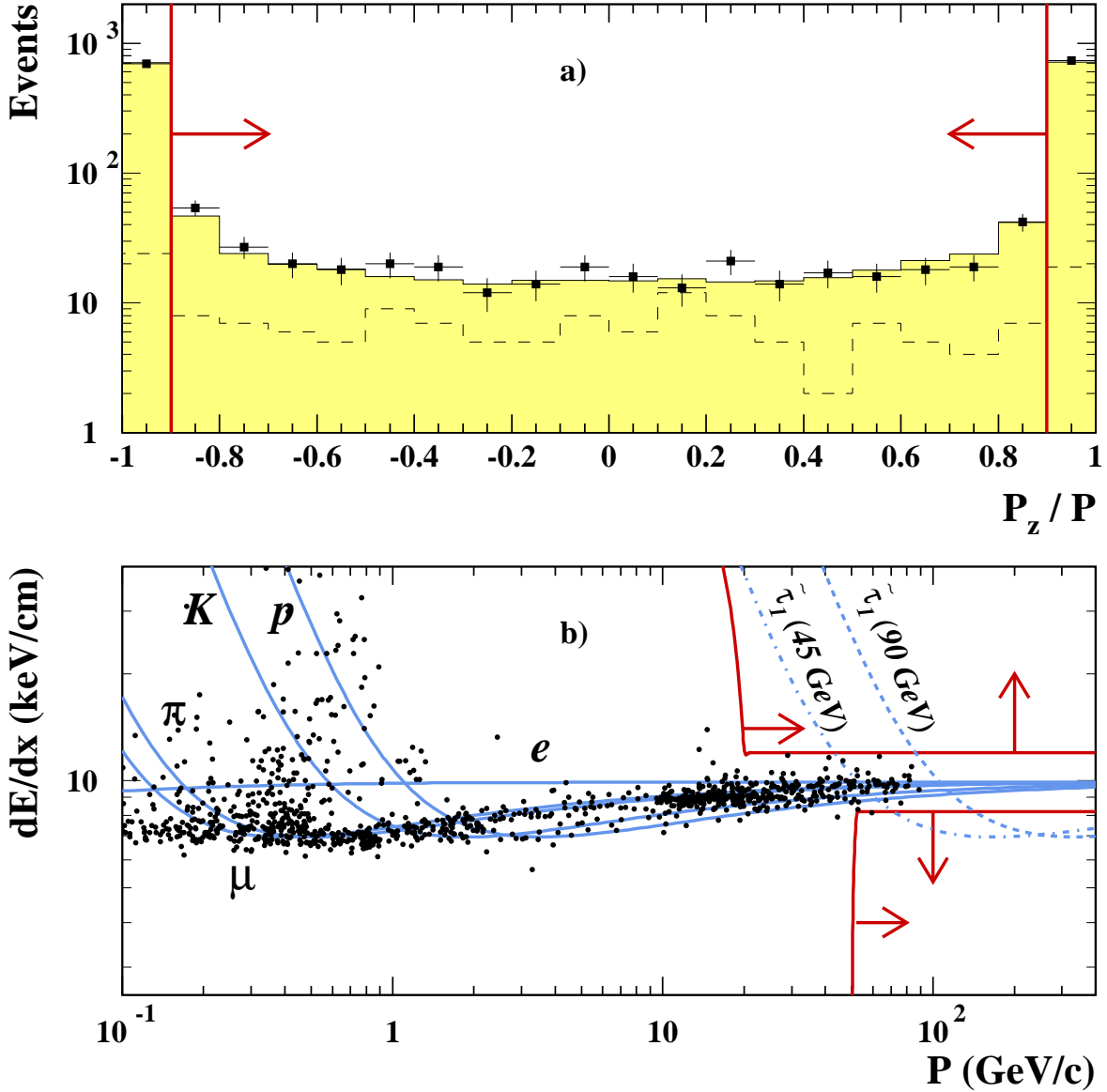}
     \end{minipage}
   \hfill
     \begin{minipage}[t]{.1cm}
       \epsfxsize=.1cm
     \end{minipage}
\caption{
\label{f:dedxcuts}
{Distributions of variables in the search for long-lived charged NLSPs 
before selection cuts.
Plot (a) shows the scaled component $P_z$ of the event momentum
$P=|\vec{P}|=|\sum\vec{p}_{\mathrm{track}}|$ along the beam axis
before cut$\,$(5), Section~\ref{s:stable2}.
The data, taken at centre-of-mass energies of 189--209$\,$GeV are represented
by dots, the simulated background from Standard Model sources by 
the solid line. 
The dashed line indicates the expected distribution
for staus with a mass of 90$\,$GeV$/c^2$, pair-produced at a centre-of-mass
energy of 208$\,$GeV.
In Plot (b) the distribution of candidate events
before cut$\,$(9) as a function of the track
momentum $P$ and the specific ionization energy loss $\dedx$ is given.
In addition the expected energy loss for electrons ($e$), muons ($\mu$),
pions ($\pi$), kaons ($K$),
protons ($p$) and long-lived staus
($\stau_1$ with $m_{\stau_1}=45$ and 90$\,$GeV$/c^2$) is indicated.
In both plots the arrows give the regions which are
selected by the analysis cuts.}
}
\end{figure}

\paragraph{(2) Event Analysis}
\begin{itemize}
\item[(1)] To reduce two--photon background events, 
           which deposit a large fraction of energy at small angles
           with respect to the beam axis, the maximum energy deposited in
           the forward calorimeters must be less than 5$\,$GeV. 
\item[(2)] The total measured relativistic invariant mass $M_{\mathrm{vis}}$
           (visible mass) must satisfy $M_{\mathrm{vis}}>10\,$GeV$/c^2$
           to reduce further the number of two--photon events.
\item[(3)] The visible energy was required to satisfy
           \mbox{$0.15<E_{\mathrm{vis}}/\sqrt{s}<1.10$},
           to reject two--photon and two--fermion events.
\item[(4)] The maximum electromagnetic energy $E_{\mathrm{ECAL}}$ was required
           to satisfy $E_{\mathrm{ECAL}}/\sqrt{s}<0.3$ to separate the 
           signal from two--fermion processes.
\item[(5)] If the sum over all track momenta in an event $P=|\vec{P}|=|\sum
           \vec{p}_{\mathrm{track}}|$ was greater than
           $10\,$GeV$/c$, the $z$ component of the sum $P_z$ must satisfy
           $|P_z/P|< 0.9$.
           This cut reduced mainly Bhabha scattering events with one 
           electron escaping through the beam pipe carrying away a significant
           fraction of the total momentum of the event.
\end{itemize}
In the data set recorded between 189$\,$GeV and 209$\,$GeV,
572 data events survived the requirements on the event properties
while 537.04 events
are expected from Standard Model
sources. Figure~\ref{f:dedxcuts}~(a) shows the distribution of the
variable $|P_z/P|$ used in cut$\,$(5) together with the distribution of
an expected signal for pair-produced staus.

\paragraph{\boldmath (3) Anomalous $\dedx$ selection}
\begin{itemize}
\item[(6)]  In case the track with momentum $p$ is associated to a cluster
            in the electromagnetic calorimeter with an energy $E_{\rm Cal}$,
            the condition $E_{\rm Cal}/p \le 0.15$ has to be satisfied.
            This cut further reduces Bhabha scattering events,
            four--fermion and two--photon background.
\item[(7)]  To avoid events with converted photons
            -- where the electron and positron tracks are unresolved
            by the tracking detectors, hence give a single track with a
            high $\dedx$ --  the track must have at least
            one hit in the high precision vertex chamber.
\item[(8)]  The track had to satisfy tightened quality criteria.
            At least 40 hits should be used for $\dedx$ measurement
            and the error on the momentum must be smaller than $10\,$GeV$/c$.
\item[(9)]  The track with momentum $p$ must feature a
            high or low specific ionization
            energy loss $\dedx$:\\
            \hbox{\ }\quad if $p > a \cdot (\dedx - b)$ then 
            $\dedx > 12.0\keVpcm$ ($a={{2}\over{17}}\,$cm, $b = 181.5\keVpcm$)
            or\\
            \hbox{\ }\quad if $ p > 52\,$GeV$/c$ then $\dedx < 8.2\keVpcm$\\
            The cut separated signal topologies from Standard Model background.
\item[(10)] Finally, to separate the signal from Standard Model events with
            taus, no other track was allowed in a cone of~$20^{\circ}$\  
            around the track with the anomalous $\dedx$.
\end{itemize}
Figure~\ref{f:dedxcuts}~(b) shows the distribution of the candidate events
before cut$\,$(9)
as a function of the track momentum and the specific ionization
energy loss. The regions in which candidate events are expected are indicated.
Out of 448 candidate events (419.33 events expected from Standard
Model sources), no data event survived selection cut$\,$(9) while
0.89 events are expected. 

The typical efficiency of the search 
ranges between 80$\,$\% and 95$\,$\% for masses of the
heavy  particle near the beam energy and between 20$\,$\% and 40$\,$\% for
particles with a $\beta\gamma$ larger than 1.5. 
The efficiency drops below 5$\,$\%
only in a small region of $\beta\gamma$,
where the ionization energy 
loss for massive particles is very similar to the energy loss of Standard
Model particles, as shown in Figure~\ref{f:dedxcuts}.
In the data, recorded at $\sqrt{s}=130\,$GeV--$209\,$GeV 
and corresponding to an integrated luminosity of $\mathcal{L}=632.1\pb^{-1}$,
no event survived after all cuts.
This is compatible with
0.78$\pm$0.38$_{\rm stat.}$$\pm$0.10$_{\rm syst.}$
events which are expected from the Standard Model sources.

\paragraph{3.4.2.1  Systematic uncertainties}
\ \\
\ \\
The main tool used in this analysis is the information from particle tracks.
Therefore, the main systematic errors of the event selection 
arise from uncertainties in the track quality 
and the modeling of the specific
energy loss in the simulation.

The systematic error arising from the track measurement 
was evaluated by smearing
simulated track resolution 
by $5\,\%$ for
$r$-$\phi$ and $20\,\%$ for $z$.
Repeating the entire analysis with modified track parameters
for signal and background determines the first contribution to
the systematic error arising from the central tracking system.
The second contribution to the systematic error is due to 
uncertainties in the modeling of the measured 
$\dedx$ value. 
These uncertainties were
evaluated by comparing muon pairs, Bhabha electrons, charged pions, 
and kaons in simulation and data, 
and found to be $10\,\%$ of the 
measurement error 
$\sigma_{\dedx}$.
The contribution to the overall systematic
error was determined by redoing the analysis
of signal and background with
the $\dedx$ value of each 
track replaced by a value ($\dedx\pm\sigma_{\dedx}$).

The studies show that 
the number of expected background 
events may vary by  $10\,\%$ from the track smearing
and by $5.0\,\%$ due to the $\dedx$ uncertainties.

\section{\boldmath Searches in the neutralino NLSP scenario}
\label{s:searches_neutnlsp}
In this section the analyses to search for indirect NLSP production
in the neutralino NLSP scenario are described. To be sensitive to all 
neutralino lifetimes, searches for the production of sleptons and charginos 
were performed. Section~\ref{s:shortneut} describes the analyses selecting
event signatures expected for short-lived neutralinos 
with lifetimes up to 10$^{-8}$s.
In Section~\ref{s:longneutral} searches for neutralinos with a long 
lifetime are presented.
A summary of the results for searches not described in other papers
is listed in Table~\ref{t:listresults}.

\subsection{Searches for short-lived neutralinos}
\label{s:shortneut}
All signatures of SUSY particles in the scenario with the neutralino
being the NLSP have in common that, for short and medium lifetimes up to
$10^{-8}\,$s, photons from the neutralino decay would be measured by
the detector.
For longer NLSP lifetimes, the decay happens outside the detector and 
typical signatures contain a significant amount of missing energy.

\subsubsection{Search for events with photons and missing energy}
\label{s:gamgam}
The search for pair-produced neutralinos, decaying in the detector volume,
selects events with two photons and large missing energy. The main
background for the search is the Standard Model process
$\ee\ra\nu\bar\nu$ + photons. The selection and the result of the search are
described in 
\cite{photons}.
No evidence is observed for new physics contributing to the expected final
state.

\subsubsection{Selection of events with isolated photons, missing energy,
               leptons and jets}
\label{s:gamgamx}

If the lightest neutralino is the short-lived NLSP, 
the production and decay of scalar leptons ($\ee\ra\slept^+_R\slept^-_R$,
$\slept^\pm_R\ra\ell\ntone$) 
or charginos ($\ee\ra\chpone\chnone$, $\chpnone\ra{\rm W}^\pm\ntone$)
lead to final states with missing energy, up to two photons
plus leptons and/or jets ($\gamma(\gamma){\emiss} $ plus
${\rm q_i \bar q_j}{\rm q_k \bar q_l}$, ${\rm q_i \bar q_j}\ell^{\pm}$
or $\ell^+\ell^-$).
For each final state, the photon energy and the missing energy
depend strongly on the mass difference 
$\dm$\,(=$M_{\rm NNLSP} - M_{\rm NLSP}$) between the pair-produced
particle (the NNLSP) and the neutralino (the NLSP).
Thus a set of analyses was designed,
using a common preselection and followed by specific series of cuts
optimized for the different $\dm$ regions:
\begin{itemize}
  \item[(A)] photons plus leptons and missing energy;
  \item[(B)] photons plus jets and missing energy
    \begin{itemize}
      \item[(B0)] small $\Delta M$ ($<20$ GeV/$c^2$),
      \item[(B1)] medium $\Delta M$ ($20<\Delta M<M_{\rm NNLSP} - 30$ GeV/$c^2$),
      \item[(B2)] large $\Delta M$ ($>M_{\rm NNLSP} - 30$ GeV/$c^2$);
    \end{itemize}
  \item[(C)] photons plus jets plus a lepton and missing energy
    \begin{itemize}
      \item[(C0)] small $\Delta M$ ($<20$ GeV/$c^2$) where analysis (B0) is used,
      \item[(C1)] medium $\Delta M$ ($20<\Delta M<M_{\rm NNLSP} - 30$ GeV/$c^2$),
      \item[(C2)] large $\Delta M$($>M_{\rm NNLSP} - 30$ GeV/$c^2$).
    \end{itemize} 
\end{itemize}
The first part of the preselection required well-contained
events from $\rm e^+e^-$ collisions and is identical to that
used in the long-lived neutralino NLSP chargino search
given in Section~\ref{s:neutcharg} and documented
in \cite{Abbiendi:2003sc}.  The lepton and jet identification
used here is also identical to that in \cite{Abbiendi:2003sc}.
All events with more than six good tracks from 
the interaction point and not originating from photon conversions
were classified as candidates for selection (B) or (C), events with
fewer tracks as candidates for selection (A).

\begin{figure}[H]
     \begin{minipage}[t]{.1cm}
       \epsfxsize=.1cm
     \end{minipage}
   \hfill
     \begin{minipage}[t]{15.5cm}
       \epsfxsize=15.5cm
       \epsfbox{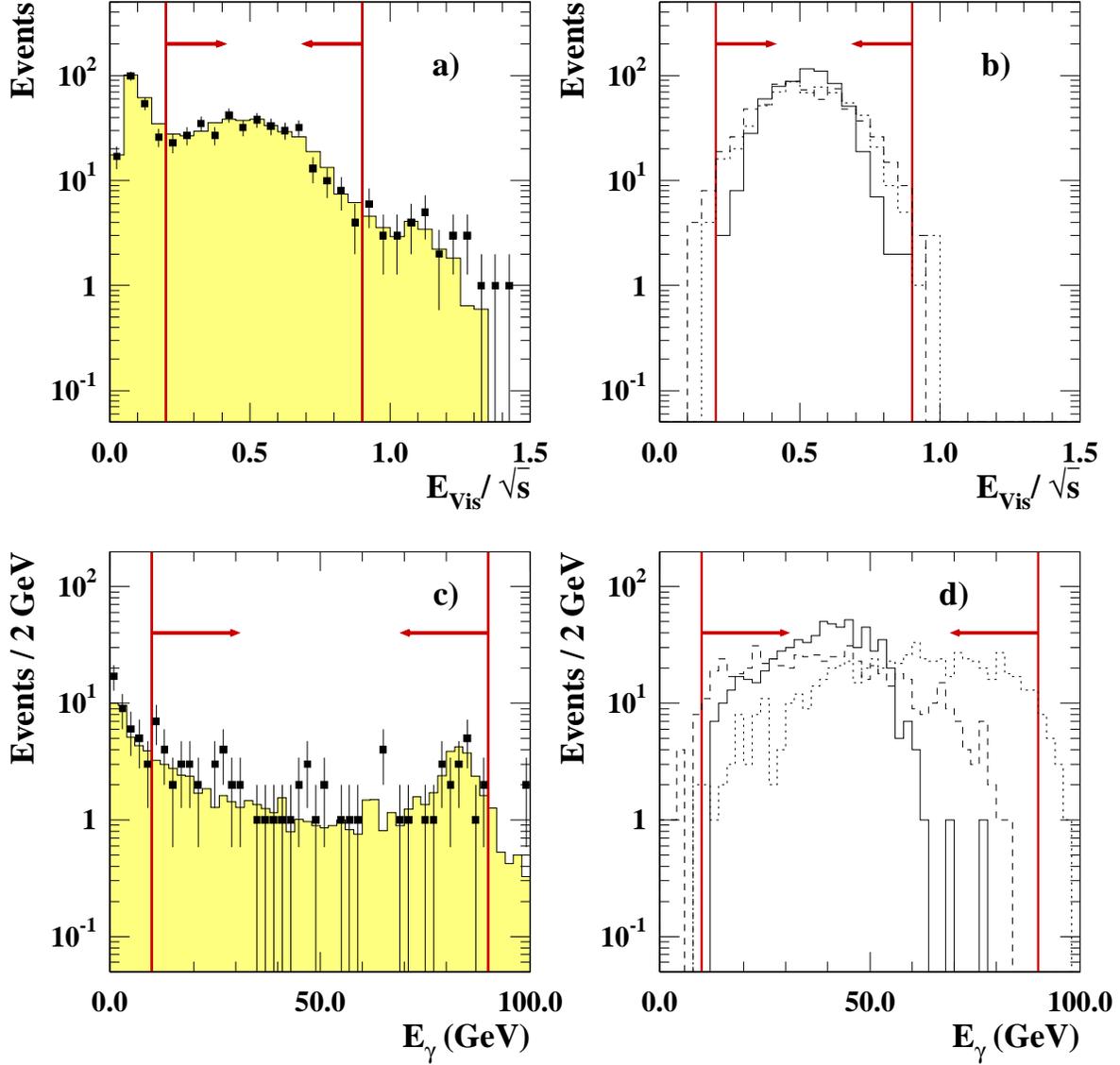}
     \end{minipage}
   \hfill
     \begin{minipage}[t]{.1cm}
       \epsfxsize=.1cm
     \end{minipage}
\caption{
\label{f:ggxxcuts}
{Distributions of variables in the search for slepton and charginos with 
very short to medium lifetime (Section~\ref{s:gamgamx}), selection$\,$(A). 
Plots (a, b) show the distribution of the visible energy in the event
as a fraction of the centre-of-mass energy, plots (c, d) the 
energies of the most energetic photon in the event.
In (a) and (c) the data recorded at $\sqrt{s}=206\,$GeV are represented by dots,
the expected Standard Model background by the shaded histogram. In (b) and (d)
the distributions of a possible signal of stau pair-production in the 
neutralino NLSP scenarios are given. The solid, dashed and dotted lines
correspond to ($\stau_1^\pm$, $\ntone$) mass combinations of (102, 51),
(80, 20) and (50, 45)$\,$GeV$/c^2$. The arrows indicate the cut positions.}
}
\end{figure}

The second part of the preselection picks up events with isolated
photons and tightens the topological requirements. 
\begin{itemize}
\item[(P1)] There have to be at least two isolated photons with:\\
           -- an electromagnetic cluster
                with an energy greater than 3$\,$GeV;\\
           -- the angle $\theta$ of the cluster with respect to the beam
                satisfying $|\cos{\theta}|<0.9$;\\
           -- in a 15$^{\circ}$ half-cone centred on the cluster the 
               sum of track momenta and additional\\
           {\color{white}--} electromagnetic energies
               not exceeding 2$\,$GeV.
\item[(P2)] The angle $\theta_{\rm miss}$ of the missing momentum
           had to satisfy $\cth < 0.95$. 
\item[(P3)] The transverse momentum of the event had to be greater than
           3$\,$\% of the centre-of-mass energy. 
\item[(P4)] The visible energy in the event had to satisfy\\
            -- $0.2\cdot\sqrt{s}<\revis<0.9\cdot\sqrt{s}$ for analysis A,\\
            -- $0.25\cdot\sqrt{s}<\revis<0.85\cdot\sqrt{s}$ for analysis B0
               and\\
            -- $0.4\cdot\sqrt{s}<\revis<0.95\cdot\sqrt{s}$ for the other
               analyses.\\
            In Figure~\ref{f:ggxxcuts}$\,$(a) the distribution
            of the visible energy in the event is shown for the data recorded
            at a centre-of-mass energy of 206$\,$GeV together
            with the expected background. After cut$\,$(P4) of 
            analysis A, 354 data events
            remained in the sample with 368.48 events expected
            in the background. Plot$\,$(b) shows the same distribution for
            a possible signal of stau pair-production at three different
            mass combinations of the stau and the neutralino. 
\item[(P5)] The acoplanarity angle $\acop$, which is defined as the
            supplementary angle in a plane perpendicular to the beam
            direction between the two vectors which are obtained by summing up
            particle momenta after splitting the event into two jets using the
            Durham algorithm, has to satisfy $\acop > 5\degree$.
\end{itemize}
After the preselection, the cuts were optimized depending on the mass
difference between the NNLSP and the NLSP as well as the number of leptons
expected in the event. 

\paragraph{Selection (A): photons plus two leptons and missing energy}
\ \\
\ \\
Events with two photons, two leptons and missing energy 
are expected from the production and decay of sleptons
($\ee\ra\slept^+_R\slept^-_R$, $\slept^\pm_R\ra\ell\ntone$,
$\ntone\ra\gamma\grav$)
and charginos
($\ee\ra\chpone\chnone$, $\chpnone\ra{\rm W}^\pm\ntone$,
$\ntone\ra\gamma\grav$) with both ${\rm W}$ bosons decaying
leptonically.
The main expected backgrounds from Standard Model sources are 
radiative return events to the ${\rm Z}$ boson, 
with $\PZz \aro \ellp\ellm$ accompanied by initial or final state radiation.
To suppress this type of background, cuts on the photon energies are
sufficient. 
The analysis does not distinguish between electrons, muons or taus
since for small mass
differences $\dm$ the lepton energy is small and they cannot
be identified efficiently.
\begin{itemize}
\item[(A-1)] The energy $E_{\gamma}$ of the most energetic photon
             had to be between
             $10\,$GeV and $90\,$GeV, and the energy of the second photon
             greater than $5\,$GeV. Almost all radiative return events
             were rejected by the requirement on the second most energetic photon.
             In Figure~\ref{f:ggxxcuts}$\,$(c) the energy distribution
             of the most energetic photon is shown for the data recorded
             at a centre-of-mass energy of 206$\,$GeV together
             with the expected background. After the cut, 74 data events
             remained in the sample with 70.63 events expected
             in the background. Plot$\,$(d) shows the same distribution for
             a possible signal of stau pair-production at three different
             mass combinations of the stau and the neutralino. 
\item[(A-2)] Surviving events with final state radiation often have
             the photon near a particle track.
             To reject them the transverse momentum
             of the photon with respect to the nearest track 
             had to be larger than $10\,$GeV$/c$ ($5\,$GeV$/c$) for the first
             (second) photon.
\end{itemize}
In data recorded at centre-of-mass energies between 189$\,$GeV and 209$\,$GeV,
3 events survived the selection with
7.74$\pm$0.47$_{\rm stat.}$$\pm$1.63$_{\rm syst.}$ 
events expected from Standard Model sources.
 
\paragraph{Selection (B0): 
           photons plus jets and missing energy for small $\dm$}
\ \\
\ \\
In the case of production and decay of charginos
($\ee\ra\chpone\chnone$, $\chpnone\ra{\rm W}^\pm\ntone$,
$\ntone\ra\gamma\grav$) with both ${\rm W}$ bosons decaying
into two jets for small mass differences between
the chargino and the neutralino NLSP, the jets contain only
a small amount of energy. The jets might not be measured well, thus the
selection depends on the two most energetic photons only.
\begin{itemize}
\item[(B0-1)] The energy of the most energetic photon has to be greater
              than $20\,$GeV and the energy of the second photon
              greater than $10\,$GeV.
\item[(B0-2)] The visible energy in the event, excluding the two most
              energetic photons, has to be less than $50\,$GeV.
\end{itemize}
After the selection there is no event left in the entire data set with
0.69$\pm$0.09$_{\rm stat.}$$\pm$0.14$_{\rm syst.}$
events expected from Standard Model sources.

\paragraph{Selection (B1): 
           photons plus jets and missing energy for medium $\dm$}
\ \\
\ \\
In case of a higher mass difference between the
chargino and the neutralino, the four jets of the
hadronically decaying $\rm W$ bosons have significant energy,
and there is no prompt lepton in the events.
The jets exclude the identified photons.
\begin{itemize}
\item[(B1-1)] The number of identified, isolated leptons had to be zero.
              The lepton was defined to be isolated if there was not more
              than 2$\,$GeV additional energy in a cone with a half-angle
              of $15^\circ$ around the lepton. 
\item[(B1-2)] To reject events with final state radiation,
              the transverse momentum
              of the photon with respect to the nearest track 
              had to be larger than $10\,$GeV$/c$ ($5\,$GeV$/c$) for the first
              (second) photon.
\end{itemize}
Different additional requirements have to be fulfilled, depending on
the number of jets in the event.
\begin{itemize}
\item[(B1-3)] 4-jets: the angle between the two reconstructed $\rm W$ bosons
               had to be less than $150^\circ$, with the two bosons
               (jet pairs) being selected by minimizing the difference
               between the jet masses and the $\rm W$ boson mass
               ($(M_{ij} - M_W)^2 + (M_{kl} - M_W)^2$). 
\item[] 3-jets: the two jets with an invariant mass closest to
                the mass of the $\rm W$ boson were paired together.
                The angle between the $\rm W$ and the remaining jet
                had to be smaller than $150\degree$.
\item[] 2-jets: the sum of the opening angles between each jet
                 and the most energetic photon had to be less than
                 $356\degree$.
\end{itemize}
After the selection 6 data events survived the cuts of the medium
$\dm$ analysis (B1) with
8.59$\pm$0.86$_{\rm stat.}$$\pm$1.80$_{\rm syst.}$
events expected from Standard Model sources.


\paragraph{Selection (B2): 
           photons plus jets and missing energy for large $\dm$}
\ \\
\ \\
Because of the large mass difference between the chargino and the neutralino,
expected signal events have less energetic photons and only a small
missing momentum.
After the selection, the remaining backgrounds are mainly from the
processes $\ee\ra\PWp\PWm$ and $\ee\ra\PZz\gamma$ with additional photons.
\begin{itemize}
\item[(B2-1)] The number of identified, isolated leptons had to be zero.
\item[(B2-2)] To reject events with final state radiation,
              the transverse momentum
              of the photon with respect to the nearest track 
              had to be larger than $5\,$GeV$/c$ ($3\,$GeV$/c$) for the first
              (second) photon.
\item[(B2-3)] The event was forced into four jets and the
              invariant mass cut of the Durham jet finding algorithm $k_T$
              had to satisfy $k_T > 4\,$GeV$/c^2$.
\end{itemize}
After the selection, 3 data events survived with
2.67$\pm$0.26$_{\rm stat.}$   
events expected from Standard Model processes.

\paragraph{Selection (C1): photons plus jets plus one lepton and
                          missing energy for medium $\dm$}
\ \\
\ \\
If one $\rm W$ boson decays into a lepton and a neutrino,
the other hadronically, two photons, two jets and one
lepton with lower energies are expected in the event.
The lepton can be identified
and candidate events had to satisfy the following constraints.
\begin{itemize}
\item[(C1-1)] There had to be at least one isolated lepton in the event.
\item[(C1-2)] To reject events with final state radiation,
              the transverse momentum
              of the photon with respect to the nearest track 
              had to be larger than $10\,$GeV$/c$ ($5\,$GeV$/c$) for the first
              (second) photon.
\item[(C1-3)] The lepton energy had to be less than $40\,$GeV and
              the invariant mass of the two jets less than $70\,$ GeV$/c^2$.
              This reduced the background from $\PWp\PWm \aro \qq\ell\nu$
              where jets and lepton are more energetic compared
              to expected signal events. 
\end{itemize}
One event survived the selection cuts
for the medium $\dm$ analysis, with
0.62$\pm$0.11$_{\rm stat.}$$\pm$0.13$_{\rm syst.}$
events expected from Standard Model sources.


\paragraph{Selection (C2): photons plus jets plus one lepton and
           missing energy for large $\dm$} 
\ \\
\ \\
In case of a large mass difference between the chargino and the neutralino
the jets and lepton have more energy.
The main expected background results from  the decay of pair-produced 
$\rm W$ bosons, $\PWp\PWm \aro \qqln$, with additional photon(s) from
the initial and(or) final state radiation. 
The signal topology is similar to the Standard Model process, but it can be
separated by requiring two isolated photons and unbalanced $\rm W$ bosons.
\begin{itemize}
\item[(C2-1)] There had to be at least one isolated lepton in the event.
\item[(C2-2)] The transverse momentum
              of the photon with respect to the nearest track or jet
              had to be larger than $5\,$GeV$/c$ ($3\,$GeV$/c$) for the first
              (second) photon.
\item[(C2-3)] The angle between the two $\rm W$ bosons, where one W is
              reconstructed from the two jets and the other by 
              the most energetic lepton and missing momentum, had to
              be less than $170^\circ$.
\end{itemize}
In the data sample, two events survived the selection cuts, with
2.94$\pm$0.31$_{\rm stat.}$   
events expected from Standard Model processes.


\paragraph{4.1.2.1 Systematic uncertainties} 
\ \\
\ \\
The uncertainty associated with the photon isolation requirement
is the dominant systematic error in these selections.
It originates from the modeling of parton-level photon emission, 
jet fragmentation and detector simulation. 
Since it is difficult to estimate
an error on each source, a total error
is obtained inclusively by comparing data
with Monte Carlo simulation using several cross-check samples.
The 
studies are limited by statistics
and a systematic error, which is at most 20\%,
is taken into account for all selections.

Uncertainty on the modeling of the other cut variables 
for simulated background
and signal was also investigated. This was done by shifting 
the cut value within
the possible error.
This is negligible for signal events but not for the background,
especially at large $\dm$, because the topology is similar to Standard Model
events and
some cuts are set near to the peak of their distributions. 
Uncertainties from 3$\,$\% for analyses (A) and (B0) up to 6$\,$\% for the
other analyses are reached.


\subsection{Searches for long-lived neutralinos}
\label{s:longneutral}
In the case that the neutralino is the NLSP and has a lifetime longer 
than 10$^{-7}\,$s,
it usually decays outside the detector volume.
The topologies are similar to
searches for supersymmetric particles in gravity-mediated
SUSY breaking scenarios in which the stable neutralino is the 
lightest supersymmetric particle. 

\subsubsection{Search for di-lepton events with missing transverse
               momentum}
\label{s:neutaccop}
If the pair-produced sleptons decay to a lepton and a long-lived neutralino, 
the signature of possible candidate events is two leptons with missing
transverse momentum.
The search for this type of events is reported in \cite{Abbiendi:2003ji}
and a short
description can be found in Section~\ref{s:accop}.

The search for di-lepton events with missing transverse momentum is also
sensitive to the pair-production of charginos. Here the chargino can
decay either to a slepton and a neutrino, followed by the
prompt decay of the slepton to a lepton and a long-lived neutralino, or 
the chargino can decay to a neutralino and a W boson with both W 
bosons decaying
leptonically.

\subsubsection{Search for charginos decaying into a neutralino and a W boson}
\label{s:neutcharg}

Approximately 438~pb$^{-1}$ of data recorded at
centre-of-mass energies of $192$--$209\,$GeV were analyzed to search for
evidence  of chargino pair-production. No evidence for a signal was
observed and the results of the search are reported in \cite{Abbiendi:2003sc}.
Search channels looking for charginos decaying into a neutralino and a 
$\W$ boson were used for the GMSB searches described in this paper.
All decays of the $\W$ boson, the hadronic or leptonic decay,
leading to topologies containing jets and missing energy, or jets with a
lepton and missing energy were considered.

In case both $\W$ bosons decay leptonically, the results of the searches
reported in \cite{Abbiendi:2003ji}, Section~\ref{s:neutaccop} were applied.

\begin{table}[htp!]
\begin{turn}{90}{
\parbox{22.cm}{
 \vspace{-.7cm}
\begin{center}
{\scalebox{.78}{
\begin{tabular}{c|l||c|c|c|c|c|c|c|c|c}
\hline
Sparticle          &      & $<\sqrt{s}>=$& $<\sqrt{s}>=$& $<\sqrt{s}>=$& $<\sqrt{s}>=$& $<\sqrt{s}>=$& $<\sqrt{s}>=$& $<\sqrt{s}>=$& $<\sqrt{s}>=$&      \\
production         & NLSP & 188.7$\,$GeV & 191.6$\,$GeV & 195.5$\,$GeV & 199.5$\,$GeV & 201.6$\,$GeV & 205.1$\,$GeV & 206.7$\,$GeV & 208.1$\,$GeV & Reference \\
(NLSP)             & lifetime  & (168$\,$pb$^{-1}$) & (30$\,$pb$^{-1}$) & (76$\,$pb$^{-1}$) & (78$\,$pb$^{-1}$) & (38$\,$pb$^{-1}$) & (74$\,$pb$^{-1}$) & (120$\,$pb$^{-1}$) & (8$\,$pb$^{-1}$) &  \\ 
\hline
\hline
$\stau_1^+\stau_1^-$ & (1) short  &1/0.93$\pm$0.56&0/0.11$\pm$0.06&1/0.25$\pm$0.13&0/0.49$\pm$0.36&0/0.18$\pm$0.28& 0/0.56$\pm$0.24&1/1.13$\pm$1.02&0/0.14$\pm$0.04& \ref{s:largeip1}\\ \cline{2-11}
($\stau_1$)          & (2) medium &0/0.71$\pm$0.31&0/0.09$\pm$0.12&0/$0.22\pm$0.33&0/0.29$\pm$0.28&0/0.17$\pm$0.18&0/0.13$\pm$0.14&0/0.21$\pm$0.22&0/0.01$\pm$0.02& \ref{s:kink2}\\ \cline{1-11}
$\ntone\ntone$       & (3) zero   &1/1.06$\pm$0.45&0/0.15$\pm$0.11&0/0.38$\pm$0.18&0/0.31$\pm$0.17&0/0.30$\pm$0.15&0/0.24$\pm$0.15&1/0.38$\pm$0.22&0/0.03$\pm$0.04& \ref{s:multil}\\ \cline{2-11}
($\stau_1$)          & (4) short  &1/1.20$\pm$0.33&1/0.23$\pm$0.07&0/0.57$\pm$0.31&0/1.25$\pm$0.90&1/0.23$\pm$0.05&0/0.91$\pm$0.31&1/1.24$\pm$0.53&0/0.08$\pm$0.04& \ref{s:largeip3}\\ \cline{2-11}
                     & (5) medium &0/0.69$\pm$0.52&0/0.00$\pm$0.01&0/0.01$\pm$0.01&1/0.06$\pm$0.04&0/0.04$\pm$0.03&0/0.09$\pm$0.18&0/0.15$\pm$0.28&0/0.01$\pm$0.02&  \ref{s:kink3}\\ \cline{2-11}
                     & (6) long   &0/0.07$\pm$0.05& --- &0/0.08$\pm$0.02&0/0.33$\pm$0.22&0/0.04$\pm$0.01&0/0.06$\pm$0.02&0/0.11$\pm$0.04&0/0.01$\pm$0.00& \ref{s:stable2}\\ \cline{1-11}
$\chpone\chnone$     & \multicolumn{10}{l}{\raisebox{-1.5ex}[-1.5ex]{
                                           short NLSP lifetime as for (1),\ \ 
                                           medium lt. as for (2),\ \ 
                                           long lt. as for (6)   }} \\ 
($\stau_1$)          & \multicolumn{10}{l}{\ } \\ \cline{1-11}
$\smu_R^+\smu_R^-$   & (7) zero   &2/1.59$\pm$0.60&0/0.22$\pm$0.13&0/0.56$\pm$0.23&1/0.46$\pm$0.20&0/0.46$\pm$0.19&1/0.35$\pm$0.20&1/0.57$\pm$0.31&0/0.05$\pm$0.05& \ref{s:multil}\\ \cline{2-11}
($\stau_1$)          & (8) short  &4/3.68$\pm$1.08&1/0.64$\pm$0.19&1/1.00$\pm$0.37&2/1.91$\pm$0.92&2/0.62$\pm$0.16&1/2.23$\pm$0.60&2/3.42$\pm$1.01&0/0.22$\pm$0.066& \ref{s:largeip4}\\ \cline{2-11}
                     & (9) medium &1/1.34$\pm$0.67&0/0.13$\pm$0.09&0/0.33$\pm$0.23&1/0.41$\pm$0.30&0/0.25$\pm$0.13&1/0.11$\pm$0.18&0/0.18$\pm$0.28&0/0.01$\pm$0.02& \ref{s:kink4}\\ \cline{2-11}
                     & \multicolumn{10}{l}{long NLSP lifetime as for (6) } \\ \cline{1-11}
$\sel_R^+\sel_R^-$   & \multicolumn{10}{l}{\raisebox{-1.5ex}[-1.5ex]{
                                           zero NLSP lifetime as for (7),\ \ 
                                           short lt. as for (8),\ \ 
                                           medium lt. as for (9),\ \ 
                                           long lt. as for (6)   }} \\
($\stau_1$)          & \multicolumn{10}{l}{\ } \\ \cline{1-11}
\hline\hline
$\sle_R^+\sle_R^-$ & (10) short  &0/0.05$\pm$0.03&0/0.01$\pm$0.01&0/0.01$\pm$0.01&0/0.02$\pm$0.01&0/0.01$\pm$0.01&0/0.01$\pm$0.01&0/0.01$\pm$0.73&0/0.01$\pm$0.01& \ref{s:largeip1}\\ \cline{2-11}
($\sle_R$)         & \multicolumn{10}{l}{medium NLSP lifetime as for (2)   } \\ \cline{1-11}
$\ntone\ntone$     & \multicolumn{10}{l}{\raisebox{-1.5ex}[-1.5ex]{
                                           zero NLSP lifetime as for (3),\ \ 
                                           short lt. as for (4),\ \ 
                                           medium lt. as for (5),\ \ 
                                           long lt. as for (6)   }} \\ 
($\sle_R$)          & \multicolumn{10}{l}{\ } \\ \cline{1-11}
$\chpone\chnone$   & \multicolumn{10}{l}{\raisebox{-1.5ex}[-1.5ex]{
                                         short NLSP lifetime as for (10),\ \ 
                                         medium lt. as for (2),\ \ 
                                         long lt. as for (6)   }} \\
($\sle_R$)          & \multicolumn{10}{l}{\ } \\ \cline{1-11}
\hline\hline
$\sle_R^+\sle_R^-$ & (11 A) short  &0/1.97$\pm$0.45&0/0.43$\pm$0.10&0/1.02$\pm$0.24&0/1.10$\pm$0.25&0/0.45$\pm$0.11&2/1.03$\pm$0.31&1/1.62$\pm$0.48&0/0.12$\pm$0.04& \ref{s:gamgamx}\\ \cline{2-11}
$[$(11) only$]$    & (12 B0) short  &0/0.13$\pm$0.05&0/0.03$\pm$0.01&0/0.03$\pm$0.01&0/0.04$\pm$0.01&0/0.03$\pm$0.01&0/0.16$\pm$0.05&0/0.25$\pm$0.09&0/0.02$\pm$0.00& \ref{s:gamgamx}\\
 and               & (12 B1) short  &0/1.71$\pm$0.40&0/0.45$\pm$0.12&1/0.99$\pm$0.26&3/1.17$\pm$0.29&0/0.43$\pm$0.18&0/1.43$\pm$0.48&2/2.25$\pm$0.75&0/0.16$\pm$0.41& \ref{s:gamgamx}\\
$\chpone\chnone$   & (12 B2) short  &0/0.60$\pm$0.15&0/0.12$\pm$0.04&0/0.27$\pm$0.09&1/0.30$\pm$0.09&1/0.12$\pm$0.04&0/0.47$\pm$0.16&1/0.74$\pm$0.25&0/0.05$\pm$0.01& \ref{s:gamgamx}\\
($\ntone$)         & (12 C1) short &0/0.12$\pm$0.08&0/0.03$\pm$0.01&0/0.06$\pm$0.03&0/0.04$\pm$0.01&0/0.05$\pm$0.02&0/0.12$\pm$0.04&1/0.19$\pm$0.06&0/0.01$\pm$0.00& \ref{s:gamgamx}\\
 $[$(11) + (12)$]$ & (12 C2) short &2/0.60$\pm$0.15&0/0.09$\pm$0.03&0/0.25$\pm$0.07&0/0.22$\pm$0.06&0/0.16$\pm$0.04&0/0.60$\pm$0.20&0/0.95$\pm$0.32&0/0.07$\pm$0.02& \ref{s:gamgamx}\\
\end{tabular}
}}
\end{center}
\caption{
\label{t:listresults}
Number of observed and expected events for all searches described in this
paper. The error on the expected number of events 
include both statistical and systematic effects.
In the interpretation,
results of analysis (6) at lower centre-of mass energies
are used in addition. Here the numbers of observed and expected events are
0/0.01$\pm$0.00 at $\sqrt{s}=130\,$GeV,
0/0.01$\pm$0.00 at $\sqrt{s}=161\,$GeV,
0/0.01$\pm$0.00 at $\sqrt{s}=171\,$GeV and
0/0.05$\pm$0.01 at $\sqrt{s}=183\,$GeV. 
The numbers given in brackets in the first row of the table give
the average recorded luminosity in each energy bin.
}
}}\end{turn}
\end{table}

\section{Analysis combination strategy}
\label{s:strategy}
For all sparticle production and decay channels 
the results from several analyses 
at several centre-of-mass energies have to be combined.
Although the different analyses are sensitive to distinct topologies, some 
correlation, an overlap between selected data, background and signal events,
is expected. To treat the overlaps between the
various analyses properly, all used the same signal Monte Carlo
samples, which were generated for each search channel, as well as the same
samples of the simulated Standard Model background.
This allows, via an event-by-event comparison of the selected signal,
expected background and selected data events, the determination
of the overlaps between all combinations of the analyses used in each channel. 

The selection efficiencies were split into exclusive selection efficiencies
and overlap efficiencies:
%
the efficiency for events selected by 
only one analysis are denoted as exclusive selection efficiencies;
events that are selected by two or more analyses are
described by overlap efficiencies.
For example, for the case of three analyses, 
there are, in total, three exclusive selection efficiencies and four overlap 
efficiencies (three for the overlaps between each two analyses and one for 
the overlap among all three analyses). 

For all search channels the signal efficiencies, i.e.~exclusive 
selection efficiencies and significant overlap efficiencies, were
determined at each generated mass--lifetime point of the signal Monte Carlo
samples.
For data as well as for the expected background events
the overlaps between the different analyses were investigated.
The overlap for the background 
was found to be negligible ($<0.01$ events) in all slepton NLSP channels.
No data event was selected by more than one analysis.
The various overlap selections were treated as 
separate search channels when combining them with the exclusive
selections in the limit calculations.

In order to avoid generating and processing excessive amounts of 
signal Monte Carlo samples and to achieve a good description of the
efficiencies over the whole mass and lifetime range 
an interpolating function was determined to calculate the
efficiencies at any given mass, lifetime point and centre-of-mass energy. 
This is particularly important in the case that the NLSP occurs at
the end of a cascade decay. In this case, it is impossible to
simulate all possible mass combinations of the SUSY particles, hence the
efficiency of the analyses has to be determined by the kinematics of the
NLSP.
In the next section such a fit function is discussed.

\subsection{The efficiency function}
\label{s:strateff1}
The general ansatz for the interpolating function uses the following
assumptions:
\begin{itemize}
\item The decay of the NLSP is described by  $\exp (-\frac{t}{\tau})$, where
      $t$ is the time since the NLSP production and $\tau$ is the NLSP
      lifetime.
\item The detector to measure the decay is a sphere with a fiducial volume
ranging from a radius \lstart\ to a radius \lstop .
\item The detection efficiency of the detector $\eps_d$ and the
efficiency of the analysis $\eps_a$,
(overall efficiency $\epsplat=\eps_d\otimes \eps_a$) are
constant within the bounds \lstart\ and \lstop . 
\item Initial and final state radiation are neglected, which is a good 
assumption for heavy particles near the kinematic limit.
\end{itemize}

The probability of detecting a particle is a function of 
its mean decay length \lmean, which is a function of 
its lifetime, $\tau$, and its boost, $\beta\gamma$.
For pair production, $\beta\gamma$ is a function of the
particle mass, $m_1$, and centre-of-mass energy $\sqrt{s}$.
Then the probability of an analysis with an
overall efficiency \epsplat\ detecting a particle decay
inside a detector with the radii \lstart\ and  \lstop\ 
is given by
\begin{equation}
\label{eq:effsingle}
  \epssing
  = \epsplat \left [ \exp \left (-\frac{\lstart}{\lmeano} \right) - 
                    \exp  \left (-\frac{\lstop}{\lmeano} \right) \right ]\,\, ,
\end{equation}
with \lmeano , the decay length of the particle, defined as
\begin{equation}
\label{eq:effevent}
  \lmeano = c \tau \beta _1 \gamma _1 = c \tau\sqrt{\frac{s}{4 m_1^2} - 1}.
\end{equation}
The values $\epssing$, $\lstart$ and $\lstop$ are effective values which are
determined by a fit of the efficiency function (\ref{eq:effsingle})
to the efficiencies obtained
for the signal Monte Carlo samples, treating each channel separately.
Depending on whether an analysis accepts events where one or both
pair-produced particles have to decay in the detector, the event selection
efficiency \epsevent\ is given by
$$
  \epsevent = 1 - [1 - \epssing]^2
{\,\,\,\,\,\,\,\rm or\,\,\,\,\,\,\,}
  \epsevent = \epssing^2\,\,{\rm respectively}.
$$
These formulae are used to describe the selection efficiency of the search
analysis as a function of the NLSP $\beta\gamma$
over a wide range of masses, lifetimes and production energies
with three parameters,
\epsplat, \lstart\ and \lstop . The input variables are
the mass of the decaying particle, its lifetime 
and the centre-of-mass energy.
For the calculation of the efficiency of the 
zero lifetime search, the parameter \lstart\ can be set to zero as
the sensitive volume starts at the interaction point. For the
heavy charged stable particle search, the parameter 
\lstop\ can be set to infinity. 

\begin{figure}[H]
     \begin{minipage}[t]{.1cm}
       \epsfxsize=.1cm
     \end{minipage}
   \hfill
     \begin{minipage}[t]{15.5cm}
       \epsfxsize=15.5cm
       \epsfbox{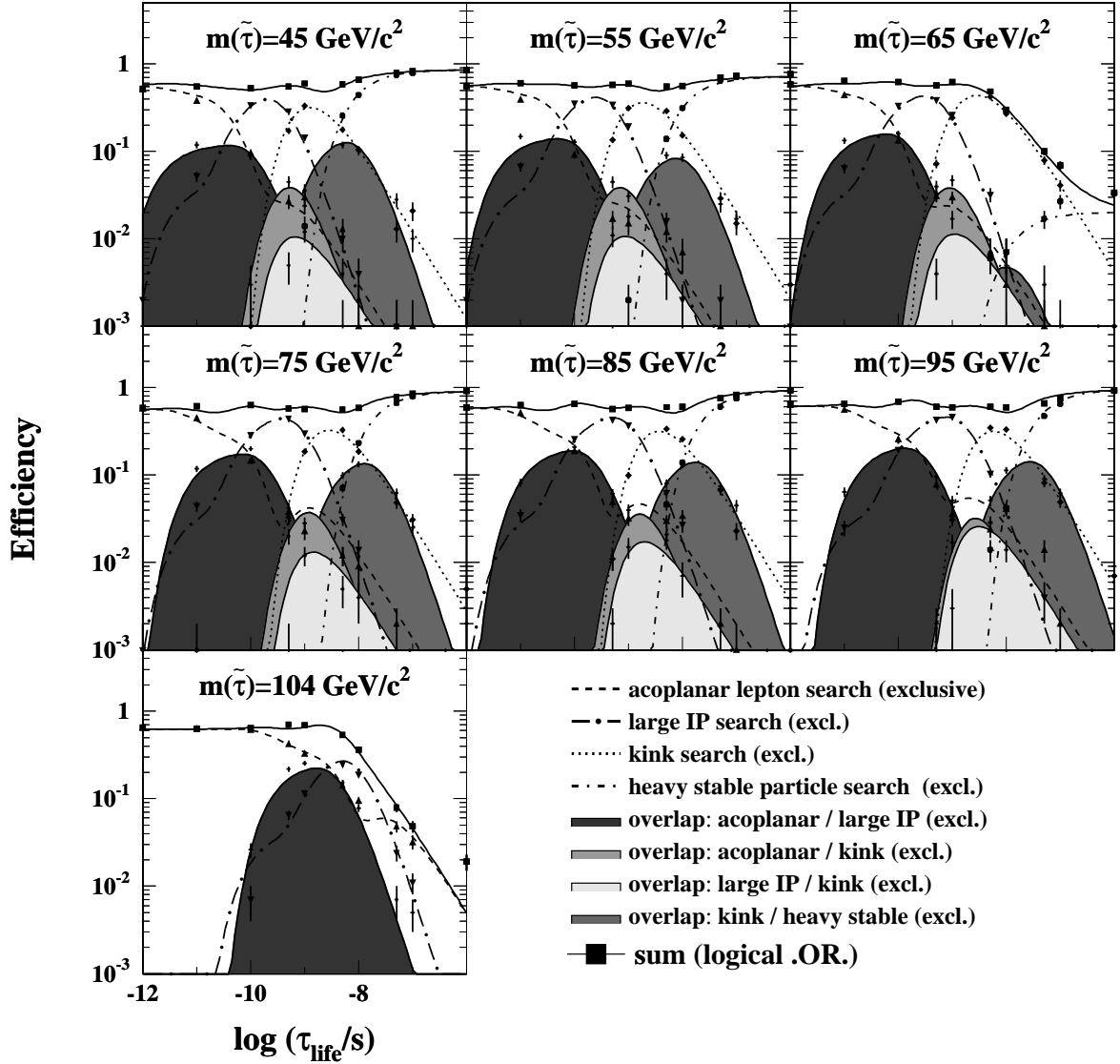}
     \end{minipage}
   \hfill
     \begin{minipage}[t]{.1cm}
       \epsfxsize=.1cm
     \end{minipage}
\caption{
\label{f:staueff}
{ Efficiencies for stau pair-production at $\sqrt{s}=208\,$GeV,
as a function of the lifetime for fixed stau masses.
The symbols represent the efficiencies for ten simulated lifetimes and
the curves show the interpolating
efficiency functions for the different searches:
the exclusive search for promptly decaying staus (dashed),
the exclusive search for large impact
parameters (long dashed-dotted), the exclusive search for kinks (dotted)
and the exclusive search for stable staus (dashed-dotted). The overlap
efficiencies between these searches
are shown as filled histograms.
The sum of all efficiencies, i.e.~the four exclusive
selection efficiencies and the four overlap efficiencies, is
shown by the squares and the corresponding solid line.}
}
\end{figure}

An example of the efficiencies described by the function is given in
Figure~\ref{f:staueff} for stau pair-production at $\sqrt{s}=208\,$GeV.
Here the search for acoplanar leptons, for tracks with large 
impact parameters or kinks, the search for particles with anomalous 
specific ionization energy loss plus all overlap analyses are included.

\subsubsection{The efficiency function for NLSP in cascade decays}
\label{s:strat_eff2}
If the particle $T_1$ with lifetime $\tau$ is not pair-produced
but a decay product
of another pair-produced particle $T_0$ with mass
$m_0$ and energy $E_0=\frac{\sqrt{s}}{2}$, the average decay length
$l\sub{mean}$
has to be replaced by
$$
 l\sub{mean} = \beta _1\gamma _1 c \tau = 
                             \frac{|p_1|}{m_1} c \tau\,.
$$
Here $|p_1|$, the average momentum of $T_1$, is calculated from the two-body
decay of $T_0$ into $T_1$ and a second decay product.

\subsubsection{Selection efficiency for long-lived, charged NLSP from cascade decays}
\label{s:effimarkus}

The use of the average momentum in the calculation of the NLSP decay length
$l\sub{mean}$ as described in the previous
section is only justified if the
efficiency, as a function of the NLSP $\beta\gamma$, is a smooth function
with moderate variation of the order of a few percent.
As the efficiency of searches for long-lived, heavy particles
from cascade decays drops drastically from about 80$\,$\% to below 5$\,$\%
for a $\beta\gamma$ of the particle between 0.7 and 1.7,
a different method has to be applied.
Instead of using the average NLSP momentum $|p|/m$,
the kinematically possible $\bg$ spectrum after the decay of a primary particle
is used, following the ansatz
$$
\epsilon=\int{{\epsilon_{0}(\beta_{1}\gamma_{1},\beta_{2}\gamma_{2})}\cdot{g(\beta_{1}\gamma_{1},\beta_{2}\gamma_{2})}\,\,\mathrm{d}\beta_{1}\gamma_{1}\mathrm{d}\beta_{2}\gamma_{2}}\,.
$$
Here, $\epsilon_{0}$ is the efficiency to select a signal event
expressed as a function of the $\bg$ values of the two possible NLSPs.
It is determined using simulated signal events.  
The normalized $\bg$ spectrum of
the heavy stable charged particles is given by the function $g$. 
The spectrum is calculated assuming a flat energy spectrum
of the stable particles in the kinematically accessible
range and neglecting ISR. 

\subsubsection{Errors on the interpolated efficiencies}
\label{s:strat_errors}



In addition to the systematic uncertainties on the selections 
described in the analysis sections,
an additional source of the systematic uncertainty 
for the signal selection
results from the
interpolation of the efficiencies to different mass and lifetimes. 
Its size was estimated
by recalculating the parameters of the efficiency function 
dropping the information of one simulated signal sample 
and
comparing the interpolated efficiency with the achieved efficiency
at this point. Deviations 
were found to be less
than 3$\,$\%.
Finally the efficiency function was varied by the errors on the fitted
parameters \epsplat, \lstart\ and \lstop. Here an uncertainty
of less than 1$\,$\% was found.
The total systematic error assigned to the efficiency function
is calculated by adding the three terms in quadrature.

The efficiency function
for long-lived, charged NLSPs in cascade decays, 
Section~\ref{s:effimarkus}, has regions of rapid
variation and an additional systematic error is 
assigned.
In the region of NLSP $\bg$ between 0.75 and 1.70
the efficiency is varying rapidly to small values,
and relative uncertainties up to $43\,$\% on the modelling
and 31$\,$\% on the interpolation of the function were found.
In the high efficiency region, the two contributions were determined to be
$11\,$\% on the modelling and $8\,$\% on the interpolation.
These dominant effects were included 
in the calculation of the limits on the production cross-section
of SUSY particles.

\section{Constraints on particle production cross-sections}
\label{s:expresults}
The absence of any significant excess of events in the data
compared to the expected
number of Standard Model events in any of the search analyses
can be translated into limits on the production cross-section of the
SUSY particles at the 95$\,$\%\ confidence level. 
The limits were calculated using the program described in \cite{limcalc},
which incorporates statistical and systematic uncertainties on the efficiency
and the expected background as well as the uncertainties on the luminosity 
of the analyzed data samples and
uncertainties on the expected cross-sections into the limits 
using a numerical convolution technique.
The results for each analysis were given at eight different bins of the
centre-of-mass energy (see Table~\ref{t:listresults}).
For all analyses, systematic and statistical uncertainties on the signal 
efficiency and on the background expectation
as well as uncertainties on the recorded luminosity were included.


Deriving cross-section limits at a single centre-of-mass
energy requires knowledge of the cross-section evolution
with $\sqrt{s}$ and,
in general, this evolution depends on the details of 
the SUSY model.  For direct NLSP pair-production, the 
GMSB model database described in Section~\ref{s:interpretation}
is scanned, and the cross-section evolution for each point
in the database is used to calculate the cross-section limit
at $\sqrt{s}=$~208~GeV for each mass and lifetime.  The maximum
excluded cross-section limit is chosen as the ``model independent''
limit for this mass and lifetime.  For ``lifetime independent'' exclusions, 
the maximum cross-section limit valid for all lifetimes is chosen.

In the channels with cascade
decays to the NLSP, this procedure does not work because there
are points in the GMSB model with vanishing NNLSP production
cross-sections and therefore the evolution is not defined.
Instead, in these cascade channels,
for spin 1/2 supersymmetric particles a
$\beta/s$ dependence of the cross-section is assumed, while for scalars
a $\beta^3/s$ dependence is used.  
The cross-section limits quoted
in the cascade channels were evaluated for the ``worst-case'' of
all intermediate particle masses, as well as NLSP lifetime.

\subsection{The stau or slepton as the NLSP} 
\label{s:expresults1}
Here the results of searches for signatures expected if the NLSP is the
lightest slepton are described. The sleptons are either produced
directly or appear as decay products of heavier pair-produced SUSY
particles. 

\subsubsection{Cross-section limits for direct NLSP (stau or slepton) pair-production}
\label{s:expresults11}
To obtain constraints on the production cross-sections for slepton 
pair-production in the slepton co-NLSP scenario and stau pair-production 
in the stau NLSP scenario the search for acoplanar
leptons (Section~\ref{s:accop}),
the search for tracks with large impact parameters (Section~\ref{s:largeip1}), 
the search for kinked tracks (Section~\ref{s:kink2})
and the search for heavy stable charged particles (Section~\ref{s:stable1})
were combined. 
A significant overlap of up to 40$\,$\% in the signal efficiency was found
between the acoplanar lepton and large impact parameter searches.
The overlap between the acoplanar lepton and kink searches is about $4\,$\%, 
between the large impact parameter and kink searches about $7\,$\%,
and between the kink and heavy stable charged particle searches
up to $20\,$\%. 
Examples of the exclusive and overlap signal efficiencies can be seen in 
Fig.~\ref{f:staueff} for the stau searches.
There is no data event selected in common, while for the expected background 
an overlap of 0.01 events was found between the large impact 
parameter and acoplanar lepton searches in the selectron and the stau channels.

\begin{figure}[H]
     \begin{minipage}[t]{.1cm}
       \epsfxsize=.1cm
     \end{minipage}
   \hfill
     \begin{minipage}[t]{16.cm}
\vspace*{1.cm}
       \epsfxsize=16.cm
       \epsfbox{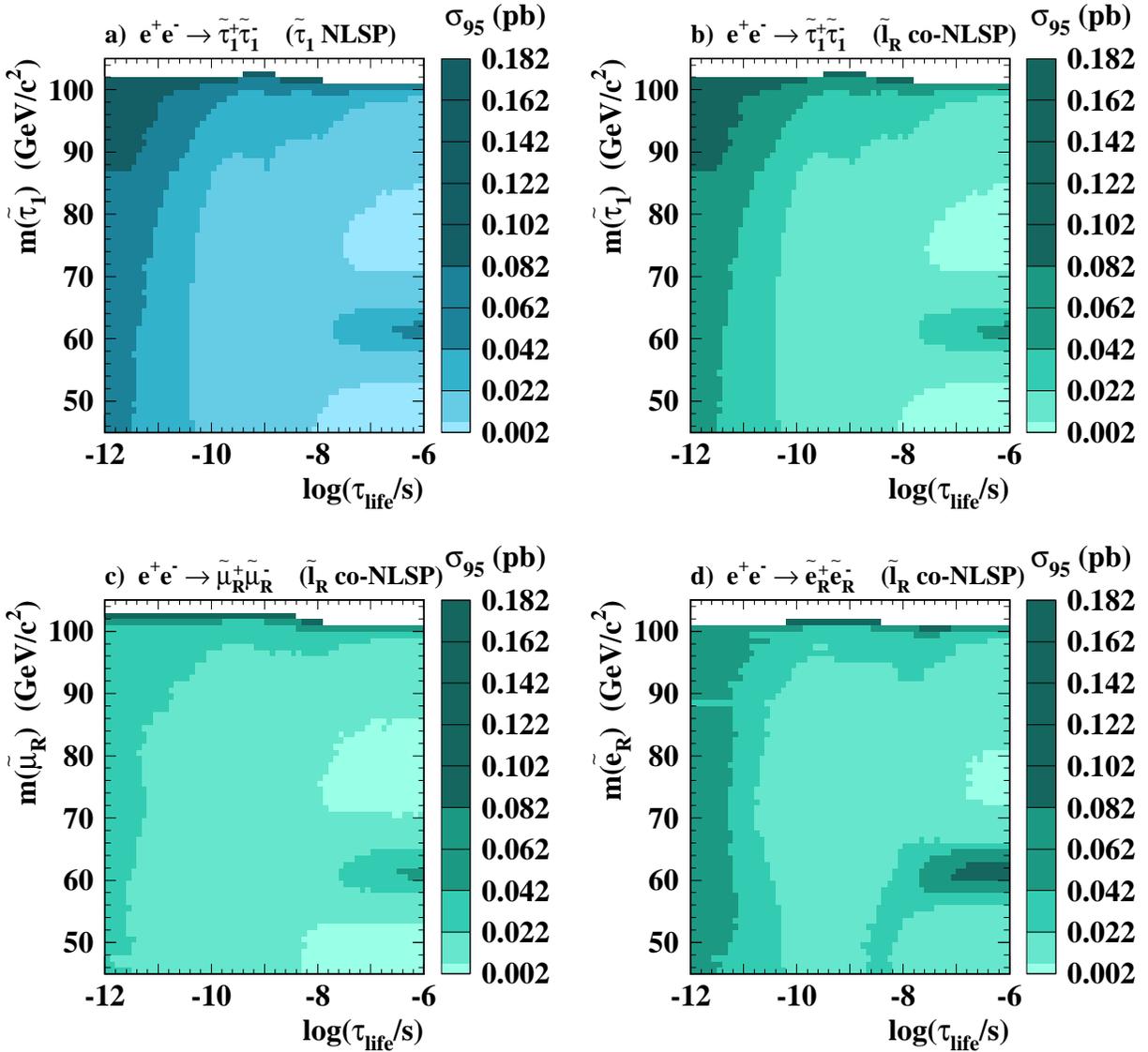}
     \end{minipage}
   \hfill
     \begin{minipage}[t]{.1cm}
       \epsfxsize=.1cm
     \end{minipage}
\caption{
\label{f:xsec_slept}
 Contours of the 95$\,$\%\ C.L. upper limits on the 
production cross-sections $\sigma_{95}$ for slepton pair-production at 
$\sqrt{s}=208\,$GeV as a function of the slepton mass and lifetime $\tau_{\rm life}$.
Shown are 
cross-section limits for (a) staus in the stau NLSP scenario, 
(b) staus, (c) smuons and (d) selectrons in the slepton co-NLSP scenario.
The shadings correspond to different ranges 
of the 
upper limit on the cross-section, as indicated by the 
scale on the right side 
.}
\end{figure}

The resulting limits at 95\,\% C.L. at $\sqrt{s}=208\,$GeV are given in
Figure~\ref{f:xsec_slept} for
staus in the stau NLSP scenario (a) and for staus (b), smuons (c) and
selectrons (d) in the slepton co-NLSP scenario.
For all flavours the 
weakest exclusion is found in the very short lifetime region 
($\tau=10^{-12}-10^{-11}\,$s), due to the 
irreducible background from $\mathrm{W^+W^-}$ production in the acoplanar 
lepton topology, whereas all other regions are almost free of expected
background.
Also, the smaller sensitivity for masses 
around $60\,$GeV$/c^2$ at long lifetimes, due to the loss of sensitivity at
d$E$/d$x$ band crossings as described in Section~\ref{s:stable1},
is visible in all channels. 
Because of the highest selection efficiencies, the best constraints
are obtained for smuons   
with an upper limit of 0.05\,pb in most of the plane, apart from the 
region with a smuon mass above 100\,GeV/$c^2$, close to the kinematic 
limit.
For selectrons the constraints are 
slightly weaker. Here cross-sections larger than 0.1\,pb can be excluded in 
most of the selectron mass--lifetime plane. The difference between 
smuons and selectrons is explained by the fact that selectrons can be 
produced in the $t$-channel, which leads to a more forward production and 
thus to lower efficiencies. For stau pair-production, cross-sections larger
than 0.1\,pb are excluded.
The stau limits in the slepton co-NLSP 
scenario and in the stau NLSP scenario are 
rather similar, with a small difference due to the slightly different 
theoretical cross-sections that are used for the combination of the results 
at various centre-of-mass energies. 

\subsubsection{Cross-section limits for neutralino pair-production}
Two scenarios exist in the theory: in the stau NLSP scenario the neutralino 
decays with 100\,\% branching ratio to $\tilde{\tau}_1\tau$, while in the 
slepton co-NLSP scenario it decays with equal branching fractions to all 
flavours. Both cases were studied.
Four analyses were combined: 
the search for events with four or more leptons plus missing
energy (Section~\ref{s:multil}),
the searches for tracks with large impact parameters (Section~\ref{s:largeip3})
or kinks (Section~\ref{s:kink3})
and the search for
tracks with an anomalous ionization energy loss in events with more than two
tracks (Section~\ref{s:stable2}).
No indication of new physics was observed in any of the analyses.
Significant overlaps exist in the signal efficiency
between the search for four leptons plus missing energy and the 
large impact parameter search (up to 
20\,\%) and between the large impact parameter and the kinked track search 
(up to 15\,\%). 
The efficiencies
in the overlap of other selections were found to be less than 0.1$\,$\%.
No data event was selected in common by two analyses.
This is
compatible with the shared expected background of 0.01 events
between the search for multi-leptons plus 
missing energy and the large impact parameter search and no expected event
for the other combinations.
The results of the different analyses
were combined assuming a $\beta/s$ dependence of the production
cross-sections at the various centre-of-mass energies. 

\begin{figure}[H]
     \begin{minipage}[t]{.1cm}
       \epsfxsize=.1cm
     \end{minipage}
   \hfill
     \begin{minipage}[t]{16.cm}
       \epsfxsize=16.cm
       \epsfbox{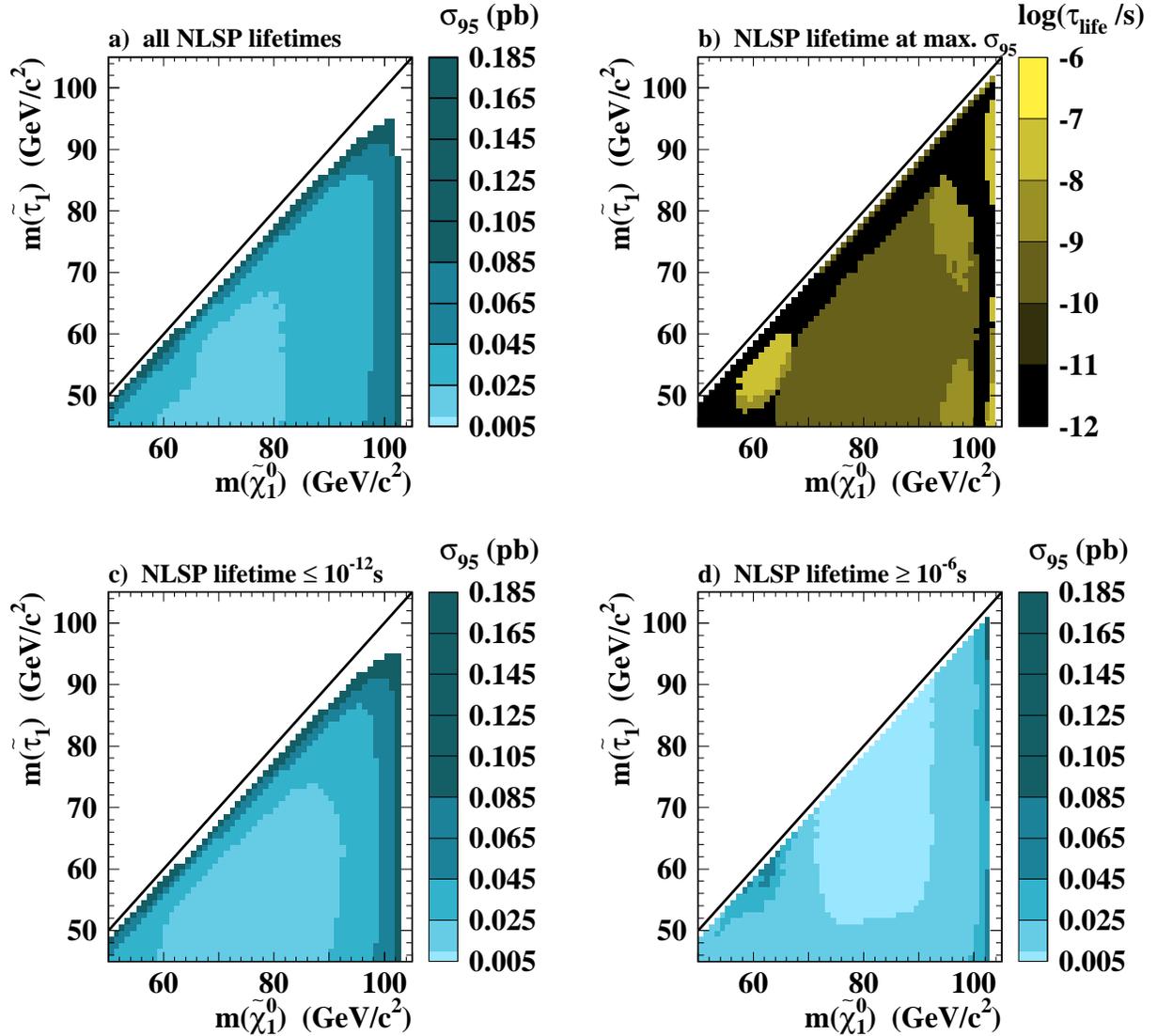}
     \end{minipage}
   \hfill
     \begin{minipage}[t]{.1cm}
       \epsfxsize=.1cm
     \end{minipage}
\caption{
\label{f:4lstaulim}
 Upper limits at 95$\,$\%\ C.L. on the production cross-section for 
neutralino pairs in the stau NLSP scenario at $\sqrt{s}=208\,$GeV.
The data analyzed, taken at different centre-of-mass energies,
were combined assuming a 
$\beta /s$ dependence of the cross-sections. 
In (a) the lifetime independent exclusions are shown as a function
of the neutralino and NLSP mass. The lifetime at which the
limit is set is given in (b).
Figures (c) and (d) give the excluded production cross-section for
neutralinos assuming a very short-lived or a stable NLSP.}
\end{figure}

Limits on the production cross-section of neutralino pairs
at $\sqrt{s}=208\,$GeV
in the stau NLSP case are shown in Figure~\ref{f:4lstaulim}.
In (a) the limit valid for all NLSP lifetimes is given. In
most regions, cross-sections larger than 0.1\,pb can be excluded,
except for low mass differences between the neutralino and the stau.
Figure~\ref{f:4lstaulim}\,(b) shows the NLSP lifetimes at which the
maximum cross-section limit is reached. For small
mass differences between the neutralino and the stau, the limit is set by
the short lifetime searches, whereas the searches for intermediate lifetimes
set the limit in the rest of the parameter space.
Figure~\ref{f:4lstaulim}\,(c) shows the excluded cross-section
for a short-lived NLSP and cross-sections larger than 0.1\,pb can
be excluded. Figure~\ref{f:4lstaulim}\,(d) shows the exclusions
of values larger than 0.05\,pb for a long-lived NLSP.

\begin{figure}[H]
     \begin{minipage}[t]{.1cm}
       \epsfxsize=.1cm
     \end{minipage}
   \hfill
     \begin{minipage}[t]{16.cm}
       \epsfxsize=16.cm
       \epsfbox{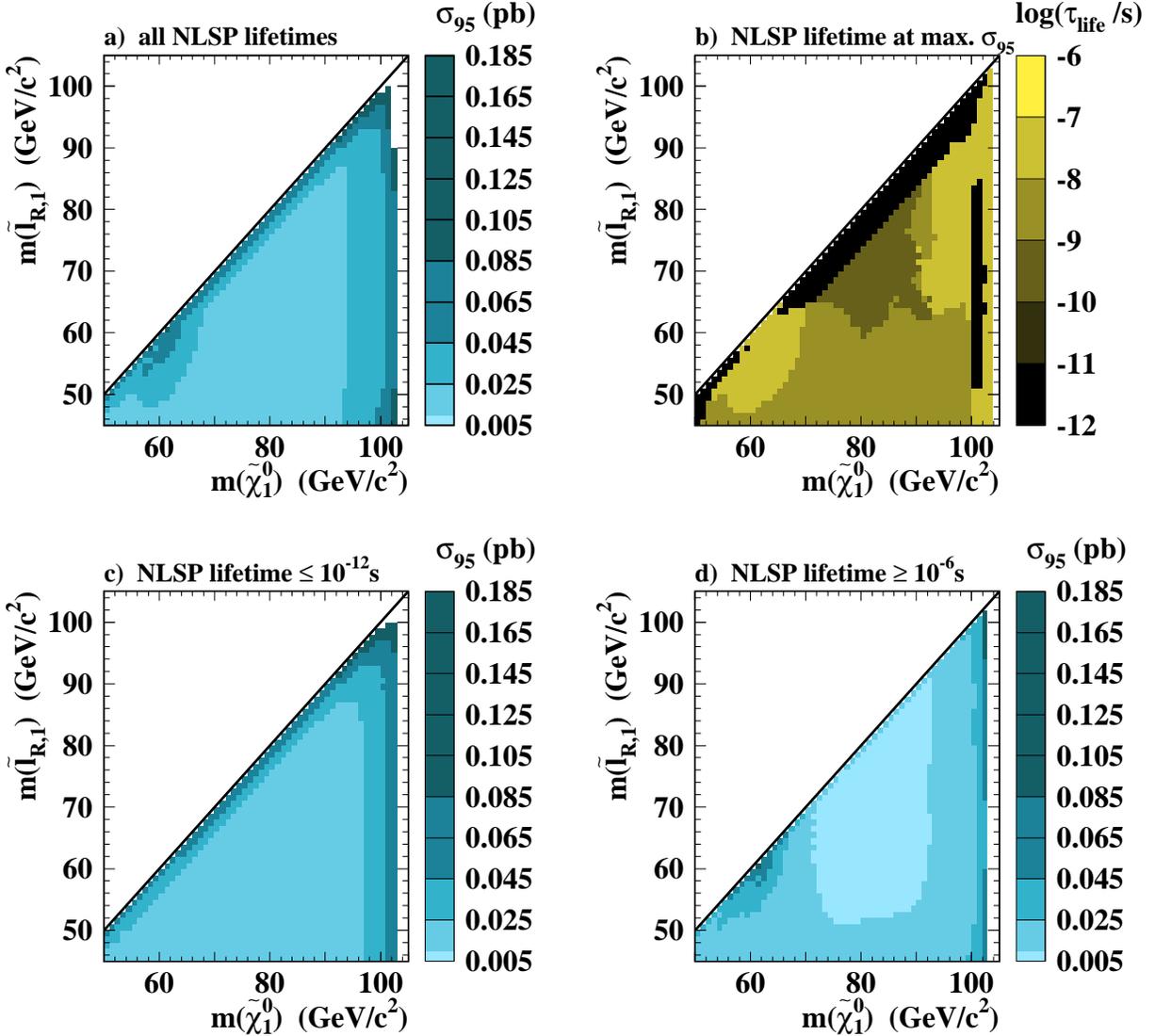}
     \end{minipage}
   \hfill
     \begin{minipage}[t]{.1cm}
       \epsfxsize=.1cm
     \end{minipage}
\caption{
\label{f:4lsleptlim}
 Upper limits at 95$\,$\%\ C.L. on the production cross-section for 
neutralino pairs in the slepton co-NLSP scenario at $\sqrt{s}=208\,$GeV.
The data analyzed, taken at different centre-of-mass energies,
were combined assuming a 
$\beta /s$ dependence of the cross-sections. 
In (a) the lifetime independent exclusions are shown as a function
of the neutralino and NLSP mass. The lifetime at which the
limit is set is given in (b).
Figures (c) and (d) give the excluded production cross-section for
neutralinos assuming a very short-lived or a stable NLSP.}
\end{figure}

In the slepton co-NLSP scenario, slightly better results are obtained because
of higher efficiencies of the short lifetime analyses for muons and electrons
compared to tau searches. As shown in
Figure~\ref{f:4lsleptlim}\,(a), cross-sections larger than 
0.1\,pb can be excluded for all NLSP lifetimes, even for small mass differences
between the neutralino and the slepton. 
Figure~\ref{f:4lsleptlim}\,(b) illustrates that
the limits are set by the medium lifetime searches in most cases,
while for small mass differences they are set by the short lifetime searches.
Figures~\ref{f:4lsleptlim}\,(c) and (d) give the excluded cross-sections
for a short-lived NLSP and a long-lived NLSP, with typical
excluded cross-sections 
larger than 0.1\,pb and 0.05\,pb, respectively.

\subsubsection{Cross-section limits for smuon and selectron pair-production (stau NLSP)}
To search for pair-produced selectrons and smuons which decay via a (virtual)
neutralino to the stau, the NLSP, four searches were combined.
Topologies with a promptly decaying NLSP were searched for using the
analysis sensitive to four or more leptons plus missing energy
(Section~\ref{s:multil}).
Candidate events for NLSPs with a medium lifetime were
identified with the large impact parameter (Section~\ref{s:largeip4})
or kink search (Section~\ref{s:kink4}), while possible stable NLSPs
were searched for by
selecting tracks with an anomalous ionization energy loss in events
with more than two tracks (Section~\ref{s:stable2}).
For all four analyses, the number of selected data events is compatible
with the number expected from Standard Model sources.

Significant overlaps for the signal efficiency exist between the search
for multi-leptons plus missing energy and the large impact parameter search 
(up to 26\,\%), between the multi-lepton search and the kink
search (up to 10\,\%) and between the 
large impact parameter and the kink search (up to 7\,\%).
For the expected background, an overlap of 0.01 events was
found between the search for multi-leptons plus missing energy and the
large impact parameter search. In the overlap regions of other
analyses no event is expected to be selected in common.

\begin{figure}[H]
     \begin{minipage}[t]{.1cm}
       \epsfxsize=.1cm
     \end{minipage}
   \hfill
     \begin{minipage}[t]{16.cm}
\vspace*{1.cm}
       \epsfxsize=16.cm
       \epsfbox{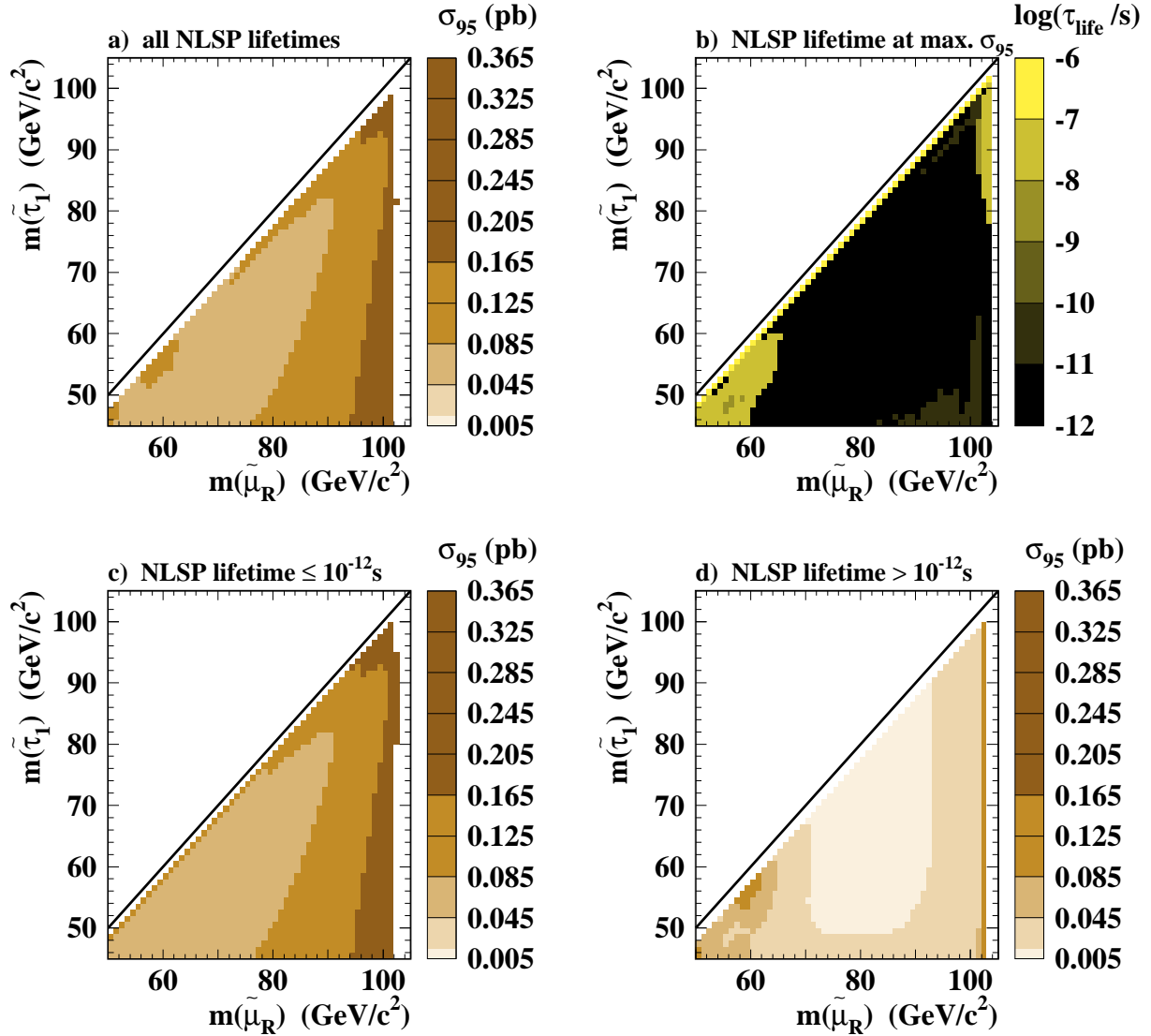}
     \end{minipage}
   \hfill
     \begin{minipage}[t]{.1cm}
       \epsfxsize=.1cm
     \end{minipage}
\caption{
\label{f:6l-smuon}
 Upper limits at 95$\,$\%\ C.L. on the production cross-section for 
smuon pairs in the stau NLSP scenario at 
$\sqrt{s}=208\,$GeV.
The data analyzed, taken at centre-of-mass energies of
$\sqrt{s}=189$--$209\,$GeV,
were combined assuming a 
$\beta^3 /s$ dependence of the cross-section. 
In (a) the lifetime independent exclusions are shown as a function
of the smuon and NLSP mass. The lifetime at which the
limit is set is given in (b).
Figures (c) and (d) give the excluded production cross-section for
smuons assuming a very short-lived or a stable NLSP.}
\end{figure}

The results for the different analyses
were combined assuming a $\beta^3/s$ dependence of the production
cross-sections at the various centre-of-mass energies.
Figure~\ref{f:6l-smuon}\,(a) shows the upper limit at 95\,\% C.L. 
on the production cross-section of smuon pairs as a function
of the NLSP mass at $\sqrt{s}=208\,$GeV for any lifetime of the NLSP.
Cross-sections larger than 0.4\,pb can be excluded,
independent of the neutralino and NLSP mass,
except for low mass differences between the smuon and the stau.
The limit is set at short NLSP lifetimes in most cases, while for 
smuon masses below 65$\,$GeV$/c^2$ the maximum excluded cross section
is given by searches for a stable NLSP (Figure~\ref{f:6l-smuon}\,(b)).
Figure~\ref{f:6l-smuon}\,(c) shows that for a short-lived NLSP
cross-sections larger than 0.4\,pb can be excluded.
For a long-lived NLSP limits on the production cross-section of smuon pairs
less than 0.1\,pb are achieved (Figure~\ref{f:6l-smuon}\,(d)).

\begin{figure}[H]
     \begin{minipage}[t]{.1cm}
       \epsfxsize=.1cm
     \end{minipage}
   \hfill
     \begin{minipage}[t]{16.cm}
\vspace*{1.cm}
       \epsfxsize=16.cm
       \epsfbox{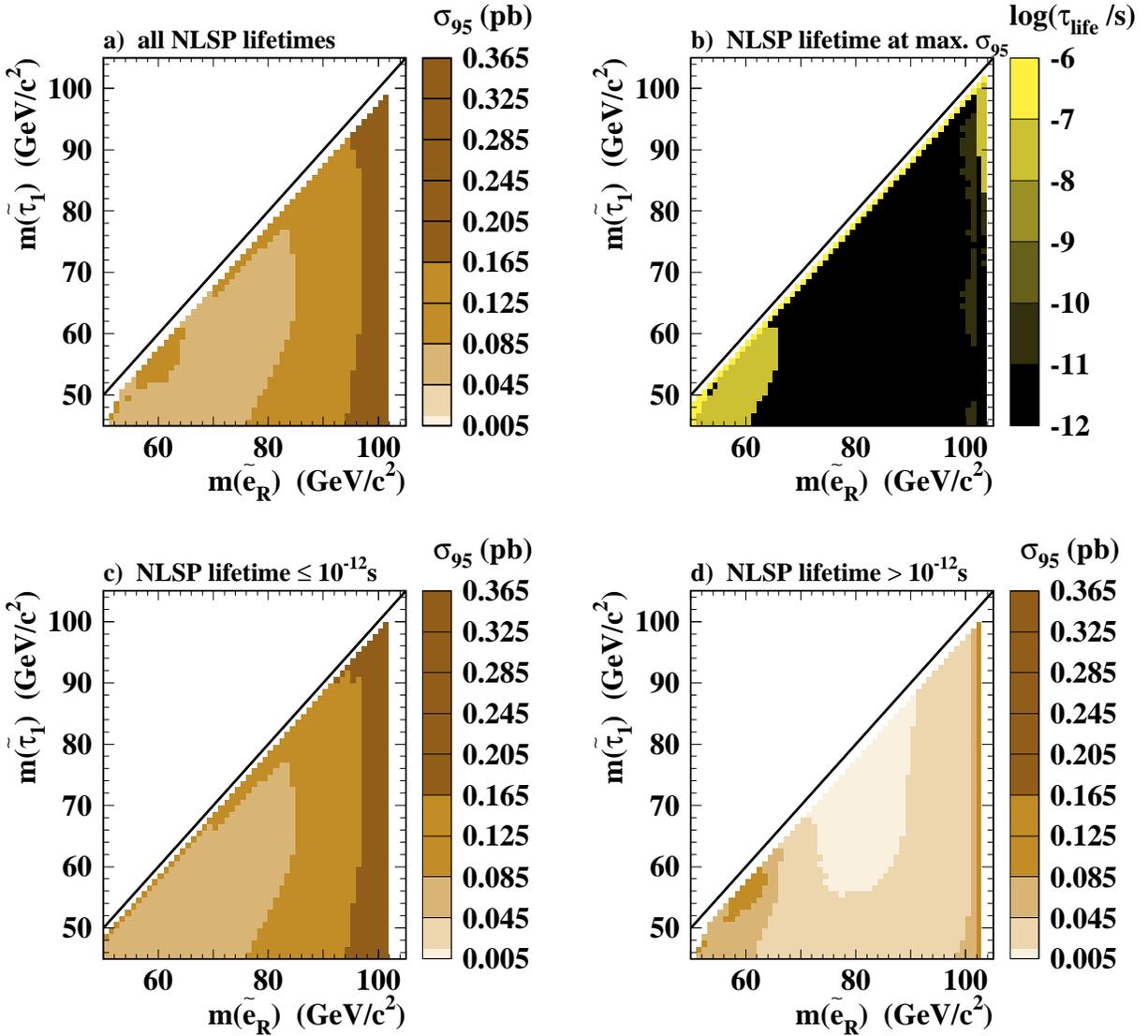}
     \end{minipage}
   \hfill
     \begin{minipage}[t]{.1cm}
       \epsfxsize=.1cm
     \end{minipage}
\caption{
\label{f:6l-selec}
 Upper limits at 95$\,$\%\ C.L. on the production cross-section for 
selectron pairs in the stau NLSP scenario at 
$\sqrt{s}=208\,$GeV.
The data analyzed, taken at centre-of-mass energies of
$\sqrt{s}=189$--$209\,$GeV,
were combined assuming a 
$\beta^3 /s$ dependence of the cross-section. 
In (a) the lifetime independent exclusions are shown as a function
of the selectron and NLSP mass. The lifetime at which the
limit is set is given in (b).
Figures (c) and (d) give the excluded production cross-section for
selectrons assuming a very short-lived or a stable NLSP.}
\end{figure}

Similar values of the excluded production cross-section are achieved for
selectrons as shown in Figure~\ref{f:6l-selec}.

\subsubsection{Cross-section limits for chargino pair-production}
In the scenarios with a stau or a slepton being the NLSP,
the charginos decay to the NLSP plus a neutrino.
Thus the signature visible in the detector
is similar to the one expected from direct slepton NLSP pair-production except
for an additional significant amount of missing energy which is taken away by
the neutrino. If the NLSP decays promptly or inside the detector volume
the same searches as for direct NLSP production can be applied.
These are the searches for
acoplanar leptons (Section~\ref{s:accop}), for tracks with large impact
parameters (Section~\ref{s:largeip1}) and for kinked tracks
(Section~\ref{s:kink2}).
In case of a long-lived NLSP the analysis described in Section~\ref{s:stable2}
has to be applied as the two NLSP tracks are no longer back-to-back.
No sign of new physics was observed by any of the four searches nor their
overlap analyses.

For the exclusive and overlap analyses, the efficiency functions for
NLSP pair-production with a stau or slepton NLSP were used,
taking into account that the NLSP is a secondary particle.
Thus, the efficiency function uses the average
$\beta\gamma$ of the NLSP which is calculated from the kinematics of a
two-body chargino decay, as described in Section~\ref{s:strat_eff2}.
It was checked using a full simulation of chargino events that the applied
interpolating functions describe the efficiencies well.

Cross-section limits were calculated both for the case of equal branching 
ratios of the chargino to all NLSP flavours in the slepton co-NLSP scenario and
100\,\% branching ratio to the stau in the stau NLSP scenario.
The data recorded at various centre-of-mass energies were combined
assuming a $\beta/s$ dependence of the cross-sections.

\begin{figure}[H]
     \begin{minipage}[t]{.1cm}
       \epsfxsize=.1cm
     \end{minipage}
   \hfill
     \begin{minipage}[t]{16.cm}
\vspace*{1.cm}
       \epsfxsize=16.cm
       \epsfbox{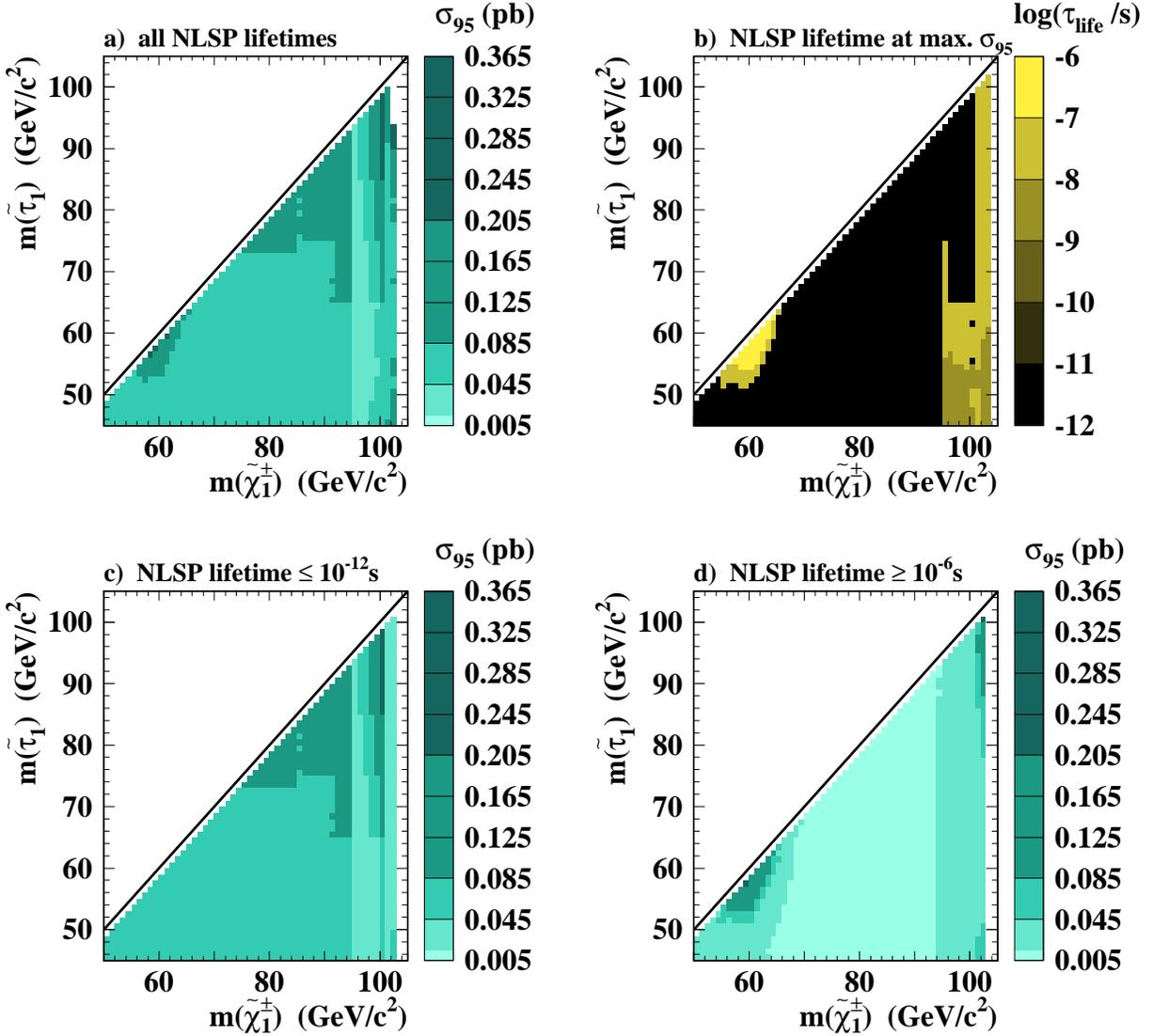}
     \end{minipage}
   \hfill
     \begin{minipage}[t]{.1cm}
       \epsfxsize=.1cm
     \end{minipage}
\caption{
\label{f:chargino-stau}
 Upper limits at 95$\,$\%\ C.L. on the production cross-section for 
chargino pairs in the stau NLSP scenario at $\sqrt{s}=208\,$GeV. 
The data analyzed, taken at different centre-of-mass energies,
were combined assuming a 
$\beta/s$ dependence of the cross-sections.
In (a) the lifetime independent exclusions are shown as a function
of the chargino and the NLSP mass. The lifetime at which the
limit is set is given in (b).
Figures (c) and (d) give the excluded production cross-section for
charginos assuming a very short-lived or a stable stau.}
\end{figure}

Figure~\ref{f:chargino-stau}\,(a) shows the upper limit at 95\,\% C.L. 
on the production cross-section of chargino pairs at $\sqrt{s}=208\,$GeV
for any lifetime of the stau NLSP.
Cross-sections larger than 0.2\,pb
can be excluded for all chargino masses, independent
of the NLSP mass. 
The limit is set at short NLSP lifetimes, except for 
chargino masses below 65$\,$GeV$/c^2$ or close to the kinematic limit 
where the maximum excluded cross section is given by searches for a
stable NLSP (Figure~\ref{f:chargino-stau}\,(b)).
Figure~\ref{f:chargino-stau}\,(c) shows that for a short-lived NLSP
cross-sections larger than 0.2\,pb can be excluded.
For a long-lived NLSP limits on the production cross-section of 
chargino pairs less than 0.1\,pb are achieved
(Figure~\ref{f:chargino-stau}\,(d)).

\begin{figure}[H]
     \begin{minipage}[t]{.1cm}
       \epsfxsize=.1cm
     \end{minipage}
   \hfill
     \begin{minipage}[t]{16.cm}
\vspace*{1.cm}
       \epsfxsize=16.cm
       \epsfbox{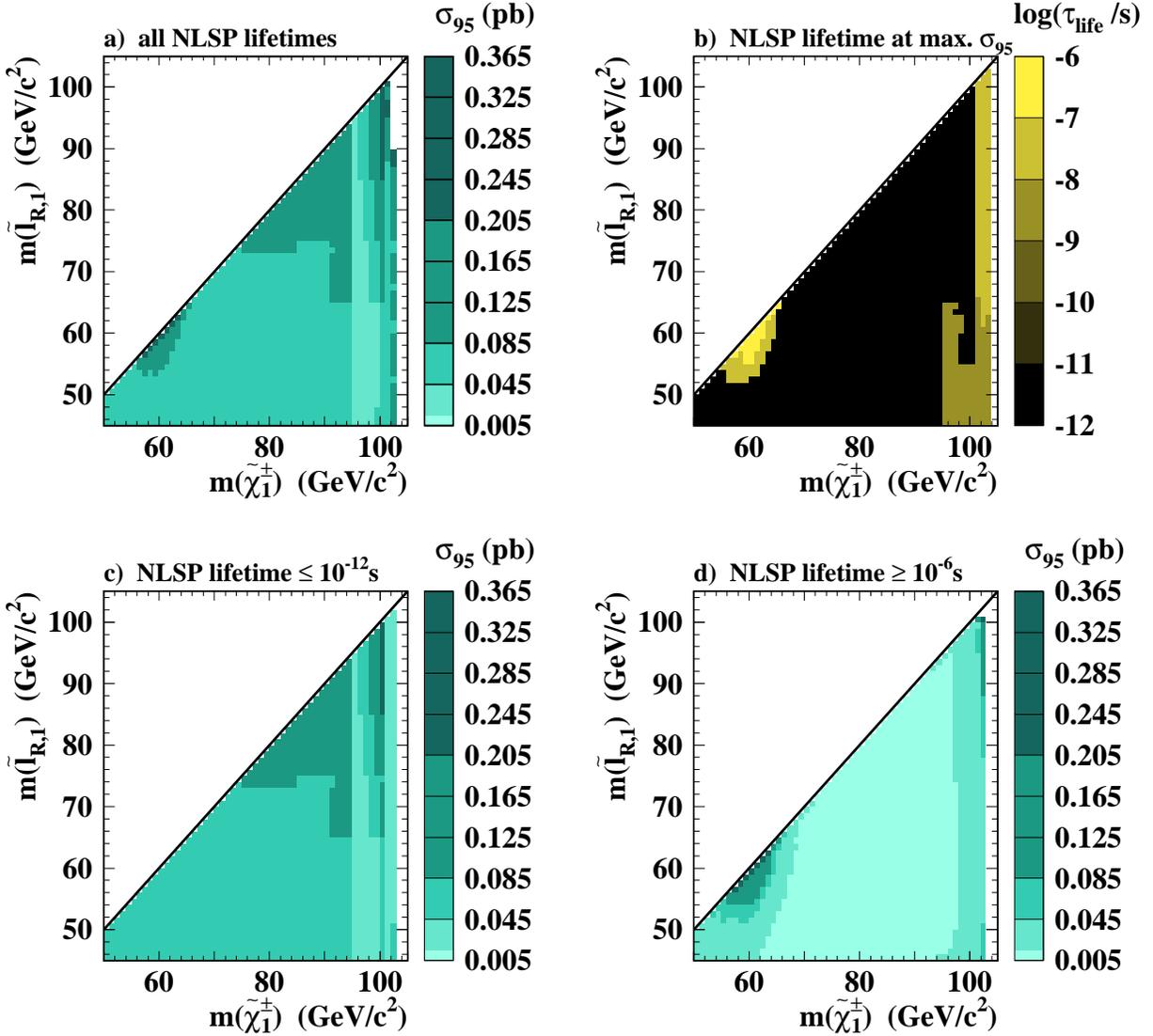}
     \end{minipage}
   \hfill
     \begin{minipage}[t]{.1cm}
       \epsfxsize=.1cm
     \end{minipage}
\caption{
\label{f:chargino-slep}
 Upper limits at 95$\,$\%\ C.L. on the production cross-section for 
chargino pairs in the slepton co-NLSP scenario at $\sqrt{s}=208\,$GeV. 
The data analyzed, taken at different centre-of-mass energies,
were combined assuming a 
$\beta/s$ dependence of the cross-sections. 
In (a) the lifetime independent exclusions are shown as a function
of the chargino and the NLSP mass. The lifetime at which the
limit is set is given in (b).
Figures (c) and (d) give the excluded production cross-section for
charginos assuming a very short-lived or a stable NLSP.}
\end{figure}

Similar values of the excluded production cross-sections are found
for
charginos in the slepton co-NLSP scenario as shown in
Figure~\ref{f:chargino-slep}.

\subsection{The neutralino as the NLSP} 

\subsubsection{Cross-section limits for neutralino pair-production}
Directly produced NLSP pairs are searched for by the acoplanar photon search,
which is sensitive to NLSP decay lengths of about $10\,$cm, corresponding to
a lifetime of $10^{-9}\,$s. If one or both neutralinos decay outside the
detector, there is no acceptance by any analysis. 

\begin{figure}[H]
     \begin{minipage}[t]{.1cm}
       \epsfxsize=.1cm
     \end{minipage}
   \hfill
     \begin{minipage}[t]{16.cm}
\vspace*{1.cm}
       \epsfxsize=16.cm
       \epsfbox{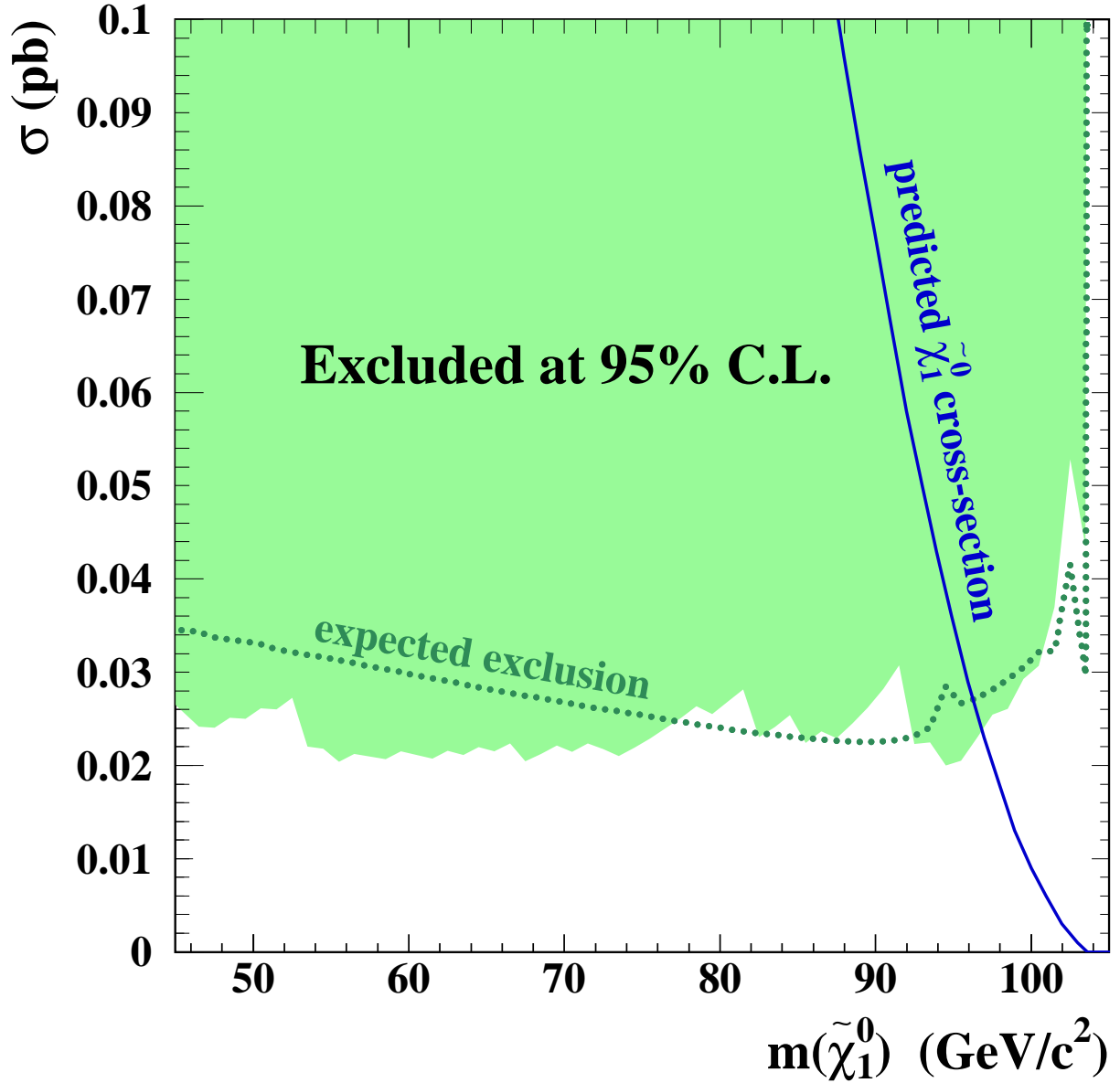}
     \end{minipage}
   \hfill
     \begin{minipage}[t]{.1cm}
       \epsfxsize=.1cm
     \end{minipage}
\caption{
\label{f:acopphoton}
The excluded production cross-section at $\sqrt{s}=208\,$GeV
for neutralino pairs in the neutralino NLSP case is given by the grey region.
The dotted line represents the expected limit.
Both limits are given at 95$\,$\%\ C.L.
The solid line shows the minimum
cross-section for neutralino pairs predicted
by the theory.
Within the model, neutralino masses between 45$\,$GeV$/c^2$ and 96.8$\,$GeV$/c^2$
(96.3$\,$GeV$/c^2$ expected) can be excluded. The result is valid
for promptly decaying NLSPs only.}
\end{figure}

Figure~\ref{f:acopphoton} gives the excluded production cross-section for
neutralino pairs at a centre-of-mass energy of $\sqrt{s}=208$\,GeV.
Cross-sections larger than 0.04$\,$pb can be excluded for masses 
of the short-lived neutralino 
ranging from 45$\,$GeV$/c^2$ up to the kinematic limit.  

\subsubsection{Cross-section limits for slepton pair-production}
To search for pair-produced sleptons in the neutralino NLSP scenario,
two analyses were applied. 
For short NLSP lifetimes with a decay length of the neutralino up to 3$\,$m 
($\tau\le 10^{-8}\,$s) the search for leptons plus
photons and missing energy (Section~\ref{s:gamgamx}) has sensitivity.
Long NLSP lifetimes ($\tau\ge 10^{-8}\,$s) were
covered by the search for acoplanar leptons (Section~\ref{s:accop}).
For each flavour of the slepton an optimized analysis was used.
The overlap of efficiencies between the analyses is of the order of 15\,\%
and a Standard Model background of less than 0.01 events is expected
for the overlap analyses.
For all searches the results from the data recorded at centre-of-mass energies
of 189$\,$GeV to 209$\,$GeV were used.
No indication for new physics in addition to the expected Standard Model
background was observed. 
The datasets at different centre-of-mass energies 
were combined assuming a $\beta^3/s$ dependence of the 
cross-sections. 

\begin{figure}[H]
     \begin{minipage}[t]{.1cm}
       \epsfxsize=.1cm
     \end{minipage}
   \hfill
     \begin{minipage}[t]{16.cm}
\vspace*{1cm}
       \epsfxsize=16.cm
       \epsfbox{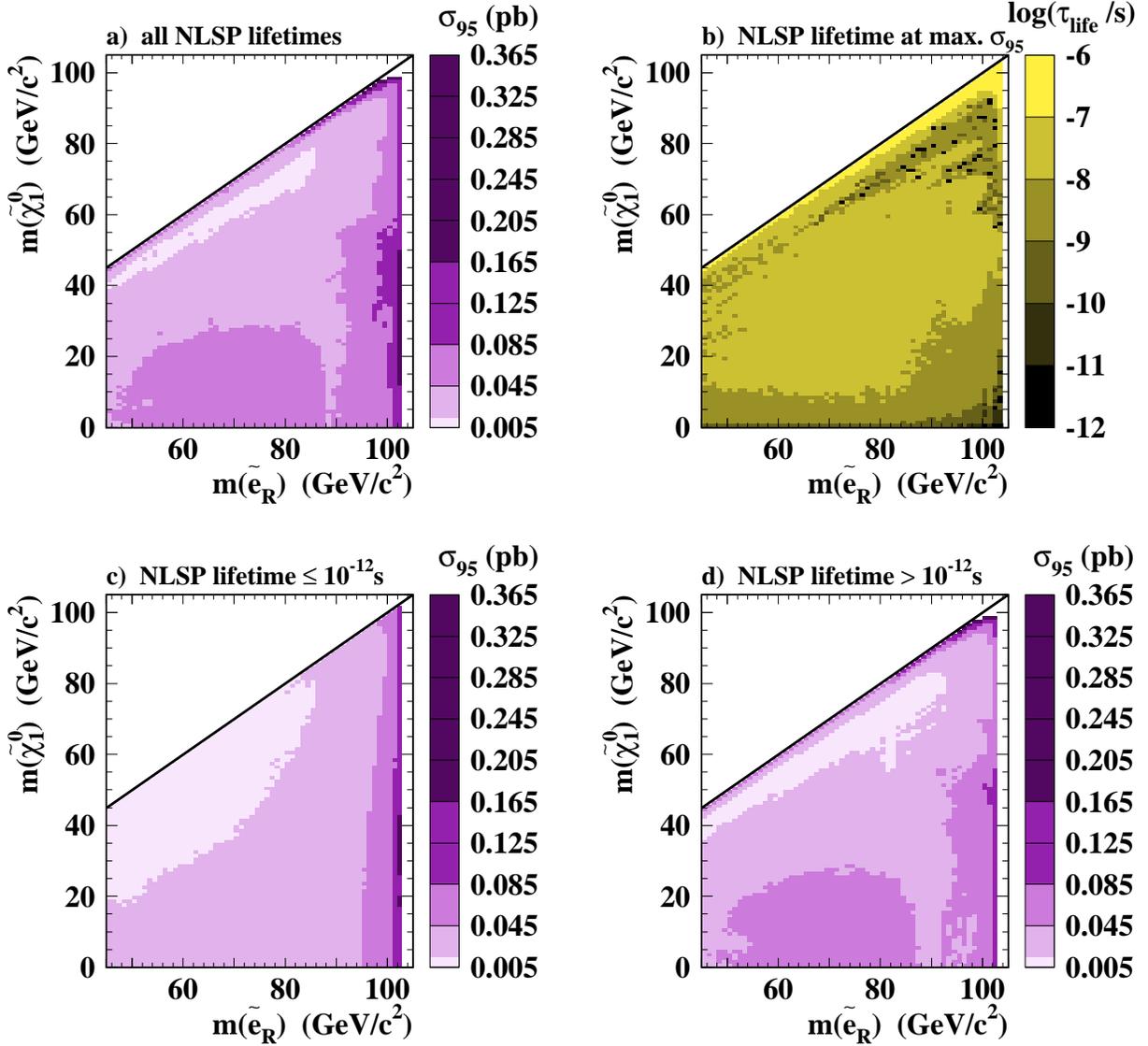}
     \end{minipage}
   \hfill
     \begin{minipage}[t]{.1cm}
       \epsfxsize=.1cm
     \end{minipage}
\caption{
\label{f:selneut}
 Contours of the cross-section limits at 95\,\% C.L. and 
$\sqrt{s}=208\,$GeV for selectron pair-production in the neutralino 
NLSP scenario. The cross-section limits in (a) are valid for any neutralino 
lifetime. Plot (b) shows the lifetimes at which the limit is set.
The 95\,\% C.L. cross-section limits 
for prompt neutralino decays, where only the 
search for photons, missing energy plus leptons/jets contributes, are plotted 
in (c). In (d) the 95\,\% C.L. cross-section limits are shown for 
very long-lived neutralinos, where only the search 
for acoplanar leptons contributes.}
\end{figure}

\begin{figure}[H]
     \begin{minipage}[t]{.1cm}
       \epsfxsize=.1cm
     \end{minipage}
   \hfill
     \begin{minipage}[t]{16.cm}
\vspace*{1.cm}
       \epsfxsize=16.cm
       \epsfbox{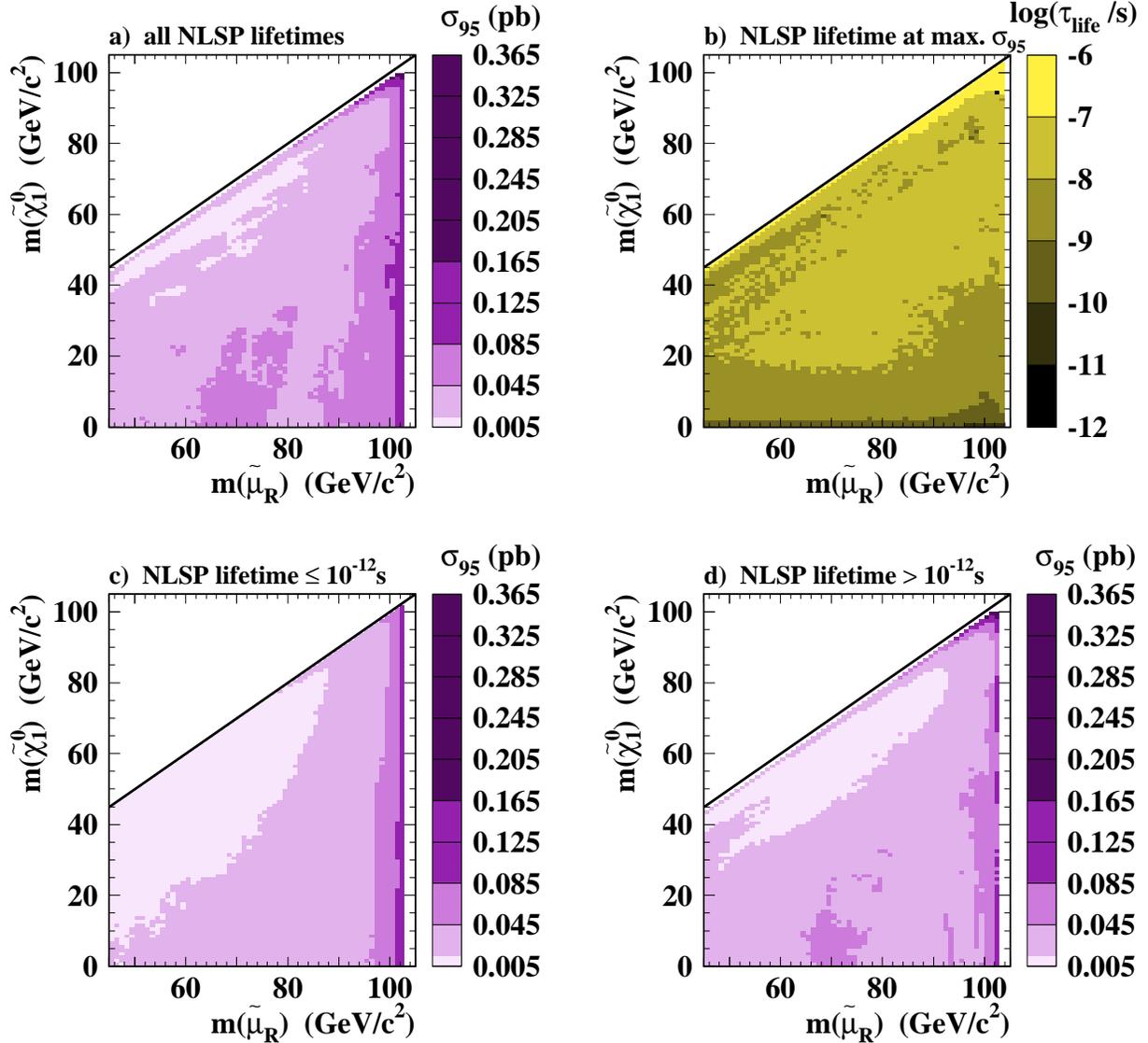}
     \end{minipage}
   \hfill
     \begin{minipage}[t]{.1cm}
       \epsfxsize=.1cm
     \end{minipage}
\caption{
\label{f:smuonneut}
 Contours of the cross-section limits at 95\,\% C.L. and 
$\sqrt{s}=208\,$GeV for smuon pair-production in the neutralino 
NLSP scenario. The cross-section limits in (a) are valid for any neutralino 
lifetime. Plot (b) shows the lifetimes at which the limit is set.
The 95\,\% C.L. cross-section limits 
for prompt neutralino decays, where only the 
search for photons, missing energy plus leptons/jets contributes, are plotted 
in (c). In (d) the 95\,\% C.L. cross-section limits are shown for 
very long-lived neutralinos, where only the search 
for acoplanar leptons contributes.}
\end{figure}

\begin{figure}[H]
     \begin{minipage}[t]{.1cm}
       \epsfxsize=.1cm
     \end{minipage}
   \hfill
     \begin{minipage}[t]{16.cm}
\vspace*{1.cm}
       \epsfxsize=16.cm
       \epsfbox{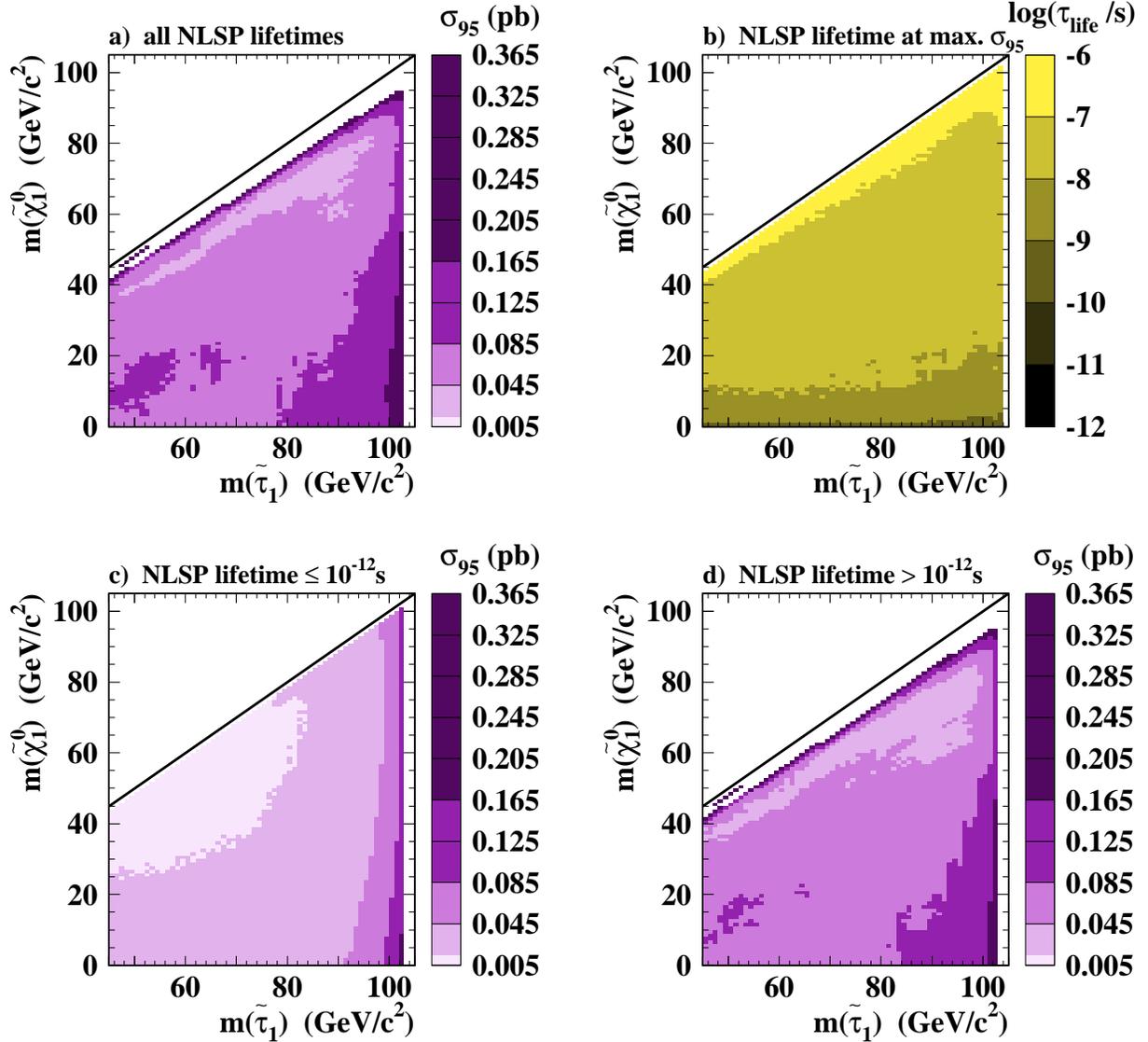}
     \end{minipage}
   \hfill
     \begin{minipage}[t]{.1cm}
       \epsfxsize=.1cm
     \end{minipage}
\caption{
\label{f:stauneut}
 Contours of the cross-section limits at 95\,\% C.L. and 
$\sqrt{s}=208\,$GeV for stau pair-production in the neutralino 
NLSP scenario. The cross-section limits in (a) are valid for any neutralino 
lifetime. Plot (b) shows the lifetimes at which the limit is set. 
The 95\,\% C.L. cross-section limits 
for prompt neutralino decays, where only the 
search for photons, missing energy plus leptons/jets contributes, are plotted 
in (c). In (d) the 95\,\% C.L. cross-section limits are shown for 
very long-lived neutralinos, where only the search 
for acoplanar leptons contributes.}
\end{figure}

Figures~\ref{f:selneut}, \ref{f:smuonneut} and \ref{f:stauneut} 
show the production cross-sections excluded at 95\,\% C.L. for
selectrons, smuons and staus, respectively.
Plot (a) of each figure gives the excluded 
cross-section valid for any NLSP lifetime as a function of the
slepton and the neutralino masses. Values larger than $0.2\,$pb
can be excluded for all slepton flavours, independent of the neutralino mass.
In plot (b) of each figure it is shown that the limit is set at NLSP 
lifetimes where the cross-over of the efficiencies of the two analyses
takes place, shifting toward higher lifetimes with higher NLSP masses.
Plots (c) and (d) of each figure give the excluded production cross-sections
assuming a short-lived or long-lived NLSP. For both
lifetime cases and for all flavours cross-sections higher
than $0.1\,$pb can be excluded.

\subsubsection{Cross-section limits for chargino pair-production}
\label{s:expresults33}
To be sensitive to all possible signatures for pair-produced charginos
in the neutralino NLSP scenario, a large set of analyses had to be combined.
For short NLSP lifetimes, the search for photons and missing energy
plus leptons or jets is applied (Section \ref{s:gamgamx}),
which is split into several selections, optimized for 
low and high track multiplicities in the event. The high multiplicity
selection itself is divided into three parts, optimized
for different mass differences between the chargino and the NLSP.
Long lifetime neutralinos in chargino candidate events are selected by 
the search for acoplanar leptons in case of leptonically decaying W bosons.
For semi-leptonic and hadronic W decays the 
chargino/neutralino searches with and without an identified lepton were 
applied.


For the acoplanar lepton search and for the search for photons and missing
energy plus leptons or jets data at centre-of-mass energies
$\sqrt{s}=189$--$209$\,GeV were analyzed.
For the chargino searches with a semi-leptonic and hadronic W decay
and a long-lived neutralino, data at $\sqrt{s}=192$--$209$\,GeV were used.
No indication for new physics was found for any of the analyses.
Limits on the production cross-section for charginos were computed,
with the results at different centre-of-mass energies combined 
assuming a $\beta/s$ dependence of the cross-sections and a pure
decay of the chargino via a W boson.

\begin{figure}[H]
     \begin{minipage}[t]{.1cm}
       \epsfxsize=.1cm
     \end{minipage}
   \hfill
     \begin{minipage}[t]{16.cm}
\vspace*{1.cm}
       \epsfxsize=16.cm
       \epsfbox{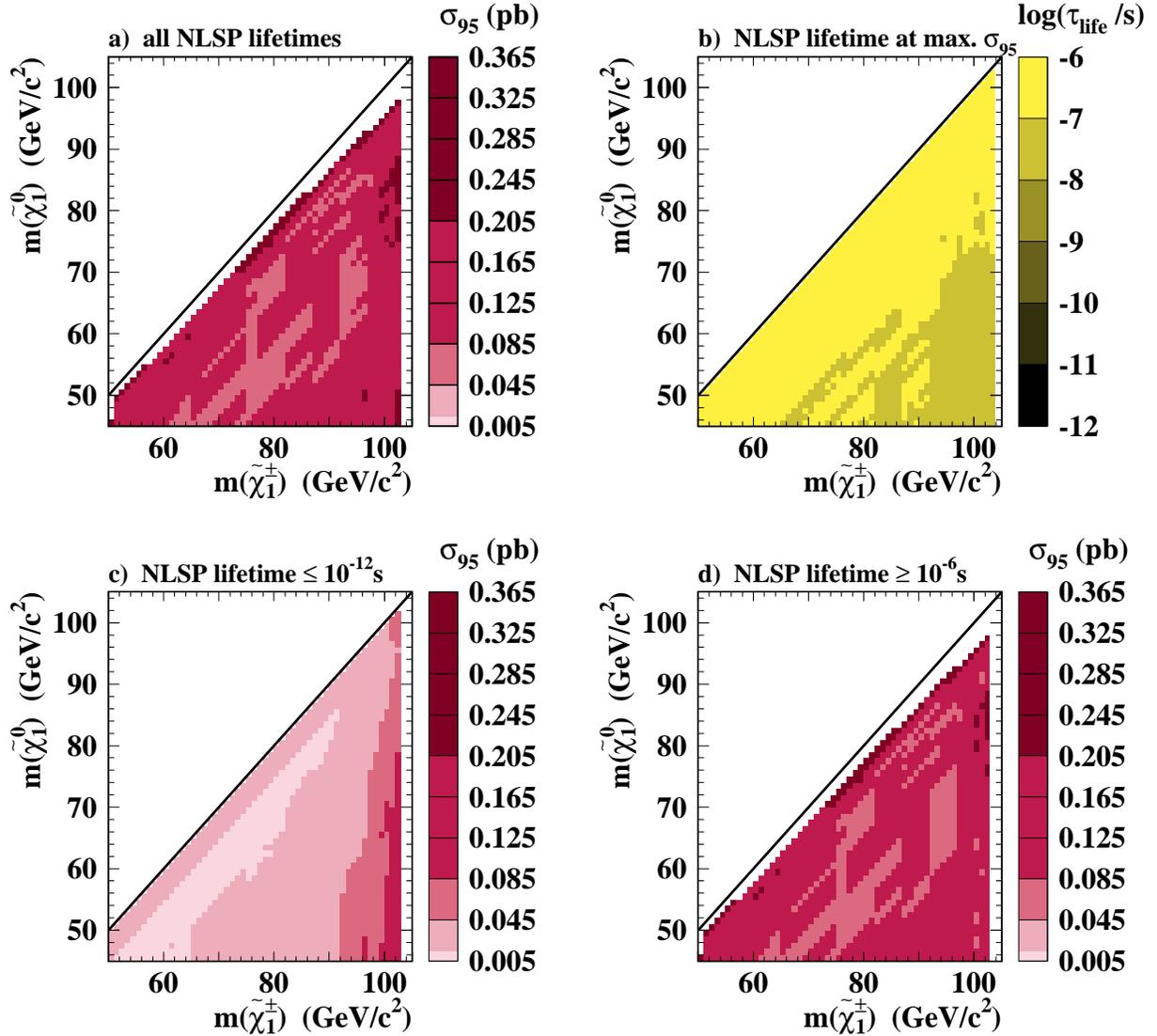}
     \end{minipage}
   \hfill
     \begin{minipage}[t]{.1cm}
       \epsfxsize=.1cm
     \end{minipage}
\caption{
\label{f:chargneut}
 Contours of the cross-section limits at 95\,\% C.L. and 
$\sqrt{s}=208\,$GeV for chargino 
pair-production in the neutralino NLSP scenario
assuming a 100\% branching faction for 
$\rm \tilde\chi^\pm \rightarrow W^{\pm *} \tilde\chi^0$.
The 
cross-section limits in (a) are valid for any neutralino lifetime. Plot (b) 
shows the NLSP lifetimes at which the maximum excluded cross-section 
is found. The results for prompt neutralino decays, 
where mainly the searches for photons and missing energy plus leptons/jets 
contribute, are plotted in (c). In (d) the limits are 
shown for very long-lived neutralinos.}
\end{figure}

Figure~\ref{f:chargneut} gives the excluded production cross-section 
at 95\,\% C.L. and for $\sqrt{s}=208$\,GeV for chargino pairs.
The limits in (a) are valid for all NLSP lifetimes; cross-sections larger
than $0.3\,$pb can be excluded for most chargino and NLSP masses.
As can be seen in (b), this limit is set at NLSP lifetimes where the
cross-over of the sensitivity of the lifetime analyses takes place, shifting
toward higher lifetimes with higher NLSP mass
(lower $\beta\gamma$ of the NLSP).
Figures (c) and (d) give the excluded production cross-section
assuming a short-lived or long-lived NLSP. For short-lived NLSPs,
cross-sections higher than $0.2\,$pb, for long-lived NLSPs,
cross-sections higher than $0.3\,$pb can be excluded.

\section{Interpretations in the framework of the GMSB model}
\label{s:interpretation}

\subsection{The GMSB scan database}
Interpreting the experimental results in terms of the GMSB model requires
a comparison with the theoretical expectations within the framework of
the model. Chosen here is the minimal version of the GMSB model
with five parameters and a sign in addition to the SM
parameters. The new parameters are the SUSY breaking scale, $\sqrt{F}$, the
messenger scale, $M$, the messenger index, $N$, the
ratio of the vacuum expectation values of the two
Higgs doublets, $\tan{\beta}$, the sign of the Higgs sector mixing
parameter, sign($\mu$), and the mass scale $\Lambda$, which determines the
SUSY particle masses at the messenger scale.
A certain point in the parameter space of
the model is excluded if the experimental upper limit on
the cross-section at $\sqrt{s}=208$~GeV discussed in Section~\ref{s:expresults}
is less than the expected cross-section
$\sigma$ at $\sqrt{s}=208$~GeV, taking into account 
the branching ratio BR ($\sigma\times$BR$^2$ for
pair-produced particles decaying to the same final state).

The parameter of the SUSY breaking scale $\sqrt{F}$ is eliminated
as experimental upper limits were calculated
on the production cross-section of SUSY
particles independent of the NLSP lifetime (Equation~\ref{eq:lifetime}).
In the phase space of the remaining parameters a scan was performed
to calculate the complete mass spectrum, production cross-sections and
branching ratios for different SUSY particles at each point considered.
For this scan the framework and formulae of~\cite{theo3}
were used and generalized to include a full mass treatment for all three
generations. The calculations are embedded in the SUSYGEN
generator.

\renewcommand{\arraystretch}{1.1}
\begin{table}[ht!]
\begin{center}
\begin{tabular}{|c|c|c|}
\hline
Parameter     & Scan points & Step size \\\hline\hline
$\Lambda$     & 5 -- $150\,$TeV/$c^2$ & 1\,TeV/$c^2$\\
$\tan{\beta}$ & 1 -- 50 & 0.2 \\
$M$           & $1.01\cdot\Lambda$, $250\,$TeV/$c^2$, $10^6\,$TeV/$c^2$ & \\
$N$           & 1, 2, 3, 4, 5 & \\
sign($\mu$)   & $+1$, $-1$ & \\
$M_{\rm top}$ & 175 GeV$/c^2$    & \\\hline
\end{tabular}
\caption{
\label{t:scan}
Scanned points in the GMSB parameter space. The parameter
$\Lambda$ sets
the overall mass scale of the SUSY particles, $\tan{\beta}$ is the
ratio of the vacuum expectation values of the two Higgs doublets, $M$
is the messenger scale, $N$ the messenger index, and
sign($\mu$) is the sign of the Higgs sector mixing parameter.
}
\end{center}
\end{table}

The model parameters as well as the range and step size considered for
them are summarized in Table~\ref{t:scan}.
The messenger scale $M$ is arbitrary in the
minimal model, but, as the mass $m_b$ of the messenger bosons is given by
$m_b=M\sqrt{1\pm\Lambda/M}$, the relation $M>\Lambda$ has to be fulfilled in
order to obtain a positive messenger boson mass squared. Both models with
$M\sim\Lambda$ and $M\gg\Lambda$ are viable; therefore, three scenarios
for the messenger scale were studied: $M$ very close to $\Lambda$
($M=1.01\cdot\Lambda$),
$M=250\,$TeV/$c^2$ and $M$ very large ($10^6$\,TeV/$c^2$). For the integer
parameter $N$, values up to five were considered. This is adequate, since
perturbativity of the gauge interactions up to the grand unification scale,
$M_{\mathrm{GUT}}$, implies $N\,\stackrel{<}{\sim}\,150\,/\ln\frac{M_{\mathrm{GUT}}}{M}$~\cite{theo1}.
Thus, for a messenger mass scale $M=100\,$TeV/$c^2$, $N\le 5$ is required.
Both signs of the parameter $\mu$ were considered.
For each of the 30 combinations of $N$, $M$ and sign$(\mu)$, a scan was
performed in
$\Lambda$ and
$\tan{\beta}$.
The upper and lower limits for this scan were
chosen according to the following considerations.
For $\Lambda>150\,$TeV/$c^2$ the Supersymmetric particles
are already very heavy and cannot be produced at LEP energies.
Both the regions with
$\tan{\beta}>50.0$ and $\Lambda<5\,$TeV/$c^2$
are theoretically forbidden. The exact shape of the theoretically inaccessible
region in the $\Lambda-\tan{\beta}$ plane depends on the other parameters.

The scan information was saved in a database for comparison with the
experimental results. 
In total, 270 such data sets were produced:
30 (with different $N$, $M$ and sign$(\mu)$) 
for each centre-of-mass energy and for nine different energy
points with $\sqrt{s}= 182.7$, $188.7$, $191.6$, $195.5$, $199.5$, $201.6$,
$205.1$, $206.7$ and $208.1\,$GeV.

A gravitino mass of $2\,$eV/$c^2$ was chosen,
corresponding to a SUSY
breaking scale of $\sqrt{F}\approx 100\,$TeV/$c^2$. 
This is
motivated by the requirement that the branching ratio of the next-to-NLSP to
the gravitino is small and only the NLSP decays to the gravitino. As long as
this is fulfilled, the cross-sections and branching
ratios do not depend on the gravitino mass. This makes it possible to 
decouple
the issue of NLSP lifetime, which depends on $\sqrt{F}$, from the
scan.  Note that $\sqrt{F}$ does not have a large effect on the
other sparticle masses.

\subsection{Direct constraints on the NLSP mass (stau or slepton)}
\label{s:const1}

Constraints on the NLSP masses were determined
from the cross-section limits obtained for NLSP pair-production in the 
slepton co-NLSP scenario and stau pair-production in the stau NLSP scenario
(Section~\ref{s:expresults11}).
This was done by comparing
the excluded cross-section at $\sqrt{s}=208$~GeV
with the production cross-section
predicted by the theory.
For pair-produced particles the expected cross-section times the branching
ratio squared, $\sigma\cdot \mathrm{BR^2}$, 
within the model has to be known, but naturally this varies
strongly within the model, depending on the parameter set chosen.
To obtain values which are valid for each of the parameter sets considered,
the following, conservative, minimization procedure was applied.

For each of the 30 parameter combinations of $M$, $N$ and sign$(\mu)$,
a scan over the parameters
$\Lambda$ and $\tan{\beta}$ was performed. In the regions
where the sleptons or stau are the NLSP
the minimum $\sigma\cdot \mathrm{BR^2}$ for each NLSP mass 
is determined within the parameter set considered.
For selectrons strong variations
were found because of additional positive and negative 
interfering $t$-channel production which
contributes differently for the various parameter sets.
This is in constrast to selectron searches in gravity mediated
SUSY models in which the selectron cross-section is always enhanced by
an interfering $t$-channel contribution. 
Finally for each NLSP mass the minimum of $\sigma\cdot \mathrm{BR^2}$ within
these 30 scenarios was calculated and the resulting minimal expected
cross-sections were compared to the experimentally achieved limit.

\begin{figure}[H]
     \begin{minipage}[t]{.1cm}
       \epsfxsize=.1cm
     \end{minipage}
   \hfill
     \begin{minipage}[t]{16.cm}
\vspace*{1.cm}
       \epsfxsize=16.cm
       \epsfbox{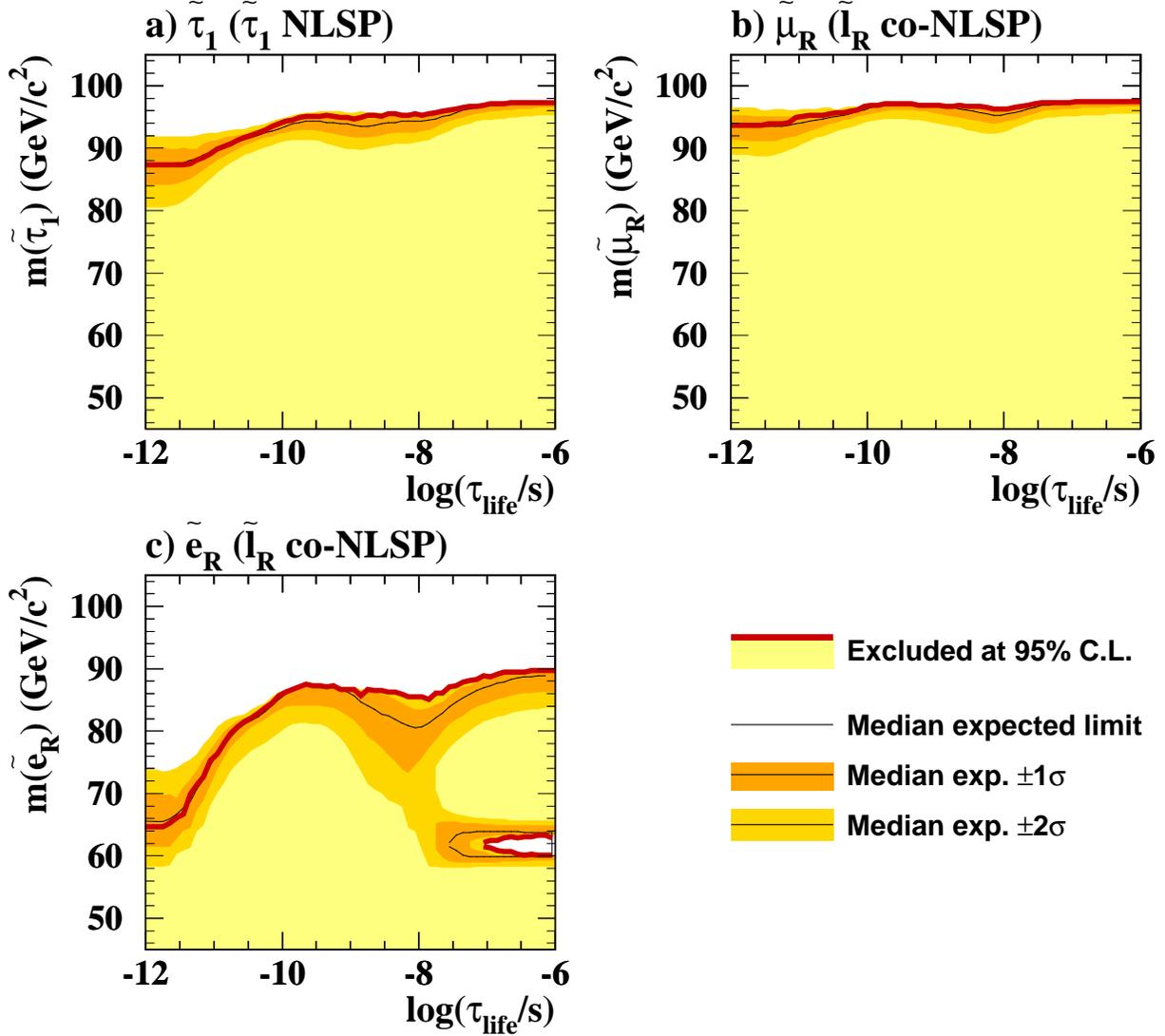}
     \end{minipage}
   \hfill
     \begin{minipage}[t]{.1cm}
       \epsfxsize=.1cm
     \end{minipage}
\caption{
\label{f:slepmass}
{ The observed lower mass limits for 
pair-produced 
staus in the stau NLSP (a) and smuons (b), selectrons (c)
in the slepton co-NLSP scenario as a function of the 
particle lifetime using the direct $\slept^+\slept^-$ search.
For staus in the slepton co-NLSP scenario the observed and expected
lower limit are identical to the limits of the stau in the stau NLSP scenario.
The mass limits are valid for a messenger index N$\leq 5$. 
For the stau NLSP and slepton co-NLSP scenarios, the
NLSP mass limits are set by the stau mass limit
($m_{\rm NLSP}>87.4\,$GeV$/c^2$ (a)) and by the
smuon mass limit ($m_{\rm NLSP}>93.7\,$GeV$/c^2$ (b)), respectively.
}
}
\end{figure}

Figure~\ref{f:slepmass} shows the mass limits for staus in the stau
NLSP scenario (a), and smuons and selectrons in the slepton co-NLSP
scenario (b,c) as a function of their lifetime.
For each 
slepton flavour, the lowest mass constraints are found for
very short slepton lifetimes, with the exception of the selectrons, where the 
region around 60\,GeV$/c^2$ cannot be excluded at long 
lifetimes. This is due to the effect of similar values of the 
ionization energy loss of heavy and light charged particles for a region
of momenta around 65$\,$GeV$/c$, thus a decreased efficiency of the
searches using the d$E$/d$x$ measurements.
In the stau NLSP 
scenario, staus with masses below $87.4\,$GeV$/c^2$ 
are excluded at 95\,\% C.L. for an expected limit from the background-only
hypothesis of
$m_{\tilde{\tau}_{\rm 1}}>87.6\,$GeV$/c^2$.
In the slepton co-NLSP scenario 
the directly excluded selectron, smuon and stau masses (and expected limits),
valid for any slepton lifetime, are
$m_{\tilde{\rm e}_{\rm R}}>60.1\,$GeV$/c^2$ ($60.0\,$GeV$/c^2$), 
$m_{\tilde{\mu}_{\rm R}}>93.7\,$GeV$/c^2$ ($93.6\,$GeV$/c^2$) and 
$m_{\tilde{\tau}_{\rm 1}}>87.4\,$GeV$/c^2$ ($88.2\,$GeV$/c^2$).
In the slepton co-NLSP scenario the sleptons are mass-degenerate;
their mass difference is -- by definition -- 
less than the $\tau$, $\mu$ or $\rm e$ mass, so that decays 
$\tilde{\ell}'\rightarrow\ell'\tilde{\ell}$ are forbidden.
Therefore the highest mass limit, the limit on the smuon mass of 
$m_{\tilde{\mu}_{\rm R}}>93.7\,$GeV$/c^2$ can be used to deduce an indirect 
common mass limit on all sleptons of
$m_{\tilde{\ell}}=m_{\tilde{\mu}_{\rm R}} - M_{\tau}>91.9\,$GeV$/c^2$ within
the slepton co-NLSP scenario.

No direct mass limits were calculated for SUSY particles heavier than the NLSP.
The expected production cross-section for neutralinos, charginos,
heavy selectrons or smuons depends strongly on the model parameters
and is suppressed in some regions of the
parameter space. Thus the minimization method gives no sensible 
general expectation which can be compared to the experimental results.

\subsection{Direct constraints on the NLSP mass (short-lived neutralino)}
For the neutralino NLSP, the same minimization method 
described in the previous section (\ref{s:const1}) was applied
to obtain an expected production cross-section valid at each
scan point.
In Figure~\ref{f:acopphoton} this prediction
is compared
to the experimentally excluded production cross-section for
neutralino pairs at a centre-of-mass energy of $\sqrt{s}=208$\,GeV.
Neutralino masses less than 96.8$\,$GeV/$c^2$ can be
excluded at 95$\,$\% C.L. for an expected limit from the background-only
hypothesis of 96.3$\,$GeV/$c^2$. The limit is valid for short-lived
neutralinos with lifetimes up to $10^{-9}\,$s only.

In the neutralino NLSP scenario no direct mass limits were computed for
charginos and sleptons.
The production cross-sections of these particles depend strongly
on the model parameters and reach zero in some parts of the GMSB
parameter space. Thus no general minimal expectation can be computed
to compare to the experimental results.

\subsection{Exclusions within the GMSB parameter space}
For each of the 30 sets of the GMSB parameters $N$, $M$ and 
sign$(\mu)$ considered in the scan, 
the exclusion in the $\Lambda$--$\tan\beta$ plane was studied. 
At each point in the plane, the cross-sections for the 
various channels as well as their branching ratios are known. A point in the 
parameter space is excluded if it is kinematically accessible and the 
expected cross-section in at least one channel is higher than the
experimentally derived 95\,\% C.L. cross-section limit in this channel,
taking into account the branching ratio(s).

\begin{figure}[H]
     \begin{minipage}[t]{.1cm}
       \epsfxsize=.1cm
     \end{minipage}
   \hfill
     \begin{minipage}[t]{15.cm}
\vspace*{-1cm}
       \epsfxsize=15.cm
       \epsfbox{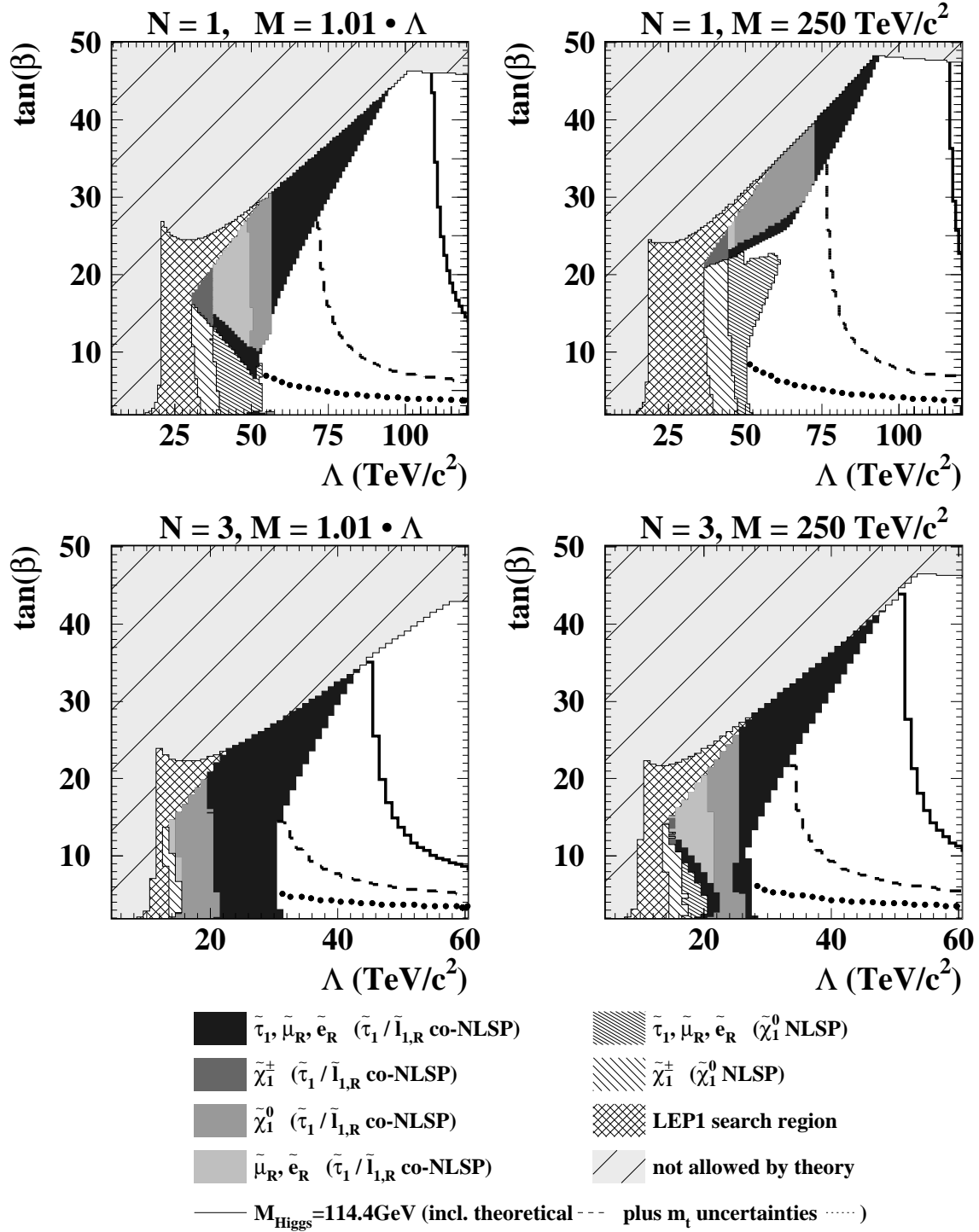}
     \end{minipage}
   \hfill
     \begin{minipage}[t]{.1cm}
       \epsfxsize=.1cm
     \end{minipage}
\caption[]{
\label{f:indirlimn} 
Examples of excluded regions in the $\Lambda-\tan{\beta}$ plane
excluded by pair-production searches for different particles,
with sign$(\mu)>0$ and valid for any NLSP lifetime
for four different sets of parameters, $N=1$ or $3$ and $M=1.01\cdot\Lambda$
or $250\,$TeV/$c^2$.
}
\end{figure}

As examples, in Figure~\ref{f:indirlimn} the excluded regions in the 
$\Lambda$--$\tan{\beta}$ plane, valid for any NLSP lifetime, are shown for 
$N=1$ and $3$, $M=1.01\cdot\Lambda$ and $250\,$TeV/$c^2$ and sign($\mu$)$>0$.
The shaded or hatched regions
correspond to different search channels, indicating the relevance 
of the various analyses in the parameter space.

In general, the stau NLSP and slepton co-NLSP scenarios
become more important
as number of messenger sets $N$ increases. Searches for
direct NLSP pair-production exclude almost completely the accessible 
region of the stau NLSP and slepton co-NLSP scenarios.
Only a few additional points are excluded by the neutralino searches,
for example regions with $\tan{\beta}\approx 12$ and
$\Lambda\approx 55\,$TeV/$c^2$ for $N=1$, $M=1.01\cdot\Lambda$ in
Figure~\ref{f:indirlimn}.

The neutralino NLSP scenario
plays an important role only for $N=1$ and in regions with high $M$ for $N=2$.
In this scenario, a large fraction of the parameter space can be excluded
by searches for slepton pair-production. Searches for
chargino pair-production exclude additional points
in regions with low $\tan{\beta}$ and high $M$.
In the chargino channel a complication arises due to the fact that the 
chargino has two decay modes which can lead to the same final state:
\begin{itemize}
\item[(1)] $\mathrm{e^+e^-\rightarrow\tilde{\chi}_1^+\,\tilde{\chi}_1^-
\rightarrow\tilde{\chi}_1^0\,W^{+*}\,\tilde{\chi}_1^0\,W^{-*}}
\rightarrow \tilde{\chi}_1^0 \ell^+\nu \tilde{\chi}_1^0 \ell^-\bar{\nu}
$,
\item[(2)] $\mathrm{e^+e^-\rightarrow\tilde{\chi}_1^+\tilde{\chi}_1^-
\rightarrow\tilde{\ell}^+\nu\,\tilde{\ell}^-\bar{\nu}\rightarrow\ell^+\nu\,
\tilde{\chi}_1^0\,\ell^-\bar{\nu}\,\tilde{\chi}_1^0}$\,.
\end{itemize}
Since in the scan for the branching ratios of the different final states no 
information about the decay chain is available, it is impossible to decide 
if the branching ratio to the $\ell^-\bar\nu\,\ell^+\nu$ 
final state is due to leptonic W decays (1), or due to the direct decay 
mode (2). For the interpretation, only results from the decay
channel (1) were used. 
The analyses sensitive to the leptonic final state are the search for 
acoplanar leptons in case of long neutralino lifetimes and the search for
photons, missing energy plus leptons for short lifetimes.
It was checked that these analyses 
have a similar efficiency for both chargino decay modes that lead to the 
leptonic final state. The branching ratio to $\ell\nu$ was 
treated as if it was purely due to channel (1). Since the efficiency for (2) 
is the same, even if this branching ratio was totally due to decay 
mode (2), the cross-section limits would not be affected. 

In general, the 
production cross-section for charginos is larger than that for 
sleptons in the regions excluded by this channel. 
Nevertheless, the slepton searches
exclude larger regions
than 
chargino searches because in GMSB models
charginos are mostly heavier than the 
(right-handed) sleptons. 
Thus, the charginos are kinematically limited and higher values of $\Lambda$ 
can be reached by the slepton searches. 

\subsection{Constraints on the SUSY particle mass scale $\Lambda$}
From the exclusions in the $\Lambda$--$\tan{\beta}$ plane 
for each of the 30 sets of the GMSB parameters $N$, $M$ and 
sign$(\mu)$ considered in the scan, lower limits were inferred
on the SUSY
particle mass scale $\Lambda$,
independent of $\tan{\beta}$.
For fixed $N$, the parameter $\Lambda$
determines the GMSB particle spectrum at the messenger scale, since the 
gaugino masses are given by
\begin{eqnarray*}
m_{\lambda_i}(M)\sim N\cdot\Lambda\frac{\alpha_i(M)}{4\pi}\,,
\end{eqnarray*}
where $\lambda_i$ are the gaugino fields of gauge group $i$ and the 
$\alpha_i$ are the GUT-scale
normalized coupling constants of these gauge groups. 
Also the scalar masses at the messenger scale are determined by~$\Lambda$:
\begin{eqnarray*}
m^2(M)\sim 2N\cdot\Lambda^2\sum_{i=1}^3{k_i\left(\frac{\alpha_i(M)}{4\pi}\right)^2}\,,
\end{eqnarray*}
where the sum is over the gauge groups $SU(3)_C$, $SU(2)_L$ and $U(1)_Y$,
and $k_i$ are constants of $\mathcal{O}$(1).

The results, valid for all NLSP lifetimes and thus independent of the
SUSY breaking scale $\sqrt{F}$, are summarized in Table~\ref{t:lambda}.
It can be seen that the constraints on $\Lambda$ decrease with larger $N$
and lower limits of typically $40\,$TeV/$c^2$ are found
for $N=1$ and $15\,$TeV/$c^2$
for $N=5$.
The constraints depend on $M$, but are almost independent of sign$(\mu)$.
The limits are set at values of $\tan{\beta}$ between 5 and 25; 
for higher
and lower $\tan{\beta}$ values they get larger.
In conclusion, constraints on $\Lambda$ of 
$\Lambda>40,\,27,\,21,\,17,\,15$\,TeV/$c^2$ were 
derived for $N=1,\,2,\,3,\,4,\,5$, respectively, for all $M$, $\tan{\beta}$,
sign$(\mu)$ and all NLSP lifetimes (all values of $\sqrt{F}$).

The constraints on $\Lambda$ imply a lower limit on the neutralino mass in
the neutralino NLSP scenario, independent of the neutralino lifetime.
Masses below 53.5$\,$GeV/$c^2$ for N=1 up to 94.0$\,$GeV/$c^2$ for N=5 
can be excluded for all neutralino lifetimes.

\begin{table}[ht!]
\begin{center}
\begin{tabular}{|l|l||c|c|c|c|c|}
\hline 
\multicolumn{7}{|c|}{Lower limit at 95\,\% C.L. on $\Lambda$ [TeV/$c^2$] for all NLSP lifetimes} \\\hline
\multicolumn{2}{|c||}{}           & $N=1$    & $N=2$    & $N=3$    & $N=4$    & $N=5$            \\\hline\hline
                        & $\mu<0$ & {\bf 40} & {\bf 27} & {\bf 21} & {\bf 17} & {\bf 15}         \\
\raisebox{1.5ex}[-1.5ex]{High $M$ ($M=10^6\,$TeV$/c^2$)} 
                        & $\mu>0$ & 43       & {\bf 27} & {\bf 21} & {\bf 17} & {\bf 15}         \\\hline
                        & $\mu<0$ & 49       & 31       & 26       & 22       & 20               \\
\raisebox{1.5ex}[-1.5ex]{Medium $M$ ($M=250\,$TeV$/c^2$)} 
                        & $\mu>0$ & 49       & 34       & 26       & 22       & 20               \\\hline
                        & $\mu<0$ & 54       & 37       & 30       & 25       & 22               \\
\raisebox{1.5ex}[-1.5ex]{Low $M$ ($M=1.01\cdot\Lambda$)} 
                        & $\mu>0$ & 52       & 37       & 30        & 25      & 22               \\\hline 
\end{tabular}
\end{center}
\caption{
\label{t:lambda}
Lower limits at 95\,\% C.L. on the SUSY particle mass scale $\Lambda$, for 
various sets of the GMSB parameters $M$, $N$ and sign$(\mu)$, valid
for all NLSP lifetimes. 
These constraints use only the direct searches for SUSY particle
production.
The minimum
value of $\Lambda$ for each $N$ is highlighted.}
\end{table}

\subsection{Impact of searches for the neutral Higgs boson}

The limit on the Standard Model Higgs boson 
reached by the four LEP experiments~\cite{higgs} can be used to place
additional constraints on the GMSB parameter space.
This is justified as in GMSB there is almost no parameter space in which the
Higgs--Z--Z coupling is suppressed compared to its Standard Model
value. Also, in GMSB the branching ratio of $H\ra{\rm b \bar b}$,
the most important channel in the SM Higgs search,
shows no strong dependence on $\tan{\beta}$ 
and is suppressed by at most 10$\,$\% in a few regions
of the phase space \cite{Ambrosanio:2001xb}.

In Figure~\ref{f:indirlimn}, in addition to the regions excluded by 
direct sparticle searches, constraints from the LEP combined
Higgs limit of 114.4~GeV$/c^2$ are also shown.  
The Higgs constraints initially appear to be much stronger than those
from the direct SUSY particle searches; however, 
for a given set of GMSB parameters, the theoretical 
uncertainty on the inferred Higgs mass is about 3~GeV$/c^2$,
and the Higgs mass uncertainty due to the uncertainty on
$M_{\rm top}$ is about 5~GeV$/c^2$ \cite{Ambrosanio:2001xb}.
The effect of these uncertainties
is also illustrated in Figure~\ref{f:indirlimn}, where it 
can be seen that the Higgs constraints can be rather weak
when the full uncertainty is taken into account.
Because of these large uncertainties, constraints on
the GMSB parameter space from the Higgs search are not 
included in our quoted limits.

\section{Conclusions}
\label{s:conclusions}
Searches have been performed for topologies predicted by GMSB models.
All possible lifetimes of the NLSP which is either the lightest neutralino
or a slepton, have been considered.

No evidence for new physics has been found in the OPAL data sample collected
at centre-of-mass energies $\sqrt{s}=189$--$209\,$GeV.
For the first time limits are presented on the production cross-section
for all search topologies.
The impact of the searches on the minimal GMSB model has been tested
by performing a scan over its parameters.
NLSP masses
below 53.5$\,$GeV/$c^2$ in the neutralino NLSP scenario,
below 87.4$\,$GeV/$c^2$ in the stau NLSP scenario and below 91.9$\,$GeV/$c^2$
in the slepton co-NLSP scenario can be excluded at 95$\,$\% C.L.
for all lifetimes of the NLSP.
The scan gives constraints on the universal SUSY mass scale $\Lambda$ 
from the direct searches for SUSY particle production of 
$\Lambda>40,\,27,\,21,\,17,\,15$\,TeV/$c^2$ for $N=1,\,2,\,3,\,4,\,5$
for all NLSP lifetimes. 

\section{Acknowledgements}

We particularly wish to thank the SL Division for the efficient operation
of the LEP accelerator at all energies
 and for their close cooperation with
our experimental group.  In addition to the support staff at our own
institutions we are pleased to acknowledge the  \\
Department of Energy, USA, \\
National Science Foundation, USA, \\
Particle Physics and Astronomy Research Council, UK, \\
Natural Sciences and Engineering Research Council, Canada, \\
Israel Science Foundation, administered by the Israel
Academy of Science and Humanities, \\
Benoziyo Center for High Energy Physics,\\
Japanese Ministry of Education, Culture, Sports, Science and
Technology (MEXT) and a grant under the MEXT International
Science Research Program,\\
Japanese Society for the Promotion of Science (JSPS),\\
German Israeli Bi-national Science Foundation (GIF), \\
Bundesministerium f\"ur Bildung und Forschung, Germany, \\
National Research Council of Canada, \\
Hungarian Foundation for Scientific Research, OTKA T-038240, 
and T-042864,\\
The NWO/NATO Fund for Scientific Research, the Netherlands.\\

\label{s:biblio}
\def\Journal#1#2#3#4{{#1} {\bf #2} (#3) #4}

\begin{figure}[H]
     \begin{minipage}[t]{.1cm}
       \epsfxsize=.1cm
     \end{minipage}
   \hfill
     \begin{minipage}[t]{15.5cm}
       \epsfxsize=15.5cm
       \epsfbox{pr409_02.epsi}
     \end{minipage}
   \hfill
     \begin{minipage}[t]{.1cm}
       \epsfxsize=.1cm
     \end{minipage}
\caption{
\label{f:licuts}
{
Distributions of variables in the search for charged NLSPs with medium
lifetime (tracks with large impact parameters)
before a specific selection cut.
The particular examples
were taken from the search for neutrino pair-production,
 Section~\ref{s:largeip3}.
The data, taken at centre-of-mass energies of 189--209$\,$GeV are represented
by dots, the simulated background from Standard Model sources by 
a shaded histogram. 
Plot (a) shows the distribution of the transverse momentum
$p_{\rm T}$ of the primary tracks before cut$\,$(3).
Events with at least two primary tracks with a momentum $p_{\rm T}$
greater than 1.0$\,$GeV/$c$ are selected. The value is indicated
by a hatched line and arrow.
Plot (c) gives the invariant mass $W_{\pi\pi}$ of the secondary tracks
assuming a decay into two charged pions before
cut$\,$(5). The arrow gives the region which is selected
by the analysis.
In (b) and (d) the solid and dashed lines indicate the distributions
with arbitrary normalization
for an expected signal which
corresponds to ($\ntone$, $\stau_1^\pm$) mass combinations of (102, 100)
and (50, 45)$\,$GeV$/c^2$.}
}
\end{figure}

\begin{figure}[H]
     \begin{minipage}[t]{.1cm}
       \epsfxsize=.1cm
     \end{minipage}
   \hfill
     \begin{minipage}[t]{15.5cm}
       \epsfxsize=15.5cm
       \epsfbox{pr409_03.epsi}
     \end{minipage}
   \hfill
     \begin{minipage}[t]{.1cm}
       \epsfxsize=.1cm
     \end{minipage}
\caption{
\label{f:kinkcuts}
{
Distributions of variables in the search for charged NLSPs with medium
lifetime (kinked tracks) before 
a specific selection cut.
The particular examples were taken from the search for direct NLSP
pair-production, Section~\ref{s:kink2}.
The data, taken at centre-of-mass energies of 189--209$\,$GeV are represented
by dots, the simulated background from Standard Model sources by 
a shaded histogram. In (b, d) the solid, dashed and dotted lines indicate
the distributions expected from a pair-produced stau with masses of 65, 80
and 95$\,$GeV$/c^2$.
Plots (a, b) show the distribution of the transverse momentum
$p_{\rm T}$ of the primary tracks before cut$\,$(6). 
Plots (c, d) give the invariant mass W$_{00}$ of the primary track
assuming a decay into two massless particles before
cut$\,$(7).
The arrows indicate the regions which are selected
by the analysis.}
}
\end{figure}

\clearpage


\clearpage


\clearpage


\clearpage 


\end{document}